\documentclass[aps,nofootinbib,longbibliography,prd,onecolumn]{revtex4}
\usepackage[a4paper,top=2cm,bottom=2cm,left=2cm,right=2cm]{geometry}
\usepackage[linktocpage=true,colorlinks,citecolor=blue,linkcolor=blue,urlcolor=blue]{hyperref}
\usepackage{graphicx}
\usepackage{xcolor}
\usepackage{amsmath,amssymb,amsfonts}
\usepackage{mathtools,mathrsfs}
\usepackage[applemac]{inputenc}
\usepackage{enumerate}
\usepackage{dsfont}
\usepackage{tabularx}
\usepackage{float}
\usepackage{comment}
\usepackage[default]{gillius}
\usepackage{MyCommands}
\usepackage[
 format = hang, 
 justification = RaggedRight, 
 textfont = it, 
 figurename = Figure,
 ]{caption}
\usepackage{setspace}
\usepackage[normalem]{ulem}
\usepackage[babel]{csquotes}

\def\Q{\mathbb{Q}}
\def\T{\mathbb{T}}
\def\M{\mathcal{M}}

\def\lie{\mathcal{L}}
\def\sol{\textsf{sol}}

\makeatletter
\renewcommand{\thetable}{\@arabic\c@table}
\makeatother

\newcommand\SKD{\bgroup\markoverwith
{\textcolor{orange}{\rule[.5ex]{2pt}{0.4pt}}}\ULon}

\newcommand{\bb}[1]{\overline{\overline{#1}}}

\begin{document}
\title{Black holes in $f(\mathbb Q)$ Gravity}
\author{Fabio D'Ambrosio}\email{fabioda@phys.ethz.ch}
\author{Shaun D.B. Fell}\email{shfell@student.ethz.ch}
\author{Lavinia Heisenberg}\email{lavinia.heisenberg@phys.ethz.ch}
\author{Simon Kuhn}\email{simkuhn@phys.ethz.ch}
\affiliation{Institute for Theoretical Physics,
ETH Zurich, Wolfgang-Pauli-Strasse 27, 8093, Zurich, Switzerland}

\begin{abstract}
\noindent We systematically study the field equations of $f(\mathbb Q)$ gravity for spherically symmetric and stationary metric-affine spacetimes. Such spacetimes are described by a metric as well as a flat and torsionless affine connection. In the Symmetric Teleparallel Equivalent of GR (STEGR), the connection is pure gauge and hence unphysical. However, in the non-linear extension $f(\Q)$, it is promoted to a dynamical field which changes the physics. Starting from a general metric-affine geometry, we construct the most general static and spherically symmetric forms of the metric and the affine connection. We then use these symmetry reduced geometric objects to prove that the field equations of $f(\Q)$ gravity admit GR solutions as well as beyond-GR solutions, contrary to what has been claimed in the literature. We formulate precise criteria, under which conditions it is possible to obtain GR solutions and under which conditions it is possible to obtain beyond-GR solutions. We subsequently construct several perturbative corrections to the Schwarzschild solution for different choices of $f(\Q)$, which in particular include a hair stemming from the now dynamical affine connection. We also present an exact beyond-GR vacuum solution. Lastly, we apply this method of constructing spherically symmetric and stationary solutions to $f(\T)$ gravity, which reproduces similar solutions but without a dynamical connection.
\end{abstract}

\maketitle
\bigskip\bigskip
\hrule
\spacing{1.7}
\tableofcontents
\singlespacing
\bigskip
\hrule
\clearpage

\section{Introduction}\label{sec:Intro}\setcounter{equation}{0}
General Relativity (GR) is by far the most successful description of gravity we have. Its predictions for gravitational effects on solar system scales and on cosmological scales fit the observations very well. However, some slight tensions, such as the $H_0$ tension, have appeared over the past few years. In addition, theoretical difficulties such as singularities, quantum gravity and a lacking explanation for the origin of dark matter and dark energy spoil the beauty of GR. It is thus fruitful to look at generalizations of GR, which might resolve both observational and theoretical issues.

One such generalization is provided by Symmetric Teleparallelism (ST), which is rooted in a different set of geometric postulates than GR. The key difference between ST and GR is the role played by the affine connection, $\Gamma^\alpha{}_{\mu\nu}$. In GR, it is postulated that the connection is torsionless and metric-compatible, which immediately implies that it is uniquely given by the Levi-Civita connection. In ST the postulate of metric-compatibility is dropped and one instead demands that $\Gamma^\alpha{}_{\mu\nu}$ is torsionless and gives rise to a vanishing Riemann tensor. As long as the connection satisfies these postulates, it can be chosen arbitrarily and, in particular, it is \textit{independent} of the metric. With curvature and torsion of $\Gamma^\alpha{}_{\mu\nu}$ postulated to be zero, the only non-trivial object left in ST which characterizes the affine geometry is the non-metricity tensor, $Q_{\alpha\mu\nu}$. This tensor can be used to construct the so-called non-metricity scalar $\Q$, which will be  defined in section~\ref{sec:SymmetricTeleparallelism}, and which in turn defines the action of ST: $\mathcal S[g, \Gamma] := \int \dd^4 x\, \sqrt{-g}\,\Q$. It is well-known~\cite{BeltranJimenez:2017, Jimenez:2018, Heisenberg:2018vsk,BeltranJimenez:2019, DAmbrosio:2020} that this action is equivalent to the Einstein-Hilbert action of GR up to a boundary term. ST therefore provides a different geometric description of gravity, which is nevertheless equivalent to GR. In particular, it can be shown that the affine connection only appears in a boundary term in the action and it is hence unphysical. More precisely, the field equations of the metric do not depend on the choice of connection and the connection field equations are identically satisfied for any choice of connection which is compatible with the postulates of ST. This means that the physical degrees of freedom reside in the metric while the connection does not carry any physical information.

This changes when one considers generalisations, such as generic theories which are quadratic in the non-metricity tensor~\cite{Dambrosio:2020b}, or extensions of ST~\cite{Jimenez:2019}. What is of interest to us in the present paper is the non-linear extension described by $\int \dd^4 x\, \sqrt{-g}\, f(\Q)$ \cite{BeltranJimenez:2017}, where $f$ is an a priori arbitrary function. Not only is this theory \textit{not} equivalent to $f(R)$ gravity, but the theory now harbours degrees of freedom in the metric \textit{and} in the affine connection, because the dependence on $\Gamma^\alpha{}_{\mu\nu}$ can no longer be absorbed in a boundary term in the action.

The connection can thus be expected to influence the metric, which describes the gravitational field. In this paper we show that this expectation is indeed realized, contrary to what has been claimed in the literature~\cite{Zhao:2021, Lin:2021}. We show this by systematically studying \textit{the most general} stationary and spherically symmetric spacetimes within $f(\Q)$ gravity.

The paper is organized as follows: Section~\ref{sec:SymmetricTeleparallelism} is dedicated to introducing symmetric teleparallelism, via the Palatini formalism, as well as $f(\Q)$ gravity. This serves the purpose to recall basic definitions and fixing notations and conventions. In section~\ref{sec:SymRed} we perform a detailed symmetry reduction of the metric and the connection. In particular, we show that there are two ways of giving an explicit parametrization of stationary, spherically symmetric, torsionless, and flat connections (subsections~\ref{ssec:Set1} and~\ref{ssec:set2}). Moreover, we show that the trivial connection (i.e., the connection in coincident gauge, $\Gamma^\alpha_{\ \mu\nu}=0$) fails to be spherically symmetric and that the connection used in~\cite{Zhao:2021, Lin:2021} belongs to the second parametrization class studied here (cf. Table~\ref{tab:SolutionSet2} for a definition of this class). In subsection~\ref{ssec:D2} we then show that the first parametrization class can be obtained from the second one by a well-defined double scaling limit. We also show that the metric can be brought into a diagonal form --with two arbitrary functions of $r$ in the first half of the diagonal and the standard metric of a $2$-sphere on the second half of the diagonal-- by means of a diffeomorphism which does not alter the structure and defining properties of the above-mentioned parametrizations of the connection.

Subsequently, in section~\ref{sec:SymRedFieldEq}, we use the diagonal metric and the two classes of connection to perform a symmetry reduction of the metric and connection field equations of $f(\Q)$ gravity. In subsection~\ref{ssec:SymRedFieldEq1}, we show that the first parametrization class cannot produce any solutions which go beyond the standard Schwarzschild-deSitter-Nordstr\"{o}m solution. The symmetry reduced field equations for the second parametrization class, which we discuss in subsection~\ref{ssec:SymRedFieldEq2}, offer more flexibility. We discuss under which conditions the field equations can produce solutions beyond the well-known GR solutions for spherically symmetric and stationary spacetimes -- and we explain why the connection used in~\cite{Zhao:2021,Lin:2021} could only produce GR solutions for arbitrary choices of the function $f$.

In section~\ref{sec:Solutions}, we use our insights to describe approximate solutions which go beyond Schwarzschild-deSitter-Nordstr\"{o}m for $f(\Q) = \Q +\alpha\,\Q^2$, assuming $\alpha$ is a small parameter. In section~\ref{ssec:ExactSolutions} we also present an exact vacuum solution which goes beyond GR for the case $f(\Q) = \Q^\kappa$, for $\kappa\in\mathbb R\setminus\{0\}$, demonstrating that such solutions exist in $f(\Q)$~gravity.

Finally, in section~\ref{sec:f(T)}, we sketch how the approach described in detail for $f(\Q)$ can be transferred to $f(\T)$ gravity, a generalization of Metric Teleparalellism. This theory of gravity is described again by a metric and a connection, but the latter is now postulated to be flat and metric-compatible, but with non-vanishing torsion. Since the construction of stationary and spherically symmetric affine geometries, as well as the analysis of the equations of motion, work in complete analogy to $f(\Q)$ gravity, one can easily construct the most general stationary, spherically symmetric, flat, and metric-compatible spacetimes of $f(\T)$ gravity. We report our results, and their relations to $f(\Q)$ gravity, and compare them to the literature \cite{Hohmann:2019,Bahamonde:2020vpb,Hohmann:2019nat,DeBenedictis:2016aze,Ruggiero:2015oka}.

We conclude the paper in section~\ref{sec:Conclusion} with a brief discussion of the main results and an outlook on future research.

\section{Symmetric Teleparallelism and $f(\Q)$ Gravity}\label{sec:SymmetricTeleparallelism}\setcounter{equation}{0}
Let $(\mathcal M, g_{\mu\nu}, \Gamma^{\alpha}_{\ \mu\nu})$ be a metric-affine geometry, where $\mathcal M$ is a four-dimensional manifold, $g_{\mu\nu}$ denotes  the components of the metric tensor of signature $(-,+,+,+)$, and $\Gamma^{\alpha}_{\ \mu\nu}$ represents an affine connection. The connection defines a notion of covariant differentiation through its action on vectors and co-vectors,
\begin{align}
	\nabla_\mu V^\alpha &= \partial_\mu V^\alpha + \Gamma^\alpha_{\ \mu\lambda} V^\lambda\notag\\
	\nabla_\mu V_\alpha &= \partial_\mu V_\alpha - \Gamma^\lambda_{\ \mu\alpha} V_\lambda,
\end{align}
and it can be used to describe three independent geometric properties of a spacetime: curvature, torsion, and non-metricity. The first two objects, curvature and torsion, are defined by
\begin{align}\label{eq:SymParaCond}
	R^\alpha_{\ \beta\mu\nu} &:= 2 \partial_{[\mu}\Gamma^\alpha_{\ \nu]\beta} + 2 \Gamma^\alpha_{[\mu|\lambda|}\Gamma^\lambda_{\ \nu]\beta}\notag\\
	T^\alpha_{\ \mu\nu} &:= 2\Gamma^\alpha_{\ [\mu\nu]},
\end{align}
and symmetric teleparallelism demands that both tensors vanish. We refer to 
\begin{equation}\label{eq:STpostulates}
	R^\alpha_{\ \beta\mu\nu} \overset{!}{=} 0 \quad\text{and}\quad T^\alpha_{\ \mu\nu} \overset{!}{=} 0
\end{equation} 
as the postulates of symmetric teleparallelism. With curvature and torsion set to zero, the non-metricity tensor is the only remaining non-trivial object. As it measures the failure of the connection to be metric-compatible, it is defined by
\begin{equation}
	Q_{\alpha\mu\nu} := \nabla_\alpha g_{\mu\nu} = \partial_\alpha g_{\mu\nu} - 2\Gamma^{\lambda}{}_{\alpha(\mu}g_{\nu)\lambda}.
\end{equation}
Notice that the Riemann and torsion tensor depend on the connection only, while the non-metricity tensor also depends on the metric. Due to the symmetry of the non-metricity tensor in its last two indices, at quadratic order there are only five independent scalars that can be built from the non-metricity tensor. Hence, a natural starting point for defining a Lagrangian which describes gravity in terms of non-metricity is a linear combination of these five terms. As it turns out~\cite{BeltranJimenez:2017,Jimenez:2018,BeltranJimenez:2019}, GR is described by a linear combination of only four of these contractions, which define the so called non-metricity scalar,
\begin{equation}\label{eq:GenNonMetricityScalar}
	\Q:= -\frac14\, Q_{\alpha\beta\gamma}Q^{\alpha\beta\gamma} + \frac12\, Q_{\alpha\beta\gamma}Q^{\beta\alpha\gamma}	 + \frac14\, Q_\alpha Q^{\alpha} - \frac12 c_5\, Q_\alpha \bar{Q}^\alpha,
\end{equation}
where $Q_\alpha:= Q_{\alpha\nu}{}^{\nu}$ and $\bar{Q}_\alpha:= Q^\nu{}_{\nu\alpha}$ denote the two independent traces of the non-metricity tensor. Since any connection can be decomposed into its torsion-, non-metricity-, and Levi-Civita-parts, it is easy to show that the non-metricity scalar can be written as
\begin{equation}\label{eq:GRequivalentidentity}
    \Q = \mathcal{D}_\mu(Q^\mu-\bar Q^\mu)+\mathcal{R},
\end{equation}
where $\mathcal D_\mu$ denotes the covariant derivative with respect to the Levi-Civita connection and $\mathcal R$ is the Ricci scalar of the Levi-Civita connection. This identity shows that symmetric teleparallelism, defined by the action $\mathcal S[g, \Gamma]:=\int\dd^4 x\,\sqrt{-g}\,\Q$, is equivalent to the Einstein-Hilbert formulation, $\mathcal S[g]:=\int\dd^4 x\,\sqrt{-g}\,\mathcal R$, up to a boundary term. 

It makes it also evident that the connection of ST is unphysical, since it is completely contained in the boundary term and only the Levi-Civita part of $\Gamma^\alpha{}_{\mu\nu}$ contributes to the metric field equations. Hence, the physical degrees of freedom are all contained in the metric and the connection can be freely chosen, as long as it satisfies the postulates of symmetric teleparallelism.

In the present work, however, we are not interested in the theory defined by $\Q$ alone. Rather, we want to consider non-linear extensions defined by the action functional \cite{BeltranJimenez:2017}
\begin{equation}\label{eq:CovAction}
	\mathcal S[g, \Gamma; \lambda, \rho] := \int_\M \dd^4 x\left(\frac12\sqrt{-g}\,f(\Q) + \lambda_\alpha{}^{\beta\mu\nu} R^\alpha{}_{\beta\mu\nu} + \rho_\alpha{}^{\mu\nu} T^\alpha{}_{\mu\nu}\right) + \mathcal S_\textsf{matter},
\end{equation} 
where the tensor densities $\lambda_\alpha{}^{\beta\mu\nu}$ and $\rho_\alpha{}^{\mu\nu}$ act as Lagrange multipliers which enforce the postulates of symmetric teleparallelism and where $f$ is an arbitrary function solely subjected to the condition $f'(\Q) := \frac{\dd f(\Q)}{\dd \Q}\neq 0$. This last requirement is necessary in order to obtain non-trivial field equations.

It is important to notice that for generic $f$, there is no identity analogous to~\eqref{eq:GRequivalentidentity}. Hence, in $f(\Q)$ gravity, the connection can in general not be absorbed into a boundary term and it has to be expected that $\Gamma^\alpha{}_{\mu\nu}$ carries degrees of freedom, in addition to the ones contained in the metric. There is evidence for this in the existing literature~\cite{Jimenez:2019} and we will show this explicitly in this paper in section~\ref{sec:SymRedFieldEq} and in section~\ref{sec:Solutions}. Moreover, the connection can no longer be arbitrarily chose, as was done in~\cite{Lin:2021,Zhao:2021}, since it has its own, non-trivial, field equations which need to be satisfied. Concretely, the field equations of $f(\Q)$ gravity are given by~\cite{Jimenez:2018}
\begin{align}\label{eq:FieldEquations}
	\M_{\mu\nu} := \frac{2}{\sqrt{-g}}\nabla_\alpha\left[\sqrt{-g}P^\alpha{}_{\mu\nu} f'(\Q)\right] + f'(\Q) q_{\mu\nu} -\frac12 f(\Q) g_{\mu\nu}- T_{\mu\nu}  &= 0\notag\\
	\mathcal C_\alpha := \nabla_\mu\nabla_\nu\left(\sqrt{-g}\,f'(\Q) P^{\mu\nu}{}_{\alpha}\right) &= 0,
\end{align}
where $T_{\mu\nu}$ denotes the stress-energy tensor (not to be confused with the torsion tensor which carries an addition contravariant index) and where we have introduced the non-metricity conjugate $P^{\alpha}{}_{\mu\nu}$ and the symmetric tensor $q_{\mu\nu}$ defined by
\begin{align}
	P^\alpha{}_{\mu\nu} &:= \frac12 \PD{\Q}{Q_\alpha{}^{\mu\nu}} = -\frac14 Q^\alpha{}_{\mu\nu} + \frac12 Q_{(\mu}{}^{\alpha}{}_{\nu)} +\frac14 g_{\mu\nu}Q^\alpha -\frac14 \left(g_{\mu\nu} \bar{Q}^\alpha + \delta^\alpha{}_{(\mu} Q_{\nu)}\right)\notag\\
	q_{\mu\nu} &:= \PD{\Q}{g^{\mu\nu}} = P_{(\mu|\alpha\beta}Q_{\nu)}{}^{\mu\nu} - 2P^{\alpha\beta}{}_{(\nu} Q_{\alpha\beta|\mu)}.
\end{align}
The metric field equations in~\eqref{eq:FieldEquations} can also be re-written in the useful and more suggestive form~\cite{Zhao:2021,Lin:2021}
\begin{equation}\label{eq:RewrittenMetricFieldEq}
	f'(\Q) G_{\mu\nu}  - \frac12 g_{\mu\nu} (f(\Q)-f'(\Q)\Q) + 2 f''(\Q) P^\alpha{}_{\mu\nu} \partial_\alpha\Q  = T_{\mu\nu},
\end{equation}
where $f''$ denotes the second derivative of $f$ with respect to $\Q$ and $G_{\mu\nu}$ is the Einstein tensor (with respect to the Levi-Civita connection and with vanishing cosmological constant). In this form, it becomes obvious that for $f(\Q) = \Q + 2\Lambda$, the metric field equations are equivalent to the Einstein field equations with cosmological constant $\Lambda$. 

In particular, in this case the theory only propagates two degrees of freedom of the metric, while the connection is pure gauge. As mentioned above, for a generic function $f$ it can be expected, and there is also evidence from a perturbative analysis~\cite{Jimenez:2019}, that the theory propagates more than two degrees of freedom and should therefore lead to potentially interesting deviations from GR. Indeed, we will see later (cf sections~\ref{sec:SymRedFieldEq} and \ref{sec:Solutions}) that under certain conditions, the connection is no longer pure gauge, but rather, one component becomes dynamical and this leads to solutions which go beyond the standard GR solutions.\\

We conclude this section by recalling that there exists a special gauge choice in which symmetric teleparallelism, where the connection can be arbitrarily chosen, can be cast in a particularly simple form: The so called coincident gauge~\cite{BeltranJimenez:2017}. In this gauge, the connection is trivial, i.e. $\Gamma^{\alpha}{}_{\mu\nu} = 0$. It is obtained by observing that the first postulate of symmetric teleparallelism, the vanishing of curvature, implies that the connection must have the form
\begin{equation}
	\Gamma^\alpha_{\ \mu\nu} = \left(\Lambda^{-1}\right)^\alpha{}_{\rho}\partial_\mu \Lambda^\rho{}_{\nu},
\end{equation}
where $\Lambda^\alpha{}_{\beta}\in GL(4,\mathbb R)$. The requirement of vanishing torsion further restricts the matrix $\Lambda^\alpha{}_{\beta}$ to have the form $\Lambda^\alpha{}_{\beta} = \partial_\beta \xi^\alpha$, for arbitrary $\xi^\alpha$, and the connection consequently becomes
\begin{equation}\label{eq:CoincidentConnection}
	\Gamma^\alpha_{\ \mu\nu} = \frac{\partial x^\alpha}{\partial \xi^\lambda}\partial_\mu\partial_\nu\xi^\lambda.
\end{equation}
Hence, the connection can be set globally to zero by the affine gauge choice $\xi^\alpha = M^{\alpha}_{\ \beta} x^\beta + \xi^\alpha_0$, where $M^{\alpha}_{\ \beta}$ is a non-degenerate matrix with constant entries and $\xi^\alpha_0$ is a constant vector~\cite{Jimenez:2018}.

\section{Symmetry reduction of the metric and the connection}\label{sec:SymRed}\setcounter{equation}{0}
Our goal is the study of the field equations of $f(\Q)$ gravity for spherically symmetric and stationary spacetimes. To that end, we first assume that the ten metric components and the $64$ components of the affine connection can be expressed in the chart $(t, r, \theta, \phi)\in \mathbb R\times\mathbb R_{> 0}\times [0,\pi]\times [0,2\pi)$. The next step is to find the most general form of the metric and the connection which respect the symmetries of the spacetime and, in the case of the connection, which is compatible with the postulates of symmetric teleparallelism.

The notion of symmetry we use is the same as the one given in~\cite{Hohmann:2019} and we recall it here for convenience: Let $G$ be a group, $\Phi:G\times\M\to\M$ the action of the group on the spacetime manifold $\M$, and denote by $\Phi_u:\M\to\M$ for $u\in G$ the induced diffeomorphism. We then say that a metric-affine geometry $(\M, g_{\mu\nu}, \Gamma^{\alpha}{}_{\mu\nu})$ is symmetric under the group action if and only if
\begin{align}
	(\Phi_u^*g)_{\mu\nu} &= g_{\mu\nu}\notag\\
	(\Phi_u^*\Gamma)^{\alpha}{}_{\mu\nu} &= \Gamma^{\alpha}{}_{\mu\nu},
\end{align}
for all $u\in G$ and where $\Phi_u^*$ denotes the pull-back of $\Phi_u$. In our case, the group $G$ will be the group of spatial rotations, $SO(3)$, and the group of time translations. Moreover, in practice, it is more convenient to consider infinitesimal actions of $G$ on the metric-affine geometry. The above symmetry conditions can then easily be re-expressed as
\begin{align}\label{eq:GeneralSymmetryConditions}
	(\lie_\xi g)_{\mu\nu} &= 0\notag\\
	(\lie_\xi \Gamma)^{\alpha}{}_{\mu\nu} &= 0,
\end{align}
where $\lie_\xi$ stands for the Lie derivative along $\xi$, which representatively stands for the generating vector fields of the Lie algebra $\mathfrak g$ of $G$. Our task is therefore to implement~\eqref{eq:GeneralSymmetryConditions} for the generator of time-translations and the generators of $SO(3)$. Of course, the most general spherically symmetric and stationary form of the metric is well-known and we can simply state the result:
\begin{equation}\label{eq:GeneralMetric}
	g_{\mu\nu} = \begin{pmatrix}
		g_{tt} & g_{tr} & 0 & 0\\
		g_{tr} & g_{rr} & 0 & 0\\
		0 & 0 & g_{\theta\theta} & 0\\
		0 & 0 & 0 & g_{\theta\theta}\,\sin^2\theta
	\end{pmatrix},
\end{equation}
where all four independent components $\{g_{tt}, g_{tr}, g_{rr}, g_{\theta\theta}\}$ only depend on $r$. In the case of the connection, we could refer to the results reported in~\cite{Hohmann:2019}, where the symmetry reduction of a general affine connection under the action of $G=SO(3)$ has been carried out. However, we will perform the symmetry reduction of the connection in detail under the \textit{additional assumptions} that it is torsionless and stationary. With the former assumption we already achieve the implementation of one of the two postulates of symmetric teleparallelism. The assumption of stationarity will play an important role in subsection~\ref{ssec:RiemannZero}, where we implement the second postulate of symmetric teleparallelism.
Note that of the three conditions the connection has to fulfill, $\mathcal{L}_\zeta\Gamma^\alpha_{\ \mu\nu}=0$, $T^\alpha_{\ \mu\nu}=0$, $R^\alpha_{\ \beta\mu\nu}=0$, the first and second are linear in $\Gamma^\alpha_{\ \mu\nu}$ and thus have unique solutions. But the Riemann tensor is quadratic in the connection, and one might thus obtain several solutions for the connection.

\subsection{Symmetry reduction of $\Gamma^{\alpha}{}_{\mu\nu}$ under the assumption that $T^{\alpha}{}_{\mu\nu} = 0$}\label{ssec:SymRed}
The torsionless condition, $T^\alpha{}_{\mu\nu}\overset{!}{=}0$, simply forces the connection to be symmetric in its lower indices, $\Gamma^\alpha{}_{[\mu\nu]} = 0$. This reduces the amount of independent connection components from $64$ to $40$. Next, we implement the condition of stationarity. Since the generating vector field of time-translations is simply given by $\mathcal T :=\mathcal T^\alpha\partial_\alpha = \partial_t$, we immediately find
\begin{equation}\label{eq:SymmetryConditionT}
	(\lie_{\mathcal T}\Gamma)^\alpha{}_{\mu\nu} = \partial_t\Gamma^\alpha{}_{\mu\nu} \overset{!}{=} 0.
\end{equation} 
In words: All $40$ components of the connection are, unsurprisingly, time-independent. Implementing spherical symmetry requires a little more work. To begin with, we recall that the generating vector fields of $SO(3)$ are\begin{align}
	\mathcal R_x &:= \mathcal R^\alpha_x \partial_\alpha = \sin\phi\,\partial_\theta +\frac{\cos\phi}{\tan\theta}\,\partial_\phi\notag\\
	\mathcal R_y &:= \mathcal R^\alpha_y \partial_\alpha= -\cos\phi\,\partial_\theta + \frac{\sin\phi}{\tan\theta}\,\partial_\phi\notag\\
	\mathcal R_z &:= \mathcal R^\alpha_z \partial_\alpha = -\partial_\phi.
\end{align}
It is easiest to start with the generator $\mathcal R_z$ since this one simply gives us
\begin{equation}\label{eq:SymmetryConditionphi}
	(\lie_{\mathcal R_z}\Gamma)^\alpha{}_{\mu\nu} = -\partial_\phi\Gamma^\alpha{}_{\mu\nu} \overset{!}{=} 0,
\end{equation}
which means that all connection components are independent of the angular coordinate $\phi$. To implement the remaining two symmetry conditions, it is convenient to study linear combinations of Lie derivatives. The first one reads
\begin{equation}\label{eq:SymmetryCondition1}
	\cos\phi\,(\lie_{\mathcal R_x} \Gamma)^{\alpha}{}_{\mu\nu} + \sin\phi\, (\lie_{\mathcal R_y}\Gamma)^{\alpha}{}_{\mu\nu} \overset{!}{=} 0
\end{equation}
and it leads to a set of equations which explicitly determine $28$ of the $40$ connection components. The solutions can be subdivided into three groups. The first group consists of $20$ components which are forced to be zero,
\begin{align}\label{eq:Zero_Components}
	\Gamma^t{}_{t\theta} &= 0 & \Gamma^t{}_{t\phi} &= 0 & \Gamma^t{}_{r\theta} &= 0 & \Gamma^t{}_{r\phi} &= 0 & \Gamma^t{}_{\theta\phi} &= 0\notag\\
	\Gamma^r{}_{t\theta} &= 0 & \Gamma^r{}_{t\phi} &= 0 & \Gamma^r{}_{r\theta} &= 0 & \Gamma^r{}_{r\phi} &= 0 & \Gamma^r{}_{\theta\phi} &= 0\notag\\
	\Gamma^\theta{}_{tt} &= 0 & \Gamma^\theta{}_{tr} &= 0 & \Gamma^\theta{}_{rr} &= 0 & \Gamma^\theta{}_{\theta\theta} &= 0 & \Gamma^\theta{}_{\theta\phi} &= 0\notag\\
	\Gamma^\phi{}_{tt} &= 0 & \Gamma^\phi{}_{tr} &= 0 & \Gamma^\phi{}_{rr} &= 0 & \Gamma^\phi{}_{\theta\theta} &= 0 & \Gamma^\phi{}_{\phi\phi} &= 0.
\end{align}
The second group contains two components which are explicitly given by trigonometric functions,
\begin{align}\label{eq:Trig_Components}
	\Gamma^\theta{}_{\phi\phi} &= -\cos\theta\ \sin\theta\notag\\
	\Gamma^\phi{}_{\theta\phi} &= \cot\theta.
\end{align}
This is an important result, since it tells us that the coincident gauge, i.e., the trivial connection $\Gamma^\alpha{}_{\mu\nu} = 0$, fails to be spherically symmetric. Any attempt to find spherically symmetric solutions to the field equations of $f(\Q)$ gravity using the coincident gauge is therefore bound to fail and one should expect inconsistencies, unless we are in symmetric teleparallelism which is described by the affine function $f(\Q)=a\,\Q + b$.\\
Finally, we find that in the third group, six components can be expressed algebraically in terms of other components:
\begin{align}\label{eq:Algebraic_Components}
	\Gamma^t{}_{\phi\phi} &= \Gamma^t{}_{\theta\theta} \sin^2\theta & \Gamma^r{}_{\phi\phi} &= \Gamma^r{}_{\theta\theta}\sin^2\theta & \Gamma^\phi{}_{t\theta} &= -\Gamma^\theta{}_{t\phi} \csc^2\theta \notag\\
	\Gamma^\theta{}_{t\theta} &= \Gamma^\phi{}_{t\phi}  & \Gamma^\phi{}_{r\theta} &= -\Gamma^\theta{}_{r\phi} \csc^2\theta & \Gamma^\theta{}_{r\theta} &= \Gamma^\phi{}_{r\phi}.
\end{align}
From the initially 40 independent connection components, we are left with $40-20-2-6 = 12$ components. These twelve independent components are
\begin{equation}\label{eq:independent_components}
	\left\{\Gamma^t{}_{tt}, \Gamma^t{}_{tr}, \Gamma^t{}_{rr}, \Gamma^t{}_{\theta\theta}, \Gamma^r{}_{tt}, \Gamma^r{}_{tr}, \Gamma^r{}_{rr}, \Gamma^r{}_{\theta\theta}, \Gamma^\phi{}_{t\phi}, \Gamma^\theta{}_{t\phi}, \Gamma^\phi{}_{r\phi}, \Gamma^\theta{}_{r\phi} \right\}
\end{equation} 
and these are functions of $r$ and $\theta$, potentially. We can further restrict the functional dependence of these components by considering the last symmetry condition which reads
\begin{equation}\label{eq:SymmetryCondition2}
	\sin\phi \, (\lie_{\mathcal R_x} \Gamma)^\alpha{}_{\mu\nu} - \cos\phi\, (\lie_{\mathcal R_y} \Gamma)^\alpha{}_{\mu\nu} \overset{!}{=} 0.
\end{equation}
This condition leads to a total of twelve first order differential equations for precisely the twelve independent components given in~\eqref{eq:independent_components}. These equations are explicitly given by
\begin{align}
	\partial_\theta \Gamma^\theta{}_{t\phi} - \Gamma^\theta{}_{t\phi}\cot\theta &= 0 & \partial_\theta\Gamma^\theta{}_{r\phi} - \Gamma^\theta{}_{r\phi}\cot\theta &= 0 & \partial_\theta\Gamma^t{}_{tt}&= 0 & \partial_\theta\Gamma^t{}_{tr}&= 0 \notag\\
	\partial_\theta\Gamma^t{}_{rr}&= 0 & \partial_\theta\Gamma^t{}_{\theta\theta}&= 0 & \partial_\theta\Gamma^r{}_{tt}&= 0 & \partial_\theta\Gamma^r{}_{tr}&= 0\notag\\
	\partial_\theta\Gamma^r{}_{rr}&= 0 & \partial_\theta\Gamma^r{}_{\theta\theta}&= 0 & \partial_\theta\Gamma^\phi{}_{t\phi}&= 0 & \partial_\theta\Gamma^\phi{}_{r\phi}&= 0.
\end{align}
The first two differential equations in the first line are easily solved and give us
\begin{equation}\label{c1c2sol}
	\Gamma^\theta{}_{t\phi} = \sin\theta\, c_1(r)\quad\text{and}\quad \Gamma^\theta{}_{r\phi} = \sin\theta\, c_2(r),
\end{equation}
where $c_1$ and $c_2$ are arbitrary functions of $r$, while the remaining ten equations tell us that the other components are only functions of $r$. This completes the symmetry reduction of the connection and we are left with the twelve independent functions
\begin{equation}\label{eq:independent_connection_components}
	\left\{c_1(r), c_2(r), \Gamma^t{}_{tt}(r), \Gamma^t{}_{tr}(r), \Gamma^t{}_{rr}(r), \Gamma^t{}_{\theta\theta}(r), \Gamma^r{}_{tt}(r), \Gamma^r{}_{tr}(r), \Gamma^r{}_{rr}(r), \Gamma^r{}_{\theta\theta}(r), \Gamma^\theta{}_{t\theta}(r), \Gamma^\theta{}_{r\theta}(r) \right\}.
\end{equation}
We will use the results obtained in this subsection in order to implement $R^\alpha{}_{\mu\nu\rho}\overset{!}{=}0$. This will be the subject of the next subsection and it will further reduce the amount of independent connection components.

\subsection{Implementation of $R^\alpha{}_{\mu\nu\rho} = 0$}\label{ssec:RiemannZero}

In the previous subsection, we already used the symmetric teleparallelism postulate that $T^\alpha{}_{\mu\nu} = 0$ and we imposed the symmetry conditions~\eqref{eq:SymmetryConditionT},~\eqref{eq:SymmetryConditionphi},~\eqref{eq:SymmetryCondition1}, and~\eqref{eq:SymmetryCondition2}. From these conditions we learned that there are only twelve independent connection components, all of which are solely functions of $r$, and that the remaining $28$ components are given by equations~\eqref{eq:Zero_Components},~\eqref{eq:Trig_Components}, and~\eqref{eq:Algebraic_Components}. We can now use these facts to simplify the equations which arise from imposing that the Riemann tensor of $\Gamma^\alpha{}_{\mu\nu}$ has to vanish. After rather long and unenlightening computations, one finds $24$ non-trivial equations. Since the Riemann tensor is quadratic in $\Gamma$ and linear in $\partial\Gamma$, and because half the connection components are zero, one can reasonably expect that these $24$ equations can be separated into non-linear algebraic equations and first-order differential equations. Indeed, one finds that there are twelve algebraic equations,
\begin{align}\label{eq:Algebraic_Equations}
	\Gamma^{t}{}_{\theta\theta}\Gamma^{\theta}{}_{t\phi} &=0\notag\\
	\Gamma^{r}{}_{\theta\theta}\Gamma^{\theta}{}_{t\phi} &= 0\notag\\
	\Gamma^{t}{}_{\theta\theta}\Gamma^{\theta}{}_{r\phi} &= 0\notag\\
	\Gamma^{r}{}_{\theta\theta}\Gamma^{\theta}{}_{r\phi} &= 0\notag\\
	\Gamma^{t}{}_{\theta\theta}\Gamma^{\theta}{}_{t\phi} + \Gamma^{r}{}_{\theta\theta}\Gamma^{\theta}{}_{r\phi} &= 0\notag\\
	1+\Gamma^{t}{}_{\theta\theta}\Gamma^{\phi}{}_{t\phi} + \Gamma^{r}{}_{\theta\theta}\Gamma^{\phi}{}_{r\phi} &= 0\notag\\
	\Gamma^{t}{}_{tr}\Gamma^{r}{}_{\theta\theta} + \Gamma^{t}{}_{\theta\theta}\left(\Gamma^{t}{}_{tt}-\Gamma^{\phi}{}_{t\phi}\right) &= 0\notag\\
	\Gamma^{t}{}_{\theta\theta}\Gamma^{r}{}_{tt}+\Gamma^{r}{}_{\theta\theta}\left(\Gamma^{r}{}_{tr}-\Gamma^{\phi}{}_{t\phi}\right) &= 0\notag\\
	\Gamma^{r}{}_{tt}\Gamma^{\theta}{}_{r\phi}+\Gamma^{\theta}{}_{t\phi}\left(\Gamma^{t}{}_{tt}-2\Gamma^{\phi}{}_{t\phi}\right) &= 0\notag\\
	\Gamma^{\theta}{}_{r\phi}\left(\Gamma^{r}{}_{tr}-\Gamma^{\phi}{}_{t\phi}\right)+\Gamma^{\theta}{}_{t\phi}\left(\Gamma^{t}{}_{tr}-\Gamma^{\phi}{}_{r\phi}\right) &= 0\notag\\
	\Gamma^{\phi}{}_{t\phi}\left(\Gamma^{t}{}_{tt}-\Gamma^{\phi}{}_{t\phi}\right) + \Gamma^{r}{}_{tt}\Gamma^{\phi}{}_{r\phi}+(\Gamma^{\theta}{}_{t\phi})^2\csc^2\theta &= 0\notag\\
	\Gamma^{t}{}_{tr}\Gamma^{\phi}{}_{t\phi} + \Gamma^{\phi}{}_{r\phi}\left(\Gamma^{r}{}_{tr}-\Gamma^{\phi}{}_{t\phi}\right) + \Gamma^{\theta}{}_{t\phi}\Gamma^{\theta}{}_{r\phi}\csc^2\theta &= 0,
\end{align}
for only ten of the twelve independent connection components. The remaining twelve equations are first-order differential equations,
\begin{align}
	\partial_r\Gamma^{\theta}{}_{t\phi} &= 0\notag\\
	\partial_r\Gamma^{\phi}{}_{t\phi} &= 0\notag\\
	\partial_r\Gamma^{r}{}_{tr} &= \Gamma^{t}{}_{rr}\Gamma^{r}{}_{tt} -\Gamma^{t}{}_{tr}\Gamma^{r}{}_{tr} \notag\\
	\partial_r\Gamma^{t}{}_{tt} &= \Gamma^{t}{}_{tr}\Gamma^{r}{}_{tr}-\Gamma^{t}{}_{rr}\Gamma^{r}{}_{tt}\notag\\
	\partial_r\Gamma^{t}{}_{\theta\theta} &= \Gamma^{t}{}_{\theta\theta}\left(\Gamma^{\phi}{}_{r\phi}-\Gamma^{t}{}_{tr}\right) - \Gamma^{t}{}_{rr}\Gamma^{r}{}_{\theta\theta} \notag\\
	\partial_r\Gamma^{r}{}_{\theta\theta} &= \Gamma^{r}{}_{\theta\theta}\left(\Gamma^{\phi}{}_{r\phi}-\Gamma^{r}{}_{rr}\right)-\Gamma^{t}{}_{\theta\theta}\Gamma^{r}{}_{tr}\notag\\
	\partial_r\Gamma^{\theta}{}_{r\phi} &= \Gamma^{\theta}{}_{r\phi}\left(\Gamma^{r}{}_{rr}-2\Gamma^{\phi}{}_{r\phi}\right) + \Gamma^{t}{}_{rr}\Gamma^{\theta}{}_{t\phi} \notag\\
	\partial_r\Gamma^{r}{}_{tt} &=\Gamma^{r}{}_{tt}\left(\Gamma^{t}{}_{tr}-\Gamma^{r}{}_{rr}\right) + \Gamma^{r}{}_{tr}\left(1-\Gamma^{t}{}_{tt}\right) \notag\\
	\partial_r\Gamma^{t}{}_{tr} &=\Gamma^{t}{}_{rr}\left(\Gamma^{t}{}_{tt}-\Gamma^{r}{}_{tr}\right) + \Gamma^{t}{}_{tr}\left(\Gamma^{r}{}_{rr}-\Gamma^{t}{}_{tr}\right) \notag\\
	\partial_r\Gamma^{\theta}{}_{t\phi} &= \Gamma^{\theta}{}_{r\phi}\left(\Gamma^{r}{}_{tr}-\Gamma^{\phi}{}_{t\phi}\right) + \Gamma^{\theta}{}_{t\phi}\left(\Gamma^{t}{}_{tr}-\Gamma^{\phi}{}_{r\phi}\right)\notag\\
	\partial_r\Gamma^{\phi}{}_{r\phi} &= \Gamma^{\phi}{}_{r\phi}\left(\Gamma^{r}{}_{rr}-\Gamma^{\phi}{}_{r\phi}\right) + \Gamma^{t}{}_{rr}\Gamma^{\phi}{}_{t\phi} + (\Gamma^{\theta}{}_{r\phi})^2\csc^2\theta\notag\\
	\partial_r\Gamma^{\phi}{}_{t\phi} &= \Gamma^{\phi}{}_{r\phi}\left(\Gamma^{r}{}_{tr}-\Gamma^{\phi}{}_{t\phi}\right) + \Gamma^{t}{}_{tr}\Gamma^{\phi}{}_{t\phi} + \Gamma^{\theta}{}_{t\phi}\Gamma^{\theta}{}_{r\phi}\csc^2\theta.
\end{align}
Let us first have a closer look at the algebraic equations~\eqref{eq:Algebraic_Equations}: Because there are more equations than independent functions, it is not clear whether the equations can even be solved and because they are non-linear, one cannot expect to find unique solutions. As it turns out, the system of equations can be solved and one finds five distinct sets of solutions. In each solution set, one can express some connection components in terms of other components in a highly non-linear fashion. However, what is remarkable, is that all five solution sets share one particularly simple solution:
\begin{equation}\label{eq:CommonAlgebraicSolutions}
	\Gamma^\theta{}_{t\phi} =0 \quad\text{and}\quad \Gamma^\theta{}_{r\phi} = 0.
\end{equation}
A quick look at equation~\eqref{c1c2sol} reveals that this is equivalent to 
\begin{equation}
	c_1(r) = 0\quad\text{and}\quad c_2(r)=0,
\end{equation}
which reduces the amount of independent connection components~\eqref{eq:independent_connection_components} from twelve to ten. Moreover, if we plug~\eqref{eq:CommonAlgebraicSolutions} back into the algebraic equations, we find the simpler system
\begin{align}\label{eq:Simplified_AlgebraicEquations}
	1+\Gamma^t{}_{\theta\theta}\Gamma^\phi{}_{t\phi} + \Gamma^r{}_{\theta\theta}\Gamma^\phi{}_{r\phi} &= 0\notag\\
	\Gamma^t{}_{\theta\theta}\left(\Gamma^t{}_{tt} - \Gamma^\phi{}_{t\phi}\right) + \Gamma^t{}_{tr}\Gamma^r{}_{\theta\theta}  &= 0 \notag\\
	\Gamma^r{}_{\theta\theta}\left(\Gamma^r{}_{tr}-\Gamma^\phi{}_{t\phi}\right) + \Gamma^t{}_{\theta\theta}\Gamma^r{}_{tt}  &= 0\notag\\
	\Gamma^\phi{}_{t\phi}\left(\Gamma^t{}_{tt}-\Gamma^\phi{}_{t\phi}\right) + \Gamma^r{}_{tt}\Gamma^\phi{}_{r\phi} &= 0 \notag\\
	\Gamma^\phi{}_{r\phi}\left(\Gamma^r{}_{tr} - \Gamma^\phi{}_{t\phi}\right) + \Gamma^t{}_{tr}\Gamma^\phi{}_{t\phi} &= 0.
\end{align}
Notice that these are five equations for eight connection components, $\left\{\Gamma^t{}_{tt}, \Gamma^t{}_{tr}, \Gamma^t{}_{\theta\theta}, \Gamma^r{}_{tt}, \Gamma^r{}_{tr}, \Gamma^r{}_{\theta\theta}, \Gamma^\phi{}_{t\phi}, \Gamma^\phi{}_{r\phi} \right\}$. Hence, it is now obvious that the system is solvable but underdetermined and clearly we obtain the same five distinct sets of solutions as before. After all, we used the algebraic equations to obtain the solution~\eqref{eq:CommonAlgebraicSolutions}.

Now we turn to the differential equations. Since no matter which of the five solution sets we use, we always find~\eqref{eq:CommonAlgebraicSolutions}, we can use this solution to simplify the twelve differential equations. Notice that~\eqref{eq:CommonAlgebraicSolutions} implies that the first, the seventh and the tenth differential equation are trivially satisfied, while the last two loose their $\csc^2\theta$ terms. Furthermore, the second equation can easily be solved and we find
\begin{equation}\label{eq:Constantc}
	\partial_r \Gamma^\phi{}_{t\phi} = 0\quad\Longleftrightarrow\quad \Gamma^\phi{}_{t\phi} = c,
\end{equation}
where $c\in\mathbb R$ is a constant (since at this stage we already know that every independent connection component is purely a function of $r$). After using~\eqref{eq:Constantc}, the last differential equation turns into an algebraic equation,
\begin{equation}\label{eq:NewAlgebraicEquation}
	\Gamma^{\phi}{}_{r\phi}\left(\Gamma^{r}{}_{tr}-c\right) + \Gamma^{t}{}_{tr}\,c =0.
\end{equation}
We can therefore update our system~\eqref{eq:Simplified_AlgebraicEquations} of algebraic equations by adding~\eqref{eq:NewAlgebraicEquation} to it. This gives us a total of six algebraic equations for eight connection components. Moreover, since we solved one differential equation, three dropped out, and one turned into an algebraic equation, we are now left with seven differential equations:
\begin{align}\label{eq:DifferentialEquations}
	\partial_r\Gamma^r{}_{tr} &= \Gamma^t{}_{rr} \Gamma^r{}_{tt} - \Gamma^t{}_{tr} \Gamma^r{}_{tr}\notag\\
	\partial_r\Gamma^r{}_{tt} &= \Gamma^r{}_{tr}\left(\Gamma^r{}_{tr} - \Gamma^t{}_{tt}\right) +  \Gamma^r{}_{tt}\left(\Gamma^t{}_{tr}-\Gamma^r{}_{rr}\right) \notag\\
	\partial_r\Gamma^\phi{}_{r\phi} &= \Gamma^\phi{}_{r\phi}\left(\Gamma^r{}_{rr}-\Gamma^\phi{}_{r\phi}\right) + c\,\Gamma^t{}_{rr}\notag\\
	\partial_r\Gamma^t{}_{tt} &= \Gamma^t{}_{tr}\Gamma^r{}_{tr} - \Gamma^t{}_{rr}\Gamma^r{}_{tt}\notag\\
	\partial_r\Gamma^t{}_{tr} &= \Gamma^t{}_{rr}\left(\Gamma^t{}_{tt}-\Gamma^r{}_{tr}\right) + \Gamma^t{}_{tr}\left(\Gamma^r{}_{rr}-\Gamma^t{}_{tr}\right)\notag\\
	\partial_r\Gamma^t{}_{\theta\theta} &= \Gamma^t{}_{\theta\theta}\left(\Gamma^\phi{}_{r\phi} - \Gamma^t{}_{tr}\right)-\Gamma^t{}_{rr}\Gamma^r{}_{\theta\theta}\notag\\
	\partial_r\Gamma^r{}_{\theta\theta} &= \Gamma^r{}_{\theta\theta}\left(\Gamma^\phi{}_{r\phi}-\Gamma^r{}_{rr}\right) -\Gamma^t{}_{\theta\theta}\Gamma^r{}_{tr}.
\end{align}
Notice that these equations allow us to re-express the $r$-derivative of seven of the ten independent connection components. The only components which do not appear on the left hand side are $\Gamma^t{}_{rr}$ and $\Gamma^r{}_{rr}$ (and $\Gamma^\phi{}_{t\phi} = c$, whose derivative is trivial). Furthermore, observe that the right hand side of the first equation in~\eqref{eq:DifferentialEquations} is equal to $(-1)$ times the right hand side of the fourth equation. This means we get the following relation between the left hand sides:
\begin{equation}
	\partial_r\Gamma^r{}_{tr} = -\partial_r\Gamma^t{}_{tt}\quad\Longleftrightarrow\quad \Gamma^t{}_{tt}=k-\Gamma^r{}_{tr},
\end{equation} 
where $k\in\mathbb R$ is a constant. This follows again from the fact that all connection coefficients we are left with are purely functions of $r$. This is again a useful relation and we are left with six differential equations. As we will see later, the six differential equations play a crucial role in determining the propagating degrees of freedom of $f(\Q)$ gravity.\\

This is all the information we can extract from the differential equations at this point. The next step is to return to the algebraic equations, supplemented by the new equation~\eqref{eq:NewAlgebraicEquation}, and study the solution sets which arise from solving these equations. Before doing so, let us briefly summarize the situation thus far:
\begin{enumerate}
	\item We started with a general affine connection $\Gamma^\alpha{}_{\mu\nu}$, which has $64$ independent components.
	\item Implementing $T^\alpha{}_{\mu\nu}\overset{!}{=}0$ in subsection~\ref{ssec:SymRed} brought this number down to $40$ independent components.
	\item The first two symmetry conditions, equations~\eqref{eq:SymmetryConditionT} and~\eqref{eq:SymmetryConditionphi}, told us that all connection components are independent of the coordinates $t$ and $\phi$.
	\item The third symmetry condition, equation~\eqref{eq:SymmetryCondition1}, told us that the $20$ components~\eqref{eq:Zero_Components} are zero. Moreover, we found that two components are given solely by trigonometric functions, equation~\eqref{eq:Trig_Components}, and we found six algebraic relations in~\eqref{eq:Algebraic_Components}. This brought the number of independent components down to $40-20-2-6=12$ and we learned that the connection in coincident gauge fails to be spherically symmetric.
	\item The fourth symmetry condition, equation~\eqref{eq:SymmetryCondition2}, gave use twelve first order differential equations. These equations tell us that all twelve independent connection components are functions of $r$ alone, and no other coordinate. (Minor exception for the components in~\eqref{c1c2sol}, but they turn out to be zero later on).
	\item We then proceeded to implement $R^\mu{}_{\nu\rho\sigma}\overset{!}{=}0$ in this subsection and we found that we get twelve non-linear algebraic equations and twelve first order differential equations. The non-linear equations all have two solutions in common: $\Gamma^\theta{}_{t\phi} = 0\text{ and }\Gamma^\theta{}_{r\phi} = 0$. These solutions eliminate $c_1(r)$ and $c_2(r)$ from the list of independent connection components. Hence, we are left with the ten independent components $\left\{\Gamma^t{}_{tt}, \Gamma^t{}_{tr}, \Gamma^t{}_{rr}, \Gamma^t{}_{\theta\theta}, \Gamma^r{}_{tt}, \Gamma^r{}_{tr}, \Gamma^r{}_{rr}, \Gamma^r{}_{\theta\theta},\Gamma^\phi{}_{t\phi}, \Gamma^\phi{}_{r\phi} \right\}$, which are all functions of $r$ and nothing else.
	\item Among the differential equations we find $\partial_r \Gamma^\phi{}_{t\phi} = 0$, which tells us that $\Gamma^\phi{}_{t\phi}$ is a constant. Moreover, we found the relation $\partial_r\Gamma^r{}_{tr} = -\partial_r\Gamma^t{}_{tt}$, which implies $\Gamma^t{}_{tt} = k - \Gamma^r{}_{tr}$. This reduces our list of independent connection components to $\left\{c, k, \Gamma^t{}_{tr}, \Gamma^t{}_{rr}, \Gamma^t{}_{\theta\theta}, \Gamma^r{}_{tt}, \Gamma^r{}_{tr}, \Gamma^r{}_{rr}, \Gamma^r{}_{\theta\theta}, \Gamma^\phi{}_{r\phi}\right\}$, where $c$ and $k$ are real constants. 
	\item Finally, there are six differential equations left. These differential equations allow us to express the $r$-derivative of the connection components $\left\{\Gamma^t{}_{tr}, \Gamma^t{}_{\theta\theta}, \Gamma^r{}_{tt}, \Gamma^r{}_{tr}, \Gamma^r{}_{\theta\theta}, \Gamma^\phi{}_{r\phi}\right\}$ in terms of the other connection components. What remains un-determined are the derivatives $\partial_r\Gamma^t{}_{rr}$ and $\partial_r\Gamma^r{}_{rr}$. This means that in the field equations, only $\Gamma^t{}_{rr}$ and $\Gamma^r{}_{rr}$ can become dynamical.
	\item We also have six non-linear algebraic equations for the eight connection components $\left\{c, k, \Gamma^t{}_{tr}, \Gamma^t{}_{\theta\theta}, \Gamma^r{}_{tt},\right.$  $\left.\Gamma^r{}_{tr}, \Gamma^r{}_{\theta\theta}, \Gamma^\phi{}_{r\phi} \right\}$. Notice that these are the same variables as in bullet point 8. (apart from $c$ and $k$). Clearly, the system is underdetermined and we should expect to get more than one solution to these equations. 
\end{enumerate}
Indeed, we find that the remaining algebraic equations now admit two solution sets. Since these solution sets look independent, we will study them separately in subsections~\ref{ssec:Set1} and~\ref{ssec:set2}. We will then show in subsection~\ref{ssec:D2} that the two sets are actually related to each other by a double scaling limit. It is nevertheless convenient to distinguish between the two sets and they both play a crucial role in the study of the symmetry reduced field equations of $f(\Q)$ gravity, which will be discussed in section~\ref{sec:SymRedFieldEq}.

\subsection{Solution set 1}\label{ssec:Set1}
As mentioned in the previous subsection, we are left with six non-linear algebraic equations, given by~\eqref{eq:Simplified_AlgebraicEquations} and~\eqref{eq:NewAlgebraicEquation}. These equations do not admit a unique solution. Rather, there are two sets of solutions. For the first solution set we find that two components are zero and three components can be expressed in terms of the constants $c, k$, and the function $\Gamma^{\phi}{}_{r\phi}$:
\begin{align}
	\Gamma^t{}_{\theta\theta} &= -\frac{1}{c} & \Gamma^r{}_{tr} &= k-c & \Gamma^t{}_{tr} &= \frac{2c-k}{c}\Gamma^\phi{}_{r\phi}\notag\\
	\Gamma^r{}_{\theta\theta} &= 0 & \Gamma^r{}_{tt} &= 0.
\end{align}
Clearly, we have to assume that $c\neq 0$ for this solution set to be well-defined and admissible. Applying this solution set to the differential equations~\eqref{eq:DifferentialEquations} reduces them to four algebraic equations,
\begin{align}
	(c-k)\Gamma^\phi{}_{r\phi} &=0\notag\\
	(c-k)(2c-k) &= 0\notag\\
	(c-k)(2c-k)\Gamma^\phi{}_{r\phi} &= 0\notag\\
	c-k &= 0,
\end{align}
and two differential equations,
\begin{align}
	  c\,\Gamma^t_{rr} + \Gamma^{\phi}{}_{r\phi}(\Gamma^{r}{}_{rr}-\Gamma^{\phi}{}_{r\phi})-\partial_r \Gamma^{\phi}{}_{r\phi} &= 0\notag\\
	 (2c-k)\left(c^2\, \Gamma^t{}_{rr}+ c\,\Gamma^r{}_{rr}\Gamma^\phi{}_{r\phi}+ (k-2c)\,(\Gamma^\phi{}_{r\phi})^2-c\,\partial_r \Gamma^\phi{}_{r\phi}\right)&= 0.
\end{align}
The algebraic equations have obviously a unique solution which is given by
\begin{equation}
	c = k \neq 0.
\end{equation}
Applying this solution to the two differential equations shows that they are actually the same and we simply get
\begin{equation}\label{eq:PDrGammaprpr}
	\partial_r \Gamma^{\phi}{}_{r\phi} = c\,\Gamma^t_{rr} + \Gamma^{\phi}{}_{r\phi}(\Gamma^{r}{}_{rr}-\Gamma^{\phi}{}_{r\phi}).
\end{equation}
This is all the information we can extract from these equations. In particular, we do not find any further conditions, constraints, or integrability conditions. What we learn thus is the following: Solution set 1 describes a stationary, spherically symmetric, torsionless, and flat connection in terms of a real constant $c\neq 0$ and the three arbitrary functions $\Gamma^{t}{}_{rr}(r), \Gamma^{r}{}_{rr}(r), \Gamma^{\phi}{}_{r\phi}(r)$. We refer to these functions as the independent components which define solution set 1, because every component which belongs to solution set 1 can be expressed in terms of these functions, the constant $c$, and trigonometric functions. Moreover, the derivative $\partial_r \Gamma^{\phi}{}_{r\phi}$ can be expressed in terms of the independent functions. The same is not true for the derivatives $\partial_r \Gamma^{t}{}_{rr}$ and $\partial_r\Gamma^{r}{}_{rr}$ of the independent components $\Gamma^{t}{}_{rr}$, $\Gamma^{r}{}_{rr}$: These derivatives remain undetermined and potentially render these components dynamical.

These are the defining properties of solution set 1 and they play a crucial role in simplifying the symmetry reduced field equations in section~\ref{sec:SymRedFieldEq}. We therefore summarize all properties of solution set 1 in the following table for later convenience and reference:
\begin{table}[h!]
\centering
	\begin{tabular}{|p{0.15\textwidth}|p{0.27\textwidth} p{0.27\textwidth} p{0.27\textwidth}|}
		\hline
		& & & \\[-1.5ex]
 		\textbf{Independent} & \multicolumn{3}{l|}{All connection components of solution set 1 can be expressed in terms of the three} \\
 		\textbf{components} & \multicolumn{3}{l|}{independent functions $\Gamma^{t}{}_{rr}(r), \Gamma^{r}{}_{rr}(r), \Gamma^{\phi}{}_{r\phi}(r)$, the real constant $c\neq 0$, and trigonometric}\\
 		& \multicolumn{3}{l|}{functions.}
 		\\[1ex] \hline 
 		& & & \\[-1.5ex]
 		\textbf{Non-zero} &  \multicolumn{3}{l|}{There are twelve non-zero components in solution set 1 (all other components vanish):}\\
 		\textbf{components} & $\Gamma^{t}{}_{rr}$ & $\Gamma^{r}{}_{rr}$ & $\Gamma^{\phi}{}_{r\phi}$ \\
 		& $\Gamma^{t}{}_{tt} = c$ & $\Gamma^{t}{}_{tr} = \Gamma^{\phi}{}_{r\phi}$ & $\Gamma^{t}{}_{\theta\theta} = -\frac{1}{c}$\\
 		& $\Gamma^{t}{}_{\phi\phi} = -\frac{\sin^2\theta}{c}$ & $\Gamma^{\theta}{}_{t\theta} = c$ & $\Gamma^{\theta}{}_{r\theta} = \Gamma^{\phi}{}_{r\phi}$ \\
 		& $\Gamma^{\theta}{}_{\phi\phi} = -\cos\theta\,\sin\theta$ & $\Gamma^{\phi}{}_{t\phi} = c$ & $\Gamma^{\phi}{}_{\theta\phi} = \cot\theta$
 		\\[1ex] \hline
 		& & & \\[-1.5ex]
 		\textbf{Derivatives of} & \multicolumn{3}{l|}{Of the three independent functions, the $r$-derivative of $\Gamma^\phi{}_{r\phi}$ can be expressed as} \\
 		\textbf{independent} & \multicolumn{3}{l|}{$\partial_r \Gamma^{\phi}{}_{r\phi} = c\,\Gamma^t_{rr} + \Gamma^{\phi}{}_{r\phi}(\Gamma^{r}{}_{rr}-\Gamma^{\phi}{}_{r\phi})$,} \\
 		\textbf{components} & \multicolumn{3}{l|}{while $\partial_r\Gamma^{t}{}_{rr}$ and $\partial_r\Gamma^{r}{}_{rr}$ cannot be expressed in terms of other components.}
 		\\[1ex] \hline
	\end{tabular}	
	\caption{A concise summary of all the properties which define solution set 1. }
	\label{tab:SolutionSet1}
\end{table}
Note that alternatively one could also define $\Gamma^\phi{}_{r\phi}$ as an arbitrary function, and in turn fix $\Gamma^t{}_{rr}$ by solving \eqref{eq:PDrGammaprpr} for it. This can always be done as $c\neq 0$, and one obtains
\begin{equation}
    \Gamma^t{}_{rr}=\frac{1}{c}\left(\partial_r\Gamma^\phi{}_{r\phi}-\Gamma^\phi{}_{r\phi}(\Gamma^r{}_{rr}-\Gamma^\phi{}_{r\phi})\right)\ .
\end{equation}
One could thus also choose $\Gamma^\phi{}_{r\phi}$ as a possible connection degree of freedom, which fixes $\Gamma^t{}_{rr}$, but this is just an issue of the freely chosen parametrization of the connection.

\subsection{Solution set 2}\label{ssec:set2}
The algebraic equations~\eqref{eq:Simplified_AlgebraicEquations} and~\eqref{eq:NewAlgebraicEquation} admit a second set of solutions, which is explicitly given by
\begin{align}
	\Gamma^r{}_{tr} &= c + c\, (2c-k)\,\Gamma^t{}_{\theta\theta} & \Gamma^t{}_{tr} &= (2c-k)\frac{\Gamma^t{}_{\theta\theta}}{\Gamma^r{}_{\theta\theta}}\left(1+c\,\Gamma^t{}_{\theta\theta}\right)\notag\\
	 \Gamma^r{}_{tt} &= -c\,(2c-k)\,\Gamma^r{}_{\theta\theta} & \Gamma^\phi{}_{r\phi} &= -\frac{1+c\,\Gamma^t{}_{\theta\theta}}{\Gamma^r{}_{\theta\theta}},
\end{align}
where we have to assume $\Gamma^r{}_{\theta\theta}\neq 0$ in order to obtain well-defined expressions. Just as in the previous subsection, we insert the obtained solutions into the differential equations~\eqref{eq:DifferentialEquations}. This leads to six complicated looking equations. None of these equations is algebraic and, moreover, they are linearly dependent. After some elementary manipulations, one finally finds the following two independent expressions:
\begin{align}\label{eq:DerivativesIndComp}
	\partial_r \Gamma^t{}_{\theta\theta} &= -\frac{\Gamma^t{}_{\theta\theta}}{\Gamma^r{}_{\theta\theta}}\left[1+\Gamma^t{}_{\theta\theta}\left(3c-k+(2c-k)\,\Gamma^t{}_{\theta\theta}\right)\right] -\Gamma^t{}_{rr}\Gamma^r{}_{\theta\theta}\notag\\
	\partial_r \Gamma^r{}_{\theta\theta} &= -1 - c\, \Gamma^t{}_{\theta\theta}\left(2+(2c-k)\,\Gamma^t{}_{\theta\theta}\right)-\Gamma^r{}_{rr}\Gamma^r{}_{\theta\theta}.
\end{align}
There are now no more equations, constraints, or integrability conditions we can exploit. Hence, we find that in solution set 2 every component of the stationary, spherically symmetric, torsionless, and flat connection can be expressed in terms of two arbitrary constants, $c,k\in\mathbb R$, the four functions $\Gamma^t{}_{rr}(r),\Gamma^t{}_{\theta\theta}(r), \Gamma^r{}_{rr}(r), \Gamma^r{}_{\theta\theta}(r)$, with $\Gamma^{r}{}_{\theta\theta}\neq 0$, and trigonometric functions. Just as in the previous subsection, we refer to the above four functions as the independent connection components which define solution set 2. The derivatives $\partial_r \Gamma^{t}{}_{rr}$ and $\partial_r \Gamma^{r}{}_{rr}$ remain undetermined, while the derivatives $\partial_r \Gamma^{t}{}_{\theta\theta}$ and $\partial_r \Gamma^{r}{}_{\theta\theta}$ can be expressed in terms of the independent connection components via~\eqref{eq:DerivativesIndComp}. Table~\ref{tab:SolutionSet2} summarizes all properties of solution set 2 for later convenience and reference. 
\begin{table}[h!]
\centering
	\begin{tabular}{|p{0.15\textwidth}|p{0.27\textwidth} p{0.27\textwidth} p{0.27\textwidth}|}
		\hline
		& & & \\[-1.5ex]
 		\textbf{Independent} & \multicolumn{3}{l|}{All connection components of solution set 2 can be expressed in terms of two} \\
 		\textbf{components} & \multicolumn{3}{l|}{arbitrary constants, $c,k\in\mathbb R$, the four functions  $\Gamma^{t}{}_{rr}(r), \Gamma^{t}{}_{\theta\theta}(r),\Gamma^{r}{}_{rr}(r), \Gamma^{r}{}_{\theta\theta}(r)$,}\\
 		& \multicolumn{3}{l|}{with $\Gamma^{r}{}_{\theta\theta}(r)\neq 0$, and in terms of trigonometric functions.}
 		\\[1ex] \hline 
 		& & & \\[-1.5ex]
 		\textbf{Non-zero} &  \multicolumn{3}{l|}{There are $16$ non-zero components in solution set 2 (all other components vanish):} \\
 		\textbf{components} & \multicolumn{3}{l|}{The four independent functions $\Gamma^{t}{}_{rr}$, $\Gamma^{t}{}_{\theta\theta}$, $\Gamma^{r}{}_{rr}$, $\Gamma^{r}{}_{\theta\theta}$ and }\\
 		& $\Gamma^{t}{}_{tt} = -c + k - c(2c-k)\Gamma^{t}{}_{\theta\theta}$ & $\Gamma^{t}{}_{tr} = \frac{(2c-k)\Gamma^{t}{}_{\theta\theta}(1+c\,\Gamma^{t}{}_{\theta\theta})}{\Gamma^{r}{}_{\theta\theta}}$ & $\Gamma^{t}{}_{\phi\phi} = \sin^2\theta\, \Gamma^{t}{}_{\theta\theta}$ \\
 		& $\Gamma^{r}{}_{tt} = -c(2c-k)\Gamma^{r}{}_{\theta\theta}$ & $\Gamma^{r}{}_{tr} = c+c(2c-k)\Gamma^{t}{}_{\theta\theta}$ & $\Gamma^{r}{}_{\phi\phi} = \sin^2\theta\, \Gamma^{r}{}_{\theta\theta}$ \\
 		& $\Gamma^{\theta}{}_{t\theta} = c$ & $\Gamma^{\theta}{}_{r\theta} = -\frac{1+c\,\Gamma^{t}{}_{\theta\theta}}{\Gamma^{r}{}_{\theta\theta}}$ & $\Gamma^{\theta}{}_{\phi\phi} = -\cos\theta\,\sin\theta$ \\
 		& $\Gamma^{\phi}{}_{t\phi} = c$ & $\Gamma^{\phi}{}_{r\phi} = -\frac{1+c\, \Gamma^{t}{}_{\theta\theta}}{\Gamma^{r}{}_{\theta\theta}}$ & $\Gamma^{\phi}{}_{\theta\phi} = \cot\theta$
 		\\[1ex] \hline
 		& & & \\[-1.5ex]
 		\textbf{Derivatives of} & \multicolumn{3}{l|}{Of the four independent functions, the $r$-derivatives of $\Gamma^{t}{}_{\theta\theta}$ and $\Gamma^{r}{}_{\theta\theta}$ can be expressed as} \\
 		\textbf{independent} & \multicolumn{3}{l|}{$\partial_r \Gamma^t{}_{\theta\theta} = -\frac{\Gamma^t{}_{\theta\theta}}{\Gamma^r{}_{\theta\theta}}\left[1+\Gamma^t{}_{\theta\theta}\left(3c-k+(2c-k)\,\Gamma^t{}_{\theta\theta}\right)\right] -\Gamma^t{}_{rr}\Gamma^r{}_{\theta\theta}$} \\
 		\textbf{components} & \multicolumn{3}{l|}{$\partial_r \Gamma^r{}_{\theta\theta} = -1 - c\, \Gamma^t{}_{\theta\theta}\left(2+(2c-k)\,\Gamma^t{}_{\theta\theta}\right)-\Gamma^r{}_{rr}\Gamma^r{}_{\theta\theta}$,}\\
 		& \multicolumn{3}{l|}{while $\partial_r\Gamma^{t}{}_{rr}$ and $\partial_r\Gamma^{r}{}_{rr}$ cannot be expressed in terms of other components.}
 		\\[1ex] \hline
	\end{tabular}	
	\caption{A concise summary of all the properties which define solution set 2. }
	\label{tab:SolutionSet2}
\end{table}

Instead of regarding $\Gamma^t{}_{rr}$ and $\Gamma^r{}_{rr}$ as being the free functions, which then fix the first derivatives of $\Gamma^t{}_{\theta\theta}$ and $\Gamma^r{}_{\theta\theta}$, one could define the latter two as free functions, and in turn determine the former by the equations
\begin{align}
\label{eq:Gammatrr}    \Gamma^t_{\ rr}&=\frac{1}{\Gamma^r_{\ \theta\theta}}\left(-\frac{\Gamma^t{}_{\theta\theta}}{\Gamma^r{}_{\theta\theta}}\left[1+\Gamma^t{}_{\theta\theta}\left(3c-k+(2c-k)\,\Gamma^t{}_{\theta\theta}\right)\right]-\partial_r \Gamma^t{}_{\theta\theta}\right)\\
\label{eq:Gammarrr}    \Gamma^r_{\ rr}&=\frac{1}{\Gamma^r_{\ \theta\theta}}\left(-1 - c\, \Gamma^t{}_{\theta\theta}\left(2+(2c-k)\,\Gamma^t{}_{\theta\theta}\right)-\partial_r \Gamma^r{}_{\theta\theta}\right).
\end{align}
Hence the (potential) connection degrees of freedom of $\Gamma^t{}_{rr}$ and $\Gamma^r{}_{rr}$ may also be described by $\Gamma^t{}_{\theta\theta}$ and $\Gamma^r{}_{\theta\theta}$, but this again just a choice of parametrization.

Finally, we observe that the spherical connection used in~\cite{Zhao:2021,Lin:2021} falls into solution set 2. Indeed, if we choose $\{c,k,\Gamma^{t}{}_{rr}, \Gamma^{t}{}_{\theta\theta}, \Gamma^{r}{}_{rr}, \Gamma^{r}{}_{\theta\theta}\} = \{0,0,0,0,0,-r\}$, then the only non-vanishing connection components of solution set 2 are
\begin{align}
	\Gamma^{r}{}_{\theta\theta} &= -r  & \Gamma^{r}{}_{\phi\phi} &= -r\,\sin^2\theta & \Gamma^{\theta}{}_{r\theta} &= \frac{1}{r} \notag\\
	\Gamma^{\theta}{}_{\phi\phi} &= -\cos\theta\,\sin\theta & \Gamma^{\phi}{}_{r\phi} &= \frac{1}{r} & \Gamma^{\phi}{}_{\theta\phi} &= \cot\theta,
\end{align}
which agrees precisely with the connection used in~\cite{Zhao:2021,Lin:2021}.

\subsection{Solution set 1 as the double scaling limit of solution set 2}\label{ssec:D2}
Solution sets 1 and 2 have certain similarities, even though solution set 2 has more free functions. We will now show that solution set 1 can indeed be obtained from solution set 2 by a double scaling limit.

The key component to consider is $\Gamma^\phi{}_{r\phi}$ of solution set 2, which  can be expressed in terms of $\Gamma^t{}_{\theta\theta}$ and $\Gamma^r{}_{\theta\theta}$ (see Table~\ref{tab:SolutionSet2}). For solution set 1, $\Gamma^\phi{}_{r\phi}$ is an arbitrary free function of $r$. These facts suggest the following parametrization of $\Gamma^t{}_{\theta\theta}$ and $\Gamma^r{}_{\theta\theta}$:
\begin{align}\label{eq:Param}
	\Gamma^t{}_{\theta\theta} &= -\frac{1}{c} + \frac{\lambda}{c}\Phi &\text{and}&& \Gamma^r{}_{\theta\theta} &= \lambda\Psi,
\end{align}
where we have to assume $c\neq 0$, $\lambda$ is a parameter, and $\Phi$ and $\Psi$ are arbitrary functions of $r$. With this parametrization one obtains in the $\lambda\to 0$ limit and under the assumption that $k=c$
\begin{equation}
	\lim_{\lambda\to 0}\Gamma^{\phi}{}_{r\phi} = -\frac{\Phi}{\Psi}.
\end{equation}
In words: The component $\Gamma^{\phi}{}_{r\phi}$ of solution set 2 is mapped to an arbitrary function of $r$, just as required by solution set 1. The parametrization~\eqref{eq:Param} also ensures that the other non-trivial components of solution set 2 are correctly mapped to their counterparts in solution set 1:
\begin{align}
	\left.\lim_{\lambda\to 0}\Gamma^{t}{}_{tt}\right|_{k=c} &= c & \left.\lim_{\lambda\to 0}\Gamma^{t}{}_{tr}\right|_{k=c} & = -\frac{\Phi}{\Psi} & \left.\lim_{\lambda\to 0}\Gamma^{t}{}_{\phi\phi}\right|_{k=c} &= -\frac{\sin^2\theta}{c}\notag\\
	\left.\lim_{\lambda\to 0}\Gamma^{r}{}_{tt}\right|_{k=c} &= 0 & \left.\lim_{\lambda\to 0}\Gamma^{r}{}_{tr}\right|_{k=c} &= 0 & \left.\lim_{\lambda\to 0}\Gamma^{r}{}_{\phi\phi}\right|_{k=c} &= 0 \notag\\
	\left.\lim_{\lambda\to 0}\Gamma^{\theta}{}_{r\theta}\right|_{k=c} &= -\frac{\Phi}{\Psi}.
\end{align}
What remains to be checked is whether the double scaling limit also allows us to recover equation~\eqref{eq:PDrGammaprpr} from the equations~\eqref{eq:DerivativesIndComp}. Indeed, when plugging the parametrization~\eqref{eq:Param} into~\eqref{eq:DerivativesIndComp}, one finds
\begin{align}
	\partial_r\Phi &= \frac{\Phi^2}{\Psi} -c\,\Gamma^{t}{}_{rr}\Psi-\lambda\frac{\Phi^3}{\Psi}\notag\\
	\partial_r\Psi &= -\Gamma^{r}{}_{rr}\Psi - \lambda\Phi^2.
\end{align}
In turn, these relations can be used to show that
\begin{align}
	\left.\lim_{\lambda\to 0}\partial_r\Gamma^{\phi}{}_{r\phi}\right|_{k=c} &= -\partial_r\left(\frac{\Phi}{\Psi}\right) = c\, \Gamma^{t}{}_{rr}+ \Gamma^{\phi}{}_{r\phi}\left(\Gamma^{r}{}_{rr}-\Gamma^{\phi}{}_{r\phi}\right).
\end{align}
That is to say: We correctly recover equation~\eqref{eq:PDrGammaprpr} of solution set 1. Thus, we have succeeded in showing that solution set 1 can be obtained from solution set 2 via a double scaling limit. We can therefore regard solution set 2 as the most general parametrization of a connection which is stationary, spherically symmetric, torsionless, and flat.\\

There is also another point of view one can take, which regards the choice of free connection variables. The main difference between solution sets 1 and 2  is the parametrization; solution set 1 is given by the free function $\Gamma^\phi{}_{r\phi}$, while in solution set 2 $\Gamma^t{}_{\theta\theta}$ is free and $\Gamma^\phi{}_{r\phi}$ is fixed. In order to see in a different way that solution set 1 can be obtained from 2, we switch the parametrization in set 2, i.e., we now leave $\Gamma^\phi{}_{r\phi}$ free and set
\begin{equation}
    \Gamma^t{}_{\theta\theta} = -\frac{1+\Gamma^r{}_{\theta\theta}\Gamma^\phi{}_{r\phi}}{c}.
\end{equation}
With this, one obtains the remaining components of the connection in solution set 2 as
\begin{align}\label{eq:ConnectionAlternativeParametrization}
	\Gamma^t{}_{tt} &= c+(2c-k) \Gamma^r{}_{\theta\theta} \Gamma^\phi{}_{r\phi}, &  \Gamma^t{}_{tr} &=\frac{2c-k}{c}(1+{\Gamma^r}_{\theta\theta}{\Gamma^\phi}_{r\phi}){\Gamma^\phi}_{r\phi}, &  	{\Gamma^r}_{tt} &=-c(2c-k){\Gamma^r}_{\theta\theta}\notag\\
	{\Gamma^r}_{tr}&=k-c-(2c-k){\Gamma^r}_{\theta\theta}{\Gamma^\phi}_{r\phi} & {\Gamma^r}_{\phi\phi} &= \sin^2\theta\, {\Gamma^r}_{\theta\theta} & {\Gamma^\theta}_{t\theta} &={\Gamma^\phi}_{t\phi}=c\notag\\
	{\Gamma^\theta}_{r\theta}&={\Gamma^\phi}_{r\phi} & {\Gamma^t}_{\theta\theta}&= -\frac{1+{\Gamma^\phi}_{r\phi}{\Gamma^r}_{\theta\theta}}{c} & {\Gamma^t}_{\phi\phi} &=-\sin^2\theta\,\frac{1+{\Gamma^\phi}_{r\phi}{\Gamma^r}_{\theta\theta}}{c}\notag\\
	{\Gamma^\theta}_{\phi\phi} & =-\cos \theta\sin\theta & {\Gamma^\phi}_{\theta\phi} & =\cot\theta
\end{align}
The flatness conditions in this parametrization becomes \eqref{eq:PDrGammaprpr} and
\begin{align}\label{eq:DerivativesIndCompAltPara}
	\partial_r \Gamma^r{}_{\theta\theta} &= -1 + \left(1+\Gamma^r_{\ \theta\theta}\Gamma^\phi_{\ r\phi}\right) \left( 2-\left(2-\frac{k}{c}\right)\left(1+\Gamma^r_{\ \theta\theta}\Gamma^\phi_{\ r\phi}\right)\right) - \Gamma^r{}_{rr}\Gamma^r{}_{\theta\theta}.
\end{align}
Now in this parametrization of solution set 2 we first observe that naivly we would now want to set $c\neq 0$; but of course this limit can be taken without a problem, as the parametrization from above shows. This demonstrates that one has to choose the correct parametrizations for certain limits. We also see in this parametrization that we can now safely set $\Gamma^r_{\ \theta\theta}=0$ and $k=c$ to obtain solution set 1, where we must now, \textit{after} taking these limits, demand $c\neq 0$. Note that if one puts $\Gamma^r_{\ \theta\theta}=0$ one has to put $k=c$ as well to have still zero curvature. This shows that solution set 1 is contained in set 2, modulo the parametrization.

\subsection{Diagonalizing the metric, canonical $\mathcal S^2$ part, and transformation behavior of the solution sets}\label{ssec:E}
Let us recall that our goal is to study the field equations of $f(\Q)$ gravity for a stationary and spherically symmetric affine geometry $(\M, g_{\mu\nu}, \Gamma^\alpha{}_{\mu\nu})$. To that end, we have performed a symmetry reduction of the connection in subsection~\ref{ssec:SymRed} and we have implemented the postulates of symmetric teleparallelism in the subsections~\ref{ssec:SymRed} and~\ref{ssec:RiemannZero}. This led us to the two solution sets studied in the subsections~\ref{ssec:Set1} and~\ref{ssec:set2}. These two sets arose from demanding that the Riemann tensor of the affine connection vanishes and both sets represent a stationary, spherically symmetric, torsionless, and flat connection. Moreover, we have seen in subsection~\ref{ssec:D2} that solution set 1 is the double scaling limit of solution set 2 and this set cannot be further simplified using symmetries or the postulates of symmetric teleparallelism.

However, what we can try to further simplify is the metric. At the beginning of section~\ref{sec:SymRed}, we have stated that the most general metric which is spherically symmetric and stationary takes the form~\eqref{eq:GeneralMetric}. This metric can be brought into a simpler form via the application of a diffeomorphism which eliminates the off-diagonal component $g_{tr}$ and which brings the $\mathcal S^2$ part of the metric into canonical form (i.e., it maps $g_{\theta\theta}$ to $r^2$). Of course, if we apply this diffeomorphism to the metric, we also need to apply it to the connection. What we will show now, is that the metric can always be brought into the described form and that the diffeomorphism which allows us to do so maps the solution sets of the connection onto themselves. In other words: The diffeomorphism which simplifies the metric preserves the structure of the solution sets.\\

To begin with, we notice that the symmetry reduced metric~\eqref{eq:GeneralMetric} describes a warped geometry. This means the following: Let $(\mathcal B, h)$ and $(\mathcal F, \sigma)$ be (pseudo-) Riemannian manifolds, where $\mathcal B$ is called the \textit{base space} and $\mathcal F$ the \textit{fiber}. Furthermore, let $f:\mathcal B\to \mathbb R_{>0}$ be a strictly positive function on the base space, called the \textit{warping factor}. A \textit{warped geometry} is then the manifold $\mathcal M := \mathcal B\times_f \mathcal F$ which is the topological space $\mathcal B\times \mathcal F$ endowed with the metric tensor $g:=h\oplus f\sigma$. Concretely, this means that the metric on the \textit{total space} $\mathcal M$ can be written as a metric tensor on $\mathcal B$ plus a metric tensor on $\mathcal F$ times a positive function which depends on the coordinates of $\mathcal B$. In our case, we have $\mathcal B=\mathbb R\times\mathbb R_{>0}$, $\mathcal F=\mathcal S^2$, and $f(t,r) = g_{\theta\theta}(r)$. Notice that it follows from the fact that the signature of~\eqref{eq:GeneralMetric} is $(-,+,+,+)$ that $g_{\theta\theta}$ is a strictly positive function of $r$. We can therefore write the metric~\eqref{eq:GeneralMetric} equivalently as the warped metric
\begin{equation}
	g = h_{tt}\,\dd t^2 + 2h_{tr}\,\dd t\,\dd r + h_{rr}\, \dd r^2 + f(r)\,\dd\Omega^2,
\end{equation}
where $\dd\Omega^2 := \dd \theta^2 + \sin^2\theta\,\dd\phi$ is the canonical metric on the unit $2$-sphere $\mathcal S^2$. Given the fact that $f$ is a strictly positive function, we can write it in the suggestive form $f(r) = \rho(r)^2$. This is merely a notational convention and we have not yet changed anything. But now we choose $\rho$ as a new coordinate and we perform the change of coordinates on the base space $\mathcal B$ from $(t, r)$ to $(t, \rho)$. This is a rather trivial manipulation and the metric tensor in the new coordinate system reads
\begin{equation}\label{eq:Canonical1}
	g = h'_{tt} \, \dd t^2 + 2 h'_{t\rho}\, \dd t\, \dd \rho + h'_{\rho\rho}\,\dd \rho^2 + \rho^2\dd \Omega^2,
\end{equation}
where the transformed metric components are given by
\begin{align}
	h'_{tt} &= h_{tt},& h'_{t\rho} &=  \left(\PD{r}{\rho}\right) h_{tr}, & h'_{\rho\rho} &= \left(\PD{r}{\rho}\right)^2 h_{rr}.
\end{align}
Hence, we can always bring the $\mathcal S^2$ part of the metric into canonical form as long as $\PD{\rho}{r} \equiv \frac{1}{2\sqrt{g_{\theta\theta}}}\PD{g_{\theta\theta}}{r} \neq 0$, which is tantamount to demanding that $g_{\theta\theta}$ is a monotonous function of $r$.\\
Let us denote the diffeomorphism which brings the metric into the form~\eqref{eq:Canonical1} by $\phi_1:\mathcal B\to\mathcal B$ and let us now ask if there exists a second diffeomorphism $\phi_2:\mathcal B\to\mathcal B$ which eliminates the off-diagonal term $h'_{t\rho}$. To that end, we assume that $\phi_2$ only acts on $t$ in the sense that it generates the new coordinate system $(\tau, \rho)$ from $(t,\rho)$. Under this assumption, the $1$-form $\dd t$ transforms as
\begin{equation}\label{eq:1Form}
	\dd t = \left(\PD{t}{\tau}\right)\dd \tau + \left(\PD{t}{\rho}\right)\dd\rho.
\end{equation}
Moreover, we demand that the $\mathcal B$-part of the metric tensor in the new coordinates takes the form
\begin{equation}\label{eq:Bpart}
	\left.g\right|_\mathcal{B} = h \overset{!}{=} \tilde{h}_{\tau\tau}\,\dd\tau^2 + \tilde{h}_{\rho\rho}\,\dd\rho^2,
\end{equation}
for some (not further specified) functions $\tilde{h}_{\tau\tau}$ and $\tilde{h}_{\rho\rho}$. By inserting~\eqref{eq:1Form} into~\eqref{eq:Bpart} we obtain the following conditions
\begin{equation}
	\begin{cases}
		h'_{tt} & \overset{!}{=}\ \ \tilde{h}_{\tau\tau} \left(\PD{\tau}{t}\right)^2\\
		h'_{t\rho} & \overset{!}{=}\ \  \tilde{h}_{\tau\tau}\left(\PD{\tau}{t}\right)\left(\PD{\tau}{\rho}\right)\\
		h'_{\rho\rho} & \overset{!}{=}\ \  \tilde{h}_{\rho\rho} + \tilde{h}_{\tau\tau}\left(\PD{\tau}{\rho}\right)^2.
	\end{cases}
\end{equation}
These are three equations for the three unknown functions $\tau(t,\rho)$, $\tilde{h}_{\tau\tau}$, and $\tilde{h}_{\rho\rho}$. Hence, the problem is in principle solvable, provided one specifies initial conditions for $\tau(t,\rho)$. We can therefore eliminate the off-diagonal term $h'_{t\rho}$.\\
All in all, we have defined two diffeomorphisms on the base space $\mathcal B$ which diagonalize the metric and bring the $\mathcal S^2$ part into canonical form. We can also combine these two diffeomorphisms into $\phi_2\circ\phi_1:\mathcal B\to\mathcal B$, with $(\phi_2\circ\phi_1)(t,r) = (\tau,\rho)$. 

In the sequel, we will drop all primes and tildes and denote the metric components again by $g_{\mu\nu}$, rather than $h_{\mu\nu}$. Also, we change notation and write $(t,r)$ for $(\tau,\rho)$ (this is for simplicity and should not cause any confusion). After applying $\phi_2\circ\phi_1$, the metric can then be written as
\begin{equation}\label{eq:SimplerMetric}
	g_{\mu\nu} = \begin{pmatrix}
		g_{tt} & 0 & 0 & 0\\
		0 & g_{rr} & 0 & 0\\
		0 & 0 & r^2 & 0\\
		0 & 0 & 0 & r^2\sin^2\theta
	\end{pmatrix}.
\end{equation}
This is the form of the metric which we shall use from now on and which will greatly simplify the field equations to be studied in section~\ref{sec:SymRedFieldEq}. However, we also need to determine how the connection transforms under $\phi_2\circ\phi_1$. In general, a connection transforms as 
\begin{equation}\label{eq:TransfConnection}
	\tilde{\Gamma}^{\alpha}{}_{\mu\nu} = \PD{\tilde{x}^\alpha}{x^\beta}\PD{x^\rho}{\tilde{x}^\mu}\PD{x^\sigma}{\tilde{x}^\nu} \Gamma^{\beta}{}_{\rho\sigma} + \PD{\tilde{x}^\alpha}{x^\lambda}\PD{^2 x^\lambda}{\tilde{x}^\mu\partial\tilde{x}^\nu}.
\end{equation} 
under coordinate transformations and this could potentially spoil the structure of the solution sets. However, notice that $\phi_2\circ\phi_1$ only acts on the base space $\mathcal B$ and that it leaves the fiber $\mathcal F = \mathcal S^2$ invariant. Intuitively, we would therefore expect that our diffeomorphism is compatible with spherical symmetry. Indeed, it is obvious that the metrics~\eqref{eq:GeneralMetric} and~\eqref{eq:SimplerMetric} possess the same Killing vectors and hence share the same isometry group. This confirms that the diffeomorphism $\phi_2\circ\phi_1$ respects spherical symmetry. Moreover, a diffeomorphism can neither produce curvature nor torsion. But if $\tilde{\Gamma}^{\alpha}{}_{\mu\nu}$, i.e., the transformed connection, is invariant under the isometries generated by the Killing vector fields and if it is torsionless and flat, it gives rise to the same two solution sets we have discussed in previous subsections. Hence, we see that the diffeomorphism $\phi_2\circ\phi_1$ maps the solution sets onto themselves. This can also be confirmed by a direct computation, see Appendix~\ref{app:A}.

The importance of this result is that it allows us to use the simpler form of the metric, given by equation~\eqref{eq:SimplerMetric}, together with solution sets 1 and 2 for the connection. The metric in the form~\eqref{eq:SimplerMetric} together with solution sets~1 and~2 for the connection constitute the simplest and yet most general metric-affine geometries which are stationary, spherically symmetric, torsionless, and flat.

\section{Symmetry reduced field equations for the metric and the connection}\label{sec:SymRedFieldEq}\setcounter{equation}{0}
In section~\ref{sec:SymRed} we have performed a detailed symmetry reduction of the metric-affine geometry $(\M, g_{\mu\nu}, \Gamma^\alpha{}_{\mu\nu})$ and we have found that the simplest --and yet most general form-- of a metric-affine geometry which is stationary, spherically symmetric, torsionless, and flat is given by the metric~\eqref{eq:SimplerMetric} and the connection has to be chosen such that it belongs either to solution set 1 (see Table~\ref{tab:SolutionSet1}) or to solution set 2 (see Table~\ref{tab:SolutionSet2}). 

We now use the metric-affine geometries described above to perform a symmetry reduction of the metric and connection field equations~\eqref{eq:FieldEquations}. We first discuss the symmetry reduced field equations for solution set 2, since, as we have seen in subsection~\ref{ssec:D2}, solution set 1 can be obtained from solution set 2 by means of a double scaling limit. Hence, once we understand the field equations for solution set 2, we can immediately derive consequences for the field equations of solution set 1.

However, before studying the symmetry reduction in detail, we have a brief look at the structure of the field equations. For both solution sets, the field equations have the following form:
\begin{align}
&\text{Structure of metric field equations: }
	& &\begin{pmatrix}
		\M_{tt} & \M_{tr} & 0 & 0\\
		\M_{tr} & \M_{rr} & 0 & 0\\
		0 & 0 & \M_{\theta\theta} & 0\\
		0 & 0 & 0 & \M_{\theta\theta}\,\sin^2\theta
	\end{pmatrix} \notag\\
&\text{Structure of connection field equations: }
	&&\begin{pmatrix}
		\mathcal C_t \\
		\mathcal C_r \\
		0 \\
		0
	\end{pmatrix}
\end{align}
This is true for the vacuum case as well as in the presence of spherically symmetric matter distributions. What we observe is that there are at most four independent metric field equations and at most two independent connection field equations. Moreover, it turns out that we can learn a lot just by looking at the off-diagonal equation $\M_{tr}$. In the following subsection we will study $\M_{tr}$ for solution set 2 and we will be able to derive necessary conditions for the existence of solutions which go beyond Schwarzschild-deSitter-Nordstr\"{o}m. We also note here for completeness the forms of the scalar $\Q$. For solution set 1 we find
\begin{equation}
    \Q=\frac{g_{tt} \left(2 g_{rr} (3 r (c\, r \Gamma^t{}_{rr}+\Gamma^{\phi}{}_{r\phi} (r \Gamma^r{}_{rr}-r \Gamma^{\phi}{}_{r\phi}+2))-2 g_{rr}-2)-3 r^2 \Gamma^{\phi}{}_{r\phi} \partial_r g_{rr}\right)+r g_{rr} \partial_r g_{tt} (3 r \Gamma^{\phi}{}_{r\phi}-4)}{2 r^2\, (g_{rr}{})^2 g_{tt}}
\end{equation}
while for solution set 2 we get
\begin{align}
    \nonumber \Q &=\frac{1}{2 r^2 g_{rr}{}^2 g_{tt}{}^2 \Gamma^r{}_{\theta \theta}{}^2}\Bigg(g_{rr}{}^2 \Gamma^r{}_{\theta\theta}{}^2 \left(c r^2 \Gamma^r{}_{\theta \theta} \partial_r g_{tt} (k-2 c)-2 c r g_{tt} (2 c-k) (c r \Gamma^t{}_{\theta\theta} (2 c \Gamma^t{}_{\theta\theta}-k \Gamma^t{}_{\theta\theta}+2)+\right.\\
    \nonumber &+ \left. \Gamma^r{}_{\theta\theta} (r \Gamma^r{}_{rr}-2)+r)+4 g_{tt}{}^2 (c \Gamma^t{}_{\theta\theta} (2 c \Gamma^t{}_{\theta\theta}-k \Gamma^t{}_{\theta\theta}+2)+\Gamma^r{}_{\theta\theta} \Gamma^r{}_{rr})-2 g_{tt} \Gamma^r{}_{\theta\theta} \partial_r g_{tt}\right)+\\
    \nonumber &+ g_{rr} g_{tt} \left(2 g_{tt} \left(\Gamma^r{}_{\theta\theta}{}^2 \left(r^2 \Gamma^t{}_{rr} (2 c \Gamma^t{}_{\theta\theta} (k-2 c)+k)-2\right)+\right.\right.\\
    \nonumber &+\left.\left.r \Gamma^r{}_{\theta\theta} (c \Gamma^t{}_{\theta \theta}+1) (r \Gamma^r{}_{rr}+2) (2 c \Gamma^t{}_{\theta\theta}-k \Gamma^t{}_{\theta\theta}-2)-\left(\Gamma^t{}_{\theta\theta}{}^2 (k-2 c)^2+2\right) (c r \Gamma^t{}_{\theta\theta}+r){}^2-\right.\right.\\
    \nonumber &-\left.\left.\Gamma^r{}_{\theta\theta}{}^3 \partial_r g_{rr}\right)+r \Gamma^r{}_{\theta\theta} \left(c r \Gamma^r{}_{\theta\theta}{}^2 \partial_r g_{rr} (2 c-k)-\partial_r g_{tt} (r (\Gamma^t{}_{\theta\theta} (c \Gamma^t{}_{\theta\theta} (k-2 c)+k)+2)+4 \Gamma^r{}_{\theta\theta})\right)\right)+\\
    &+r^2 g_{tt}{}^2 \Gamma^r{}_{\theta\theta} \partial_r g_{rr} (c \Gamma^t{}_{\theta\theta}+1) (\Gamma^t{}_{\theta\theta} (k-2 c)+2)\Bigg).
\end{align}

\subsection{Off-diagonal metric field equation for solution set 2}\label{ssec:SymRedFieldEq2}
For solution set 2, the off-diagonal metric field equation takes the form
\begin{equation}\label{eq:OffDiagSolset2}
	\M_{tr} = \frac12 \left(k - 2 c\,(2c-k)\Gamma^{t}{}_{\theta\theta}\right) \partial_r\Q\, f''(\Q) = 0,
\end{equation}
which is valid in the vacuum as well as in the electro-vacuum case. This equation can be solved in three distinct ways which we will discuss in turn. The three different possible solutions are
\begin{enumerate}
	\item $\partial_r \Q = 0$;
	\item $f''(\Q) = 0$;
	\item $\frac12 \left(k - 2 c\,(2c-k)\Gamma^{t}{}_{\theta\theta}\right) = 0$.
\end{enumerate}
The first option is tantamount to saying that the non-metricity scalar, when evaluated on a solution of the field equations, is constant. But this cannot yield a solution to the field equations which goes beyond Schwarzschild-deSitter-Nordstr\"{o}m. This can be readily seen from the alternative form~\eqref{eq:RewrittenMetricFieldEq} of the metric field equations. Let us re-write these equations here for convenience:
\begin{equation}\label{eq:AlternativeForm}
	f'(\Q) G_{\mu\nu}  + \frac12 g_{\mu\nu} (f(\Q)-f'(\Q)\Q) + 2 f''(\Q) P^\alpha{}_{\mu\nu} \partial_\alpha\Q  = T_{\mu\nu}.
\end{equation}
If we assume that a solution to the connection and metric field equations exists such that $\Q_\sol =\textsf{const}.$ - we will see in subsection~\ref{sec:Solutions} that such solutions indeed exist, but one can also use that $\Q$ is linear in $\Gamma^r_{\ rr}$, hence one can solve the equations $\Q=\Q_\text{sol}$ for $\Gamma^r_{\ rr}$ for any $\Q_\text{sol}\in\mathbb{R}$ to obtain such solutions by a choice of the connection - then the last term on the left hand side of~\eqref{eq:AlternativeForm} vanishes and we are left with
\begin{equation}
	G_{\mu\nu} + \Lambda_\textsf{eff} \, g_{\mu\nu} = \bar{T}_{\mu\nu},
\end{equation}
where, as we recall, $G_{\mu\nu}$ is the standard Einstein tensor with respect to the Levi-Civita connection and where we have defined
\begin{align}
	\Lambda_\textsf{eff} &:= \frac{1}{2}\frac{f(\Q_\textsf{sol})-f'(\Q_\textsf{sol})\Q_\textsf{sol}}{f'(\Q_\textsf{sol})}\notag\\
	\bar{T}_{\mu\nu} &:= \frac{1}{f'(\Q_\textsf{sol})} T_{\mu\nu}.
\end{align}
The first term, $\Lambda_\textsf{eff}$, is clearly constant and simply represents an effective cosmological constant, while $\bar{T}_{\mu\nu}$ is just a re-scaled energy-momentum tensor. Notice that dividing by $f'$ is allowed since we need to assume $f'\neq 0$ in order to obtain non-trivial and self-consistent field equations. We therefore reach the conclusion that assuming $\Q_\textsf{sol}=\text{const.}$ can only produce the Schwarzschild-deSitter-Nordstr\"{o}m solution for arbitrary $f$.\\

Let us now consider the second option, namely that the off-diagonal field equation $\M_{tr}$ is solved by $f''(\Q) = 0$. It is important to notice that this equation has to hold \textit{on a solution} of all field equations, i.e., we should write $f''(\Q_\sol) = 0$. We can now easily show that when $f''(\Q_\sol) = 0$ \textit{and} $\Q_\sol \neq$ const., then it follows that $f$ is necessarily of the form $f(\Q) = a\, \Q + b$, where $a$ and $b$ are real constants and this form of $f$ holds everywhere, not just on solutions. To see this, let us first assume that $f(\Q)$ does \textit{not} have the above form. Then it follows that $f''(\Q_\sol) = 0$ is an equation which will be satisfied for at least one $\Q_\sol$ (at least one solution needs to exist, otherwise $f''(\Q_\sol)\neq 0$, but the solution needs not be unique). But if we fix $\Q_\sol$ via this equation, we find that $\Q_\sol$ is a constant! Hence, we fall back into the GR regime for arbitrary $f$, which we have already discussed above. If, on the other hand side, we have $f(\Q) = a\, \Q + b$, then $f''(\Q_\sol) = 0$ is trivially satisfied and $\Q_\sol$ is not a constant. This is what we wanted to prove.\\

We reach the conclusion that solving $\M_{tr} = 0$ via $f''(\Q) = 0$ automatically leads us into the symmetric teleparallelism sector of $f(\Q)$ gravity. In particular, this means that we can only get the Schwarzschild-deSitter-Nordstr\"{o}m solution and nothing else. There is only one little caveat, which also applies to solving $\M_{tr}$ via $\partial_r \Q = 0$: We have only used the metric field equations but completely neglected the connection field equations. However, it is easy to see that these equations will not give rise to any constraints or inconsistencies which would alter the conclusions we have reached thus far. In fact, the connection field equations can be written in the schematic form
\begin{align}\label{eq:HigherOrderPoly}
	\mathcal C_t &= A\,\left(\partial_r\Q\right)\, f''(\Q) + B\, \left(\partial^2_r\Q\right)\, f''(\Q) + C\, \left(\partial_r\Q\right)^2\, f^{(3)}(\Q) = 0\notag\\
	\mathcal C_r &= D\,\left(\partial_r\Q\right)\, f''(\Q) + E\, \left(\partial^2_r\Q\right)\, f''(\Q) + F\, \left(\partial_r\Q\right)^2\, f^{(3)}(\Q) = 0,
\end{align}
where $A$, $B$, $C$, $D$, $E$, and $F$ are complicated functions of the metric components, their derivatives, and connection components. What matters is that the connection equations can be written as polynomials in $\partial_r\Q$, $f''(\Q)$, and their higher order derivatives, as indicated by equation~\eqref{eq:HigherOrderPoly}. This makes it obvious that the solutions $\partial_r\Q = 0$ or $f''(\Q)=0$ to $\M_{tr}$ automatically also satisfy the connection field equations. Hence, no further constraints appear and the equations are self-consistent. We can therefore confirm that solving $\M_{tr} = 0$ via $\partial_r \Q = 0$ or $f''(\Q) = 0$ only gives the Schwarzschild-deSitter-Nordstr\"{o}m solution (for arbitrary $f$ in the case of $\partial_r \Q = \textsf{const}$, while $f''(\Q)$ simply reduces $f(\Q)$ to standard symmetric teleparallelism, as one would expect). \\

What remains to be examined is the third option. Namely, that $\M_{tr} = 0$ is solved by
\begin{equation}\label{eq:AWayOut}
	k - 2 c\,(2c-k)\Gamma^{t}{}_{\theta\theta} = 0.
\end{equation}
This is a \textit{constraint equation for the connection} and we will see shortly that it does indeed eliminate one of two \textit{potential} degrees of freedom of the connection. To that end, we solve~\eqref{eq:AWayOut} for $\Gamma^{t}{}_{\theta\theta}$, which gives us
\begin{align}\label{eq:Option1}
	\Gamma^{t}{}_{\theta\theta} &= \frac{k}{2c(2c-k)}\quad\text{for } c\neq 0\text{ and } k\neq 2c.
\end{align}
Notice that since $c$ and $k$ are constants, $\Gamma^{t}{}_{\theta\theta}$ is now constrained to be a constant. In particular, this means that its derivative vanishes. But its derivative can also be written in terms of other connection components (cf. last row of Table~\ref{tab:SolutionSet2}). We therefore obtain the condition
\begin{equation}\label{eq:FixedGamma}
	0 = -\frac{k(8c^2+2ck-k^2)}{8c^2(2c-k)^2\Gamma^{r}{}_{\theta\theta}}-\Gamma^{t}{}_{rr}\Gamma^{r}{}_{\theta\theta}\quad\Longleftrightarrow\quad \Gamma^{t}{}_{rr} = -\frac{k(8c^2+2ck-k^2)}{8c^2(2c-k)^2(\Gamma^{r}{}_{\theta\theta})^2}.
\end{equation}
This expression for $\Gamma^{t}{}_{rr}$ is well defined since solution set 2 requires $\Gamma^{r}{}_{\theta\theta}\neq 0$ for its very definition and we already had to assume $c\neq 0$ and $k\neq 2c$ in order to obtain~\eqref{eq:Option1}. Hence, what we find is that we can fix two of the four free functions of solution set 2 in terms of other connection components. More importantly, the constraint equation~\eqref{eq:AWayOut} allowed us to fix the component $\Gamma^{t}{}_{rr}$, whose derivative was up to now not known. This means that up to now, $\Gamma^{t}{}_{rr}$ was a potential degree of freedom since its derivative could in principle be determined by one of the field equations. But since $\Gamma^{t}{}_{rr}$ is now given by~\eqref{eq:FixedGamma}, whose derivatives can all be determined \textit{without} using the field equations, it cannot become dynamical and it is therefore not a physical degree of freedom.
Hence, the constraint equation~\eqref{eq:AWayOut} has effectively removed a potential degree of freedom. The only connection component which can now become dynamical is $\Gamma^{r}{}_{rr}$ - or alternatively $\Gamma^r_{\ \theta\theta}$ - because its derivative is undetermined and can therefore not be eliminated from the field equations.

There is just a little caveat: Our considerations only hold as long as $c\neq 0$ and $k\neq 2c$. To remedy this shortcoming, i.e., to see what happens for $c=0$, we can solve the constraint equation~\eqref{eq:AWayOut} for $k$:
\begin{equation}\label{eq:Option2}
	k = \frac{4 c^2\,\Gamma^{t}{}_{\theta\theta}}{1+2c\, \Gamma^{t}{}_{\theta\theta}}\quad\text{for } 1+2c\,\Gamma^{t}{}_{\theta\theta} \neq 0.
\end{equation}
Just as before, we can compute the derivative of~\eqref{eq:Option2} and express the derivative of $\Gamma^{t}{}_{\theta\theta}$ in terms of other connection components, using the relation shown in Table~\ref{tab:SolutionSet2}. This gives us again an equation which we can solve for a connection component and we obtain
\begin{equation}\label{eq:Option2Gammatrr}
	\Gamma^{t}{}_{rr} = -\frac{\Gamma^{t}{}_{\theta\theta}+ 5c\, (\Gamma^{t}{}_{\theta\theta})^2+ 4 c^2\, (\Gamma^{t}{}_{\theta\theta})^3}{(1+2c\, \Gamma^{t}{}_{\theta\theta})(\Gamma^{r}{}_{\theta\theta})^2},
\end{equation}
which is again well-defined under the assumptions for which~\eqref{eq:Option2} is valid. The equations~\eqref{eq:Option2} and~\eqref{eq:Option2Gammatrr} have the advantage that they hold for $c=0$ and for $k=2c$. Both options imply
\begin{equation}
	c = k = 0\quad\text{and}\quad \Gamma^{t}{}_{rr} = -\frac{\Gamma^{t}{}_{\theta\theta}}{(\Gamma^{r}{}_{\theta\theta})^2}.
\end{equation}
Hence, it is true in full generality\footnote{Notice that~\eqref{eq:Option2} demands $1+2\,c\,\Gamma^{t}{}_{\theta\theta}\neq 0$ and it seems that we have to treat this case separately. But this is not true: If $\Gamma^{t}{}_{\theta\theta} = -\frac{1}{2c}$, then the constraint equation~\eqref{eq:AWayOut} reduces to $c=0$, which leads to an inconsistency. Hence, $1+2\,c\,\Gamma^{t}{}_{\theta\theta}= 0$ is not admissible and we have therefore already found the most general solution to the constraint equation.} that the constraint equation~\eqref{eq:AWayOut} removes one degree of freedom and only $\Gamma^{r}{}_{rr}$ is left as a \textit{candidate} for a propagating degree of freedom stemming from the connection. \\

Before concluding this subsection, we remark that the connection field equations can be simplified by using the constraint equation~\eqref{eq:AWayOut}. In fact, one can show that the equation $\mathcal C_t$ can be written as
\begin{equation}
	\mathcal C_t = A\, \left(k-2c(2c-k)\Gamma^{t}{}_{\theta\theta}\right) - B\, \left(2c(2c-k)\partial_r\Gamma^{t}{}_{\theta\theta}\right) = 0,
\end{equation}
where $A$ and $B$ are functions of metric components and their derivatives. The coefficient of $A$ is simply the constraint equation~\eqref{eq:AWayOut}, while the coefficient of $B$ is the derivative of the constraint equation. This means that once we have choosen a connection which satisfies the constraint equation~\eqref{eq:AWayOut}, the connection field equation $\mathcal C_t$ is automatically satisfied. The equation $\mathcal C_r$ is \textit{not} trivially satisfied and it contains derivatives of $\Gamma^r{}_{rr}$. We therefore find that after imposing~\eqref{eq:AWayOut}, the field equations have the following structure:
\begin{align}\label{eq:NewStructure}
&\text{New structure of metric field equations:}
	& &\begin{pmatrix}
		\M_{tt} & 0 & 0 & 0\\
		0 & \M_{rr} & 0 & 0\\
		0 & 0 & \M_{\theta\theta} & 0\\
		0 & 0 & 0 & \M_{\theta\theta}\,\sin^2\theta
	\end{pmatrix} \notag\\
&\text{New structure of connection field equations:}
	&&\begin{pmatrix}
		0 \\
		\mathcal C_r \\
		0 \\
		0
	\end{pmatrix} 
\end{align}
This means there are at most three independent metric field equations and one connection field equation for the dynamical variables $\{g_{tt}, g_{rr}, \Gamma^{r}{}_{rr}\}$. Alternatively, one can trade $\Gamma^r_{\ rr}$ for $\Gamma^r_{\ \theta\theta}$ and regard the latter as the dynamical degree of freedom stemming from the connection.

The next natural step is to check the internal consistency of these equations. This is necessary since the number of field equations is larger than the number of dynamical variables and it is therefore not clear whether these equations can be solved consistently. In subsection~\ref{ssec:RemainingEqs} we show the self-consistency of the field equations and we will see on an abstract level that they produce solutions which go beyond Schwarzschild-deSitter-Nordstr\"{o}m.

However, before doing so, in subsection~\ref{ssec:SymRedFieldEq1} we briefly discuss the implications of $\M_{tr} = 0$ for solution set 1. In particular, we show that solution set 1 is \textit{not} viable when looking for solutions to $f(\Q)$ gravity which go beyond GR.

\subsection{Off-diagonal metric field equation for solution set 1}\label{ssec:SymRedFieldEq1}
In the previous subsection we have found that the off-diagonal metric field equation can be solved in three distinct ways. The different options which solve $\M_{tr} = 0$ can be summarized as follows:
\begin{enumerate}
	\item $\partial_r\Q_\sol = 0$: If $\M_{tr} = 0$ is solved by $\partial_r\Q_\sol = 0$, or, in other words, if $\Q$ is constant when evaluated on a solution of the field equations - again one can see from the form of $\Q$ that this can be achieved by solving $\Q=\Q_\text{sol}$ for $\Gamma^r_{\ rr}$ - then it follows that one can only find the Schwarzschild-deSitter-Nordstr\"{o}m solution for \textit{arbitrary} $f$ (as long as $f$ satisfies $f'\neq 0$, which is required in order to obtain non-trivial field equations).
	\item $f''(\Q_\sol) = 0$: If the second derivative of $f$ is zero when evaluated on a solution of the field equations, and assuming that $\partial_r\Q_\sol \neq 0,$ then it follows that $f$ is an affine function; $f(\Q) = a\,\Q + b$. This means we are in the GR sector of $f(\Q)$ gravity and it naturally follows that the only solution is Schwarzschild-deSitter-Nordstr\"{o}m.
	\item Constraint~\eqref{eq:AWayOut}: The constraint equation eliminates two of the four free functions of solution set $2$ and we are left with one potential degree of freedom stemming from the connection; $\Gamma^r{}_{rr}$. This is the only option which allows us to get solutions to $f(\Q)$ gravity which go beyond GR. But from section \ref{ssec:D2} we know how to obtain solution set 1 from set 2, in particular we have to set $\Gamma^t_{\ \theta\theta}=-1/c$ and $k=c$ with now $c\neq 0$. \eqref{eq:AWayOut} then becomes the equation $3c=0$, which cannot be fulfilled. This is thus not an option.
\end{enumerate}
Only the first two options are viable. Hence, solution set 1 can only give rise to the Schwarzschild-deSitter-Nordstr\"{o}m solution.

\subsection{Self-consistency of the remaining field equations for solution set 2}\label{ssec:RemainingEqs}
In subsection~\ref{ssec:SymRedFieldEq2} we have seen that the constraint equation~\eqref{eq:AWayOut} is the only possibility to solve $\M_{tr}=0$ which does not immediately force the Schwarzschild-deSitter-Nordstr\"{o}m solution on us. In particular, we have seen that there are two options to solve the constraint~\eqref{eq:AWayOut}. Both options and the assumptions which go into them are summarized in Table~\ref{tab:Options} for convenience.
\begin{table}[h!]
\centering
	\begin{tabular}{|p{0.15\textwidth}|p{0.27\textwidth} p{0.27\textwidth} p{0.27\textwidth}|}
		\hline
		& & & \\[-1.5ex]
		& \multicolumn{3}{c}{\textbf{The most general solution to equation~\eqref{eq:AWayOut}}}
		\\[1ex] \hline 
 		& & & \\[-1.5ex]
 		& \multicolumn{3}{l|}{$\Gamma^{t}{}_{\theta\theta} = \frac{k}{2c(2c-k)}$}\\
 		\textbf{Option 1} & & & $\text{for } c\neq 0\text{ and } k\neq 2c$\\
 		 & \multicolumn{3}{l|}{$\Gamma^{t}{}_{rr} = -\frac{k(8c^2+2ck-k^2)}{8c^2(2c-k)^2(\Gamma^{r}{}_{\theta\theta})^2}$}
 		\\[1ex] \hline 
 		& & & \\[-1.5ex]
 		&  \multicolumn{3}{l|}{$c = k = 0$}\\
 		\textbf{Option 2} & & & \\
 		&  \multicolumn{3}{l|}{$\Gamma^{t}{}_{rr} = -\frac{\Gamma^{t}{}_{\theta\theta}}{(\Gamma^{r}{}_{\theta\theta})^2}$}
 		\\[1ex] \hline
	\end{tabular}	
	\caption{A summary of the different choices of connection which satisfy the constraint equation~\eqref{eq:AWayOut}.}
	\label{tab:Options}
\end{table}

What we need to do now is to show that the remaining field equations are self-consistent and that they can indeed produce solutions which go beyond GR. The consistency of the equations is not immediately obvious since we have potentially four independent equations for the three degrees of freedom $\{g_{tt}, g_{rr}, \Gamma^{r}{}_{rr}\}$. Moreover, we wish to re-write the equations in the simplest possible form so that they can, at least in principle, be integrated. This needs to be done for each option in Table~\ref{tab:Options} individually, but the strategy to get to the final results is always the same. This method works in particular when including the electro-vacuum.
\begin{enumerate}
	\item Choose one of the options from Table~\ref{tab:Options} and simplify all equations and the non-metricity scalar $\Q$ using the chosen relations.
	\item Solve the $\Q$ scalar for $\Gamma^{r}{}_{rr}$. (This is the only weak point of the strategy as there are scenarios in which this step fails. We discuss the ramifications of this failure in the next subsection.)
	\item Replace every $\Gamma^{r}{}_{rr}$ in the metric and connection field equations by the expression obtained in 2. In other words: We trade $\Gamma^{r}{}_{rr}$ in each equation for $\Q$ and we treat $\Q$ as a degree of freedom. This step will produce much more compact expressions.
	\item For each option, the metric field equations $\M_{tt}$ and $\M_{rr}$ only contain first order derivatives of the metric. Solve these equations for $\partial_r g_{tt}$ and $\partial_r g_{rr}$. This will produce expressions of the form\footnote{Strictly speaking, the functions $1$ and $2$ also depend on the matter fields as well as $f$ and its first two derivatives. However, we will suppress these dependencies for all appearing functions here and in the sequel for the sake of readability.}
		\begin{align}\label{eq:FirstOrderSol}
			\partial_r g_{tt} &= \text{function}_1(g_{tt}, g_{rr},\Q, \partial_r\Q, CC)\notag\\
			\partial_r g_{rr} &= \text{function}_2(g_{tt}, g_{rr},\Q, \partial_r\Q, CC),
		\end{align}
		where $CC$ stands for ``Connection Components''.
	\item The metric field equation $\M_{\theta\theta}$ contains first order derivatives of the metric and the second order derivative $\partial^2_r g_{tt}$. Use~\eqref{eq:FirstOrderSol} to eliminate the first oder derivatives and then solve $\M_{\theta\theta}$ for $\partial^2_r g_{tt}$. This leads to an expression of the form
		\begin{align}\label{eq:SecondOrderSol}
			\partial^2_r g_{tt} &= \text{function}_3(g_{tt}, g_{rr}, \Q, \partial_r \Q, CC).
		\end{align}
	\item Since we solved $\Q$ for $\Gamma^{r}{}_{rr}$ and used the resulting expression to simplify all field equations, the connection equation $\mathcal C_r$ is a second order differential equation for $\Q$. Moreover, it contains first order derivatives of the metric. Use~\eqref{eq:FirstOrderSol} to get rid of the first order derivatives of the metric and solve the connection equation for $\partial^2_r\Q$.
	\item We can derive a consistency condition for the field equations by looking at the equations~\eqref{eq:FirstOrderSol} and~\eqref{eq:SecondOrderSol}. In fact, the $r$-derivative of $\partial_r g_{tt}$ should be equal to $\partial^2_r g_{tt}$ and this leads to the consistency condition
		\begin{equation}\label{eq:ConsistencyCond}
			\partial_r \text{function}_1(g_{tt},g_{rr}, \Q, \partial_r\Q, CC) \overset{!}{=} \text{function}_3(g_{tt},g_{rr}, \Q, \partial_r\Q, CC).
		\end{equation}
		Notice that $\text{function}_1$ and $\text{function}_3$ depend on the same arguments, but the derivative operator on the left hand side generates an expression of the form
		\begin{equation}
			\partial_r\text{function}_1 = \text{function}_4(g_{tt},g_{rr}, \Q, \partial_r\Q, CC, \partial_r g_{tt}, \partial_r g_{rr}, \partial^2_r\Q, \partial_r CC).
		\end{equation}
		The first order derivatives of the metric can be eliminated using~\eqref{eq:FirstOrderSol} and the derivatives of the connection components are all known: A quick look at Table~\ref{tab:SolutionSet2} reveals that we know how to express all derivatives of connection components, except $\partial_r\Gamma^{t}{}_{rr}$ and $\partial_r\Gamma^{r}{}_{rr}$. But the options described in Table~\ref{tab:Options} all allow us to express $\Gamma^{t}{}_{rr}$ in terms of other connection components, and hence we also know how to express its derivative. Moreover, $\Gamma^{r}{}_{rr}$ has been traded for $\Q$ in all equations, hence there are no derivatives of $\Gamma^{r}{}_{rr}$ which we need to worry about and the consistency condition~\eqref{eq:ConsistencyCond} becomes
		\begin{equation}\label{eq:ConsistencyCond2}
			\text{function}_4(g_{tt},g_{rr}, \Q, \partial_r\Q, \partial^2_r\Q, CC) \overset{!}{=} \text{function}_3(g_{tt},g_{rr}, \Q, \partial_r\Q, CC).
		\end{equation}
	\item Observe that the left hand side of the consistency condition~\eqref{eq:ConsistencyCond2} depends on $\partial^2_r\Q$, while this term does not appear on the right hand side. At this point, we need to use the connection field equation, which we solved for $\partial^2_r\Q$ in step 6. Once we eliminate $\partial^2_r\Q$ by the expression found in 6., we find that the consistency condition~\eqref{eq:ConsistencyCond2} is satisfied for both options given in Table~\ref{tab:Options}.
\end{enumerate}
This strategy uses all field equations and it makes extensive use of the properties of solution set 2. What it shows is that the field equations are self-consistent and that there are only three independent field equations, not four. The relevant field equations are $\M_{tt}$, $\M_{rr}$ and $\mathcal C_{r}$ , while $\M_{\theta\theta}$ is trivially satisfied - as in GR - when the other three equations are satisfied. Moreover, these equations determine the dynamical variables $\{g_{tt}, g_{rr}, \Gamma^{r}{}_{rr}\}$, or, alternatively, $\{g_{tt}, g_{rr}, \Q\}$ or $\{g_{tt}, g_{rr}, \Gamma^{r}{}_{\theta\theta}\}$, as we will see shortly. What remains to be done is to write out explicitly the equations for $\partial_r g_{tt}$, $\partial_r g_{rr}$ and $\partial^2_r \Q$ derived from $\M_{tt}$, $\M_{rr}$ and $\mathcal C_{r}$ for both options in Table~\ref{tab:Options}. For option 1 we find
\begin{align}\label{eq:FEQ1}
	\partial_r g_{tt} &= -\frac{g_{tt}\left(2+g_{rr}\left(\Q\, r^2-2-\frac{f(\Q)\,r^2}{f'(\Q)}\right)\right)}{2r} \notag\\
	&\phantom{=} + \frac{\left(\frac{(k-4c)^2 g_{tt}\, r^2}{c(2c-k)\Gamma^{r}{}_{\theta\theta}}-4\Gamma^{r}{}_{\theta\theta}\,g_{rr} (2 g_{tt}-c(2c-k)r^2)\right)}{8\, r\, f'(\Q)} (\partial_r\Q)\, f''(\Q)\notag\\
	\partial_r g_{rr} &=	 \phantom{+}\frac{g_{rr}\left(2+g_{rr}\left(\Q\, r^2 -2 - \frac{f(\Q)\, r^2}{f'(\Q)}\right)\right)}{2\, r} \notag\\
	&\phantom{=} +  \frac{g_{rr}\left(16r+\frac{(k-4c)^2\, r^2}{c(2c-k)\Gamma^{r}{}_{\theta\theta}} + \frac{4\Gamma^{r}{}_{\theta\theta}\,g_{rr}\,(2g_{tt}+c(2c-k)r^2)}{g_{tt}}\right)}{8\, r\, f'(\Q)}  (\partial_r\Q)\, f''(\Q)
\end{align}

\begin{align}
	\partial^2_r \Q &= \frac{(\partial_r \Q)}{2 (4 c (2 c-k) \Gamma^{r}{}_{\theta\theta}^2 g_{rr} (c r^2 (2 c-k)-2
   g_{tt})+r^2 (k-4 c)^2 g_{tt})^2}\times\notag\\
   &\times\bigg(\bigg\{128 c^2 (k-2 c)^2 (\Q r^2-2) \Gamma^{r}{}_{\theta\theta}^3 g_{tt}
   g_{rr}^2 (c r^2 (2 c-k)-2 g_{tt})\notag\\
   &+16 c^3 r (2 c-k)^3
   \Gamma^{r}{}_{\theta\theta}^4 g_{rr}^2 ((\Q r^2-2)
   g_{rr}+6) (c r^2 (2 c-k)-2 g_{tt})\notag\\
   &+8 c r (2 c-k) (k-4 c)^2
   \Gamma^{r}{}_{\theta\theta}^2 g_{tt} g_{rr} (c r^2 (2
   c-k) ((\Q r^2-2) g_{rr}-6)+\notag\\
   &g_{tt} ((6-3 \Q r^2)
   g_{rr}+6))+r^3 (k-4 c)^4 g_{tt}^2 ((\Q r^2-2)
   g_{rr}-2)\bigg\}\notag\\
   & -g_{rr}\, r^3\,\left((4c-k)^2g_{tt}-4c^2 (2c-k)^2(\Gamma^{r}{}_{\theta\theta})^2\, g_{rr}\right) \frac{f(\Q)}{f'(\Q)} \notag\\
	&\phantom{=} + 2\left(2(4c-k)^2\Gamma^{r}{}_{\theta\theta}\, g_{tt}\,g_{rr}\, r + (4c-k)^2\,g_{tt}\, r^2 + 4c (2c-k)(\Gamma^{r}{}_{\theta\theta})^2\, g_{rr}\left(2 g_{tt}-c(2c-k)\, r^2\right)\right)\frac{f''(\Q)}{f'(\Q)}(\partial_r \Q)\notag\\
	&\phantom{=} -2\left((4c-k)^2g_{tt}\, r^2 - 4c(2c-k)(\Gamma^{r}{}_{\theta\theta})^2\,g_{rr}\left(2 g_{tt} - c(2c-k)r^2\right)\right) \frac{f^{(3)}(\Q)}{f''(\Q)} (\partial_r\Q)\bigg).
\end{align}

Observe that the only connection coefficients which appear in these equations are $c$, $k$, and $\Gamma^{r}{}_{\theta\theta}$. The constants $c$ and $k$ are freely specifiable, but the function $\Gamma^{r}{}_{\theta\theta}$ is fixed, up to an integration constant, by the differential equation
\begin{equation}\label{eq:DiffEqOpt1}
	\partial_r \Gamma^{r}{}_{\theta\theta} = \frac12 + \frac{k}{4c} - \frac{3}{2c-k} - \Gamma^{r}{}_{rr} \Gamma^{r}{}_{\theta\theta}.
\end{equation}
This differential equation follows from the last row of Table~\ref{tab:SolutionSet2} when specialized to option 1 of Table~\ref{tab:Options}. Also, recall that we traded $\Gamma^{r}{}_{rr}$ for $\Q$. Hence, if we replace $\Gamma^{r}{}_{rr}$ by its expression in terms of $\Q$ we obtain a highly non-linear differential equation, which nonetheless determines $\Gamma^{r}{}_{\theta\theta}$ (at least in principle) in terms of the dynamical variables $\{g_{tt}, g_{rr}, \Q\}$ and in terms of an integration constant. Hence, for the field equations~\eqref{eq:FEQ1} to produce a solution it is necessary to choose three constants, $c$, $k$, and the integration constant in~\eqref{eq:DiffEqOpt1}, as well as initial conditions for the dynamical degrees of freedom. Since the metric field equations are first order, we need a total of two initial conditions for the metric and we need two initial conditions for $\Q$, given that its differential equation is second order.

The field equations derived from option~2 have a simpler and more compact form and they require less specifications in order to be solved:
\begin{align}\label{eq:FEQ2}
	\partial_r g_{tt} &= -\frac{g_{tt}\left(2+g_{rr}\left(\Q\, r^2-2-\frac{f(\Q)\,r^2}{f'(\Q)}\right)\right)}{2r} +\frac{g_{tt}\left(r^2-g_{rr}(\Gamma^{r}{}_{\theta\theta})^2\right)}{\Gamma^{r}{}_{\theta\theta}\,r\, f'(\Q)}(\partial_r\Q)\, f''(\Q)\notag\\
	\partial_r g_{rr} &=	 \phantom{+}\frac{g_{rr}\left(2+g_{rr}\left(\Q\, r^2 -2 - \frac{f(\Q)\, r^2}{f'(\Q)}\right)\right)}{2\, r} +  \frac{g_{rr}\left((\Gamma^{r}{}_{\theta\theta})^2\,g_{rr} + 2\,r\, \Gamma^{r}{}_{\theta\theta}+ r^2\right)}{\Gamma^{r}{}_{\theta\theta}\, r\, f'(\Q)}(\partial_r\Q)\, f''(\Q)
\end{align}

\begin{align}
	\partial^2_r \Q &= \frac{(\partial_r \Q)}{2\left[(\Gamma^{r}{}_{\theta\theta})^2\,g_{rr}-r^2\right]}\left(2 r + g_{rr}\left(4\Gamma^{r}{}_{\theta\theta}\left(1+\Gamma^{r}{}_{rr}\Gamma^{r}{}_{\theta\theta}\right)+2r-\Q\, r^3\right) + g_{rr}\, r^3\frac{f(\Q)}{f'(\Q)}\right.\notag\\
	&\phantom{=} \left.-2\left((\Gamma^{r}{}_{\theta\theta})^2\,g_{rr} + 2 \Gamma^{r}{}_{\theta\theta}\, g_{rr}\, r + r^2\right) \frac{f''(\Q)}{f'(\Q)} (\partial_r\Q)- \left((\Gamma^{r}{}_{\theta\theta})^2\,g_{rr}-r^2\right)\frac{f^{(3)}(\Q)}{f''(\Q)}(\partial_r\Q)\right)
\end{align}
Since option~2 demands $c=k=0$, the only connection component which now appears in the field equations is $\Gamma^{r}{}_{\theta\theta}$, which is implicitly fixed by the differential equation
\begin{equation}
	\partial_r \Gamma^{r}{}_{\theta\theta} = -1 - \Gamma^{r}{}_{rr}\Gamma^{r}{}_{\theta\theta}.
\end{equation}
This equation is again derived from the last row of Table~\ref{tab:SolutionSet2} when specialized to option~2 of Table~\ref{tab:Options}. After replacing $\Gamma^{r}{}_{rr}$ by its expression in terms of $\Q$ we find the highly non-linear differential equation
\begin{equation}\label{eq:DiffEqOpt2}
	\partial_r\Gamma^{r}{}_{\theta\theta} = -\frac{\Gamma^{r}{}_{\theta\theta}\left(f(\Q)\, g_{rr}\, r^2(\Gamma^{r}{}_{\theta\theta}+r)+(2r-g_{rr}(2\Gamma^{r}{}_{\theta\theta}+r)(\Q\, r^2-1))f'(\Q)\right)}{2\left((\Gamma^{r}{}_{\theta\theta})^2\, g_{rr}- r^2\right)f'(\Q)}
\end{equation}
In principle, this equation determines $\Gamma^{r}{}_{\theta\theta}$ in terms of $\{g_{tt}, g_{rr}, \Q\}$ and an integration constant. The origin of the integration constant can also be understood in a different way: Instead of treating $\Gamma^{r}{}_{rr}$ or $\Q$ as a degree of freedom, we can regard $\Gamma^{r}{}_{\theta\theta}$ as the degree of freedom stemming from the connection. By solving the above differential equation for $\Q$ and plugging the result into the connection field equation for option 2, we obtain a differential equation for $\Gamma^{r}{}_{\theta\theta}$. This differential equation is now third order, and hence we need to specify three initial conditions for $\Gamma^{r}{}_{\theta\theta}$.

We therefore find that for option 2 we need to specify initial conditions for $\{g_{tt}, g_{rr}, \Q\}$ ($1+1+2$) and an integration constant for~\eqref{eq:DiffEqOpt2}, or, alternatively, provide initial conditions for $\{g_{tt}, g_{rr}, \Gamma^{r}{}_{\theta\theta}\}$ ($1+1+3$). This latter point of view with a third order differential equation for $\Gamma^{r}{}_{\theta\theta}$ will be particularly useful in section~\ref{sec:Solutions}, where we derive approximate and exact solutions for $f(\Q)$ gravity. 

Before doing so, however, we show that there is also a different strategy to tackle the symmetry reduced field equations. More precisely, we show that the caveat alluded to in step 2 of the above strategy opens up a rout to solve the field equations in a different way. Namely, the connection can be determined through a constraint equation, while the metric remains dynamical. This will nevertheless lead to solutions which go beyond GR.

\subsection{Reduction of the field equations by a constraint on the connection}\label{ssec:AdditionalConstraints}
In the previous subsection we saw that there are two dynamical equations for the metric and one dynamical equation for the connection. However, this conclusion hinges crucially on the validity of step 2, which requires us to solve $\Q$ for $\Gamma^{r}{}_{rr}$. A closer examination of the non-metricity scalar $\Q$ reveals that this is only possible if the coefficient in front of $\Gamma^{r}{}_{rr}$ is different from zero. Concretely, one finds that $\Q$ takes the schematic form
\begin{equation}
	\Q = \begin{cases}
		\frac{1}{4}\left(\frac{(4c-k)^2}{c\,(2c-k)\,g_{rr}}-\frac{4(\Gamma^{r}{}_{\theta\theta})^2\left(2g_{tt}-c\,(2c-k)\,r^2\right)}{g_{tt}\, r^2}\right)\frac{\Gamma^{r}{}_{rr}}{\Gamma^{r}{}_{\theta\theta}} + \textsf{other terms} & \text{for option 1}\\
		2\left(\frac{1}{g_{rr}}-\frac{(\Gamma^{r}{}_{\theta\theta})^2}{r^2}\right)\frac{\Gamma^{r}{}_{rr}}{\Gamma^{r}{}_{\theta\theta}} + \textsf{other terms} & \textsf{for option 2}.
	\end{cases}
\end{equation}
If we now impose the constraint that the factor in front of $\Gamma^{r}{}_{rr}$ vanishes, i.e., if we impose the condition that  $\Gamma^{r}{}_{\theta\theta}$ is of the form\footnote{The signature of the metric demands that $g_{tt}<0$ and hence $\sqrt{-g_{tt}}$ is real for option 1.}
\begin{equation}\label{eq:ConstrOnGamma}
	\Gamma^{r}{}_{\theta\theta} = \begin{cases}
		\pm\frac{|4c-k|\,\sqrt{-g_{tt}}\, r}{\sqrt{4c\,(2c-k)\,g_{rr}\,(c\,(2c-k)\,r^2-2g_{tt})}}& \textsf{for option 1}\\
		\pm\frac{r}{\sqrt{g_{rr}}}& \textsf{for option 2} 
	\end{cases}
\end{equation}
then it is no longer possible to trade $\Gamma^{r}{}_{rr}$ for $\Q$ and the strategy presented in the previous subsection does not work anymore. However, this does not mean that the field equations become inconsistent. In fact, if $\Gamma^{r}{}_{\theta\theta}$ has been chosen to have one of the forms of~\eqref{eq:ConstrOnGamma}, then it can be shown that the connection field equation $\mathcal C_r$ is trivially satisfied. We are thus left with the metric field equations and they turn out to be self-consistent. The strategy to show self-consistency is as follows, which again holds when including the electro-vacuum.
\begin{enumerate}
	\item Choose either option 1 or option 2 from Table~\ref{tab:Options} and fix $\Gamma^{r}{}_{\theta\theta}$ by choosing the appropriate expression from~\eqref{eq:ConstrOnGamma} for any choice of sign.
	\item Solve $\M_{tt}$ and $\M_{rr}$ for $\partial_r g_{tt}$ and $\partial_r g_{rr}$. This gives rise to expressions of the form
		\begin{align}
			\partial_r g_{tt} &= \textsf{function}_1(g_{tt}, g_{rr}, c, k)\notag\\
			\partial_r g_{rr} &= \textsf{function}_2(g_{tt}, g_{rr}, c, k, \partial_r \Q),
		\end{align}
		where $c$ and $k$ only appear for option 1 and where we have suppressed the dependence of the two functions on matter fields as well as the function $f$ and its derivatives. Observe that only $\partial_r g_{rr}$ depends on $\partial_r \Q$.
	\item Take the expression for $\Q$ and simplify it using the relations of option 1/option 2, the corresponding form of $\Gamma^{r}{}_{\theta\theta}$ from~\eqref{eq:ConstrOnGamma} as well as the expressions obtained in step 2. This leads to an expression of the form
		\begin{equation}
			\Q = \textsf{function}_3(g_{tt}, g_{rr}, \partial_r g_{tt},c, k).
		\end{equation}
	\item Observe that only $\partial_r g_{tt}$ appears in $\textsf{function}_3$. Hence, if we replace $\partial_r g_{tt}$ by $\textsf{function}_1$ and then take the $r$-derivative of $\Q$, we obtain an equation which we can solve for $\partial_r \Q$. This leads to the schematic expression
		\begin{equation}
			\partial_r\Q = \textsf{function}_4(g_{tt}, g_{rr}, c, k),
		\end{equation}
		which in turn can be used to replace $\partial_r\Q$ in $\textsf{function}_2$. This gives us
		\begin{align}
			\partial_r g_{tt} &= \textsf{function}_1(g_{tt}, g_{rr}, c, k)\notag\\
			\partial_r g_{rr} &= \widetilde{\textsf{function}}_2(g_{tt}, g_{rr}, c, k).
		\end{align}
	\item Plugging the above expressions for $\partial_r g_{tt}$ and $\partial_r g_{rr}$ into the remaining metric field equation, $\M_{\theta\theta}$, does not yield anything new. The equation $\M_{\theta\theta}$ is trivially satisfied. With this, we have exhausted all field equations and we have shown their self-consistency even in the case where the connection component $\Gamma^{r}{}_{\theta\theta}$ is fixed by a constraint equations, rather than by a dynamical field equation.
\end{enumerate}
In summary, we find that we can fix the connection component $\Gamma^{r}{}_{\theta\theta}$ through one of the constraints in~\eqref{eq:ConstrOnGamma} and we are then left with two dynamical equations for the metric. To solve these equations, we need to specify initial data for $g_{tt}$ and $g_{rr}$ and, in the case of option 1, we also need to specify $c$ and $k$.
 
Also, notice that if we choose to solve the field equations using the constraints~\eqref{eq:ConstrOnGamma}, we need less initial data than when we let the connection be dynamical. 

In the next section we will derive approximate and exact solutions for both, a dynamically determined connection and one fixed by the constraints. We will focus on option~2 because the equations are more compact and simpler\footnote{We have not found any (approximate) solutions when including $c$ or $k$, but it would be interesting to see what role these constants play and how they affect solutions.} due to the absence of $c$ and $k$.

\section{Approximate and exact solutions}\setcounter{equation}{0}\label{sec:Solutions}
Having shown that we have a self-consistent set of field equations, it is now time to look for solutions. To that end it is sensible to choose option 2 from Table~\ref{tab:Options} since the field equations~\eqref{eq:FEQ2} have a simpler form than~\eqref{eq:FEQ1} and they do not depend on an arbitrary choice of $c$ and $k$ -- they only need the specification of initial conditions in order to produce well-defined solutions. Moreover, we want to use the spherical connection considered in~\cite{Zhao:2021,Lin:2021} as a partial guide line toward finding solutions beyond the GR solutions. The idea is that the spherical connection of~\cite{Zhao:2021,Lin:2021} is already known to produce the GR solutions for arbitrary $f$ and a controlled deformation of that connection could therefore lead to deformations of the GR solutions for a given choice of $f$. We will explain how this can be achieved in a perturbative fashion in the next subsection.

However, before doing so, we want to consider the exact spherical connection used in~\cite{Zhao:2021,Lin:2021} in order to explicitly show that it can only produce GR solutions for arbitrary $f$ and because this provides a concrete example for the fact that solutions with $\Q_\sol = \textsf{const}$. do exist.\\

As already noted in subsection~\ref{ssec:set2}, the spherical connection is obtained by setting $\{c,k,\Gamma^{t}{}_{rr}, \Gamma^{t}{}_{\theta\theta}, \Gamma^{r}{}_{rr}, \Gamma^{r}{}_{\theta\theta}\} = \{0,0,0,0,0,-r\}$. The only non-zero connection coefficients are then explicitly given by
\begin{align}
	\Gamma^{r}{}_{\theta\theta} &= -r  & \Gamma^{r}{}_{\phi\phi} &= -r\,\sin^2\theta & \Gamma^{\theta}{}_{r\theta} &= \frac{1}{r} \notag\\
	\Gamma^{\theta}{}_{\phi\phi} &= -\cos\theta\,\sin\theta & \Gamma^{\phi}{}_{r\phi} &= \frac{1}{r} & \Gamma^{\phi}{}_{\theta\phi} &= \cot\theta.
\end{align}
This means that the spherical connection falls into solution set 2 and that it corresponds to option 2 in Table~\ref{tab:Options}. Hence, we can simply insert this connection into the field equations~\eqref{eq:FEQ2} and we obtain
\begin{align}\label{eq:EqsForSphericalC}
	\partial_r g_{tt} &= -\frac{g_{tt}\left(2+g_{rr}\left(\Q\, r^2-2-\frac{f(\Q)\,r^2}{f'(\Q)}\right)\right)}{2r} -g_{tt}\left(1-g_{rr}\right)(\partial_r\Q)\, \frac{f''(\Q)}{f'(\Q)}\notag\\
	\partial_r g_{rr} &=	 \phantom{+}\frac{g_{rr}\left(2+g_{rr}\left(\Q\, r^2 -2 - \frac{f(\Q)\, r^2}{f'(\Q)}\right)\right)}{2\, r} +  g_{rr}\left(1-g_{rr}\right)(\partial_r\Q)\, \frac{f''(\Q)}{f'(\Q)}\notag\\
	\partial^2_r\Q &= \frac12(\partial_r \Q)\left(\frac{g_{rr}^2(\Q\, r^2 - 2) + g_{rr} (4+\Q\, r^2)-2}{(g_{rr}-1)^2\, r} + \frac{g_{rr}\, r}{g_{rr}-1}\frac{f(\Q)}{f'(\Q)} + 2\frac{f''(\Q)}{f'(\Q)}(\partial_r\Q) - 2\frac{f^{(3)}(\Q)}{f''(\Q)}(\partial_r\Q)\right).
\end{align}
If we multiply the first equation by $g_{rr}$ and the second one by $g_{tt}$ and then add the two multiples together we obviously obtain
\begin{align}
	g_{rr}\partial_r g_{tt} + g_{tt}\partial_r g_{rr} &= 0 &\Longleftrightarrow & & g_{tt} g_{rr} &= \textsf{const}.
\end{align}
This is already a first indication that we will obtain the GR solution. If we now use the spherical connection to compute the non-metricity scalar $\Q$ we find
\begin{equation}\label{eq:QSpherical}
	\Q = -\frac{g_{rr} - 1}{g_{tt}\, g_{rr}^2\, r}\left(g_{rr}\partial_r g_{tt} + 	g_{tt}\partial_r g_{rr}\right).
\end{equation}
Hence, if we evaluate $\Q$ on a solution of the field equations, which imply that $g_{rr}\partial_r g_{tt} + 	g_{tt}\partial_r g_{rr} = 0$, we obtain $\Q_\sol = 0$ from~\eqref{eq:QSpherical}. In turn this implies that the connection equation in~\eqref{eq:EqsForSphericalC} is trivially satisfied. Moreover, the metric field equations in~\eqref{eq:EqsForSphericalC} reduce to
\begin{align}
	\partial_r g_{tt} &= -\frac{g_{tt}\left(1-g_{rr}\left(1+\frac{f(0)\,r^2}{2f'(0)}\right)\right)}{r}\notag\\
	\partial_r g_{rr} &=	 \phantom{+}\frac{g_{rr}\left(1 - g_{rr}\left(1 + \frac{f(0)\, r^2}{2f'(0)}\right)\right)}{r}.
\end{align}
As anticipated, this shows explicitly that solutions with $\Q_\sol=\textsf{const}.$ exist and that they can only produce the Schwarzschild-deSitter-Nordstr\"{o}m solution\footnote{Our considerations also hold for the electro-vacuum but we stuck to the pure vacuum case for simplicity.} for an arbitrary choice of $f$. Indeed, upon integrating the above differential equations one finds
\begin{align}
	g_{tt} &= c_2 + \frac{c_1\, c_2}{r} + \frac{c_2}{6}\frac{f(0)}{f'(0)} r^2 \equiv c_2 + \frac{c_1\, c_2}{r} + \frac{c_2\,\Lambda_\textsf{eff}}{3}r^2\notag\\
	g_{rr} &= \frac{1}{c_2\, g_{tt}}, 
\end{align}
where we have used that $\frac12\frac{f(0)}{f'(0)}=:\Lambda_\textsf{eff}$ acts as an effective cosmological constant, as we have explained in subsection~\ref{ssec:SymRedFieldEq2}. This example will also be the point of departure to construct approximate solutions to $f(\Q)$ gravity which go beyond GR but which reduce to the GR solutions in an appropriate limit. The key observation is that the metric field equations for option 2 (cf. equation~\eqref{eq:FEQ2}) imply that
\begin{align}
	\frac{g_{rr}\partial_r g_{tt} + 	g_{tt}\partial_r g_{rr}}{g_{tt}g_{rr}} = \frac{\partial_r(g_{tt} g_{rr})}{g_{tt}g_{rr}} = 2\frac{\Gamma^{r}{}_{\theta\theta}+r}{\Gamma^{r}{}_{\theta\theta}}\frac{f''(\Q)}{f'(\Q)}(\partial_r\Q).
\end{align}
This equation can easily be integrated and one obtains
\begin{align}
	g_{tt} g_{rr} = c_1\,\exp\left(2\int\dd r\,\frac{\Gamma^{r}{}_{\theta\theta}+r}{\Gamma^{r}{}_{\theta\theta}}\frac{f''(\Q)}{f'(\Q)}(\partial_r\Q) \right),
\end{align}
where $c_1$ is an integration constant chosen\footnote{It follows from $\det g\neq 0$ and the signature of the metric that $g_{tt}g_{rr}$ is strictly negative.} such that $g_{tt}g_{rr}<0$. This equation shows again that the spherical connection, which imposes $\Gamma^{r}{}_{\theta\theta} = -r$, reproduces the GR relation $g_{tt}\propto\frac{1}{g_{rr}}$. Unsurprisingly, it also shows that $\partial_r \Q = 0$ or $f''(\Q)$ produce this relation. However, what is more important to us, is that this equation suggests that we can consider a deformation of the spherical connection where $\Gamma^{r}{}_{\theta\theta} = -r + \gamma(r)$. This would lead to a deviation of the typical $g_{tt}\propto\frac{1}{g_{rr}}$ behavior of GR (although we will not actually see this in all of the approximate solution derived in the following subsections because the difference between $g_{tt}$ and $-1/g_{rr}$ are sometimes hidden in higher order perturbations). In the next subsection we will make this idea more precise and show that it is possible to obtain an approximate solution for the ansatz $f(\Q) = \Q + \alpha\, \Q^2$, where $\alpha$ is assumed to be a small parameter.

We also report some results for $f(\Q)=\Q+\alpha\,\Q^\kappa$ for integer $\kappa\geq 2$ in Appendix~\ref{app:B}, which generalize some of the results derived below.

\subsection{Approximate vacuum solution beyond GR for $f(\Q) = \Q + \alpha\, \Q^2$}\label{ssec:ApproxSolutions}
Since we wish to consider the deformation $\Gamma^{r}{}_{\theta\theta} = -r + \gamma(r)$, it is convenient to solve the differential equation~\eqref{eq:DiffEqOpt2} for $\Q$ and plugging the resulting expression into the differential equation for $\partial^2_r \Q$. This results in a third order differential equation for $\gamma(r)$. Moreover, this equation also contains terms proportional to $f^{(3)}(\Q)$. Given the complexity of this equation, it is sensible to first consider the ansatz $f(\Q) = \Q + \alpha\,\Q^2$, which gets rid of terms proportional to $f^{(3)}(\Q)$. Furthermore, we consider the pure vacuum case with vanishing cosmological constant.

In the sequel we wish to consider $\alpha$ as being a small parameter, i.e., $|\alpha|\ll 1$. We can therefore expect that this ansatz will lead to small deviations from the GR solutions and we choose the ans\"{a}tze 
\begin{align}
	g_{tt} &= g_{tt}^{(0)} + \alpha\,g_{tt}^{(1)} + \alpha^2\, g_{tt}^{(2)}\notag\\
	g_{rr} &= g_{rr}^{(0)} + \alpha\,g_{rr}^{(1)} + \alpha^2\, g_{rr}^{(2)}\notag\\
	\Gamma^{r}{}_{\theta\theta} &= -r + \alpha\,\gamma^{(1)} + \alpha^2\, \gamma^{(2)},
\end{align}
where $g_{tt}^{(0)}$ and $g_{rr}^{(0)}$ are given by the Schwarzschild solution:
\begin{align}\label{eq:SchwarzschildSolution}
	g_{tt}^{(0)} &= -\left(1-\frac{2M}{r}\right) &\text{and} & & g_{rr}^{(0)} &= -\frac{1}{g_{tt}^{(0)}}.
\end{align}
We include second order terms in $\alpha$ because, as we will see shortly, in the zeroth and first order equations, the metric and the connection equations decouple from each other. Only at second order do we find coupled equations which imply that the connection influences the metric and leads to what we dub a ``connection hair''.

The next step is to plug these ans\"{a}tze into the field equations and to solve them order by order. At zeroth order we find, unsurprisingly, that the metric as well as the connection field equations are trivially satisfied. At first order in $\alpha$ we find that the metric field equations reduce to
\begin{align}
	\partial_r g_{tt}^{(1)} &= \frac{2 M\,r\, g_{tt}^{(1)} - (r-2M)^2\, g_{rr}^{(1)}}{(r-2M)\, r^2}\notag\\
	\partial_r g_{rr}^{(1)} &= \frac{(2M+r)\, g_{rr}^{(1)}}{(2 M - r)\,r}
\end{align}
while the connection equation is identically fulfilled\footnote{Notice that at zeroth order in $\alpha$ we have symmetric teleparallelism, where the connection and the metric or not only independent, but the connection is completely arbitrary. That is why at zeroth order there is no equation for the connection, there is only an identity. At first order we do get an equation for the connection, but this equation is identically satisfied for $\gamma^{(0)} =-r$. Only at second order can we expect something interesting for the connection.}.
As anticipated, the metric field equations do not depend on the connection and the connection field equations do not depend on the metric. The above equations can easily be integrated and one finds
\begin{align}\label{eq:FirstOrderSols}
	g_{tt}^{(1)} &= \frac{c_2 + c_1\,(r-2M)}{r}\notag\\
	g_{rr}^{(1)} &= \frac{c_2 \, r}{(r-2M)^2},
\end{align}
where $c_i$ are real integration constants. We set $c_1=0$ in order to obtain an asymptotically flat solution in the sense that $\lim_{r\to \infty}g_{tt} = -1$ and $\lim_{r\to\infty} g_{rr} = 1$. At this point one notices that the only effect of the  perturbations is to renormalize the mass, in the sense that we can write the full metric at first order in $\alpha$ as
\begin{align}
    g_{tt}&=-1+\frac{2M_\text{ren}}{r}\notag\\
    g_{rr}&=-\frac{1}{g_{rr}},
\end{align}
where the renormalized mass is defined as
\begin{equation}
    2M_\text{ren} := 2M+\alpha\, c_2.
\end{equation}
Since the mass $M$ appears in the Schwarzschild solution only as an integration constant, the only physically observable mass is $M_\text{ren}$. We now move to the second order and we find that the metric equations now do depend on $\gamma^{(1)}$. After inserting the solutions for $g_{tt}^{(1)}$, and $g_{rr}^{(1)}$ into the second order equations we find that they can be written as
\begin{align}
	\partial_r g_{tt}^{(2)} &= -\frac{-\frac{2Mr^5g_{tt}^{(2)}}{r-2M}+g_{rr}^{(2)}r^4(r-2M)-\frac{c_2 r^5(c_2+c_2(r-2M))}{(r-2M)^2}+16M^2(4\gamma^{(1)}-4r\partial_r{\gamma^{(1)}}+r^2\partial_r^2{\gamma^{(1)}})}{r^6}\notag\\
	\partial_r g_{rr}^{(2)} &= \frac{-c_2^2r^5-g_{rr}^{(2)}(r+2M)r^3(r-2M)^3-16M^2(r-2M)^2(4\gamma^{(1)}-4r\partial_r{\gamma^{(1)}}+r^2\partial_r^2{\gamma^{(1)}})}{r^4(r-2M)^4},
\end{align}
while for the connection field equations at second order we find the non-trivial equation
\begin{equation}
	16 \gamma^{(1)} - 16 r\, \partial_r \gamma^{(1)}+ 6 r^2\, \partial^2_r \gamma^{(1)}-r^3 \,\partial^3_r \gamma^{(1)} = 0.
\end{equation} 
The connection field equation can indeed be integrated and we find the solution
\begin{equation}
    \gamma^{(1)} = r\,\left(c_5+c_6\, r^3+c_7\, r^3\ln(r)\right).
\end{equation}
After plugging the above solution for $\gamma^{(1)}$ into the metric field equations, we are able to integrate them and we find the second order contributions to the metric:
\begin{align}
	g_{tt}^{(2)} &=\frac{-c_2c_1+c_3-2Mc_4-48M^2c_6-16M^2c_7-48M^2c_7\ln(r)+c_4r}{r}\notag\\
	g_{rr}^{(2)} &= \frac{r(c_2^2+(c_3-16M^2(3c_6+c_7))(r-2M)-48M^2c_7\ln(r)(r-2M)}{(r-2M)^3}.
\end{align}
Notice that $c_5$ does not appear in the metric. In order to maintain asymptotic flatness, i.e., $\lim_{r\to\infty}g_{tt}= -1$ and $\lim_{r\to\infty} g_{rr} = 1$, we need to set $c_4=0$. In particular, we obtain a new beyond GR logarithmic correction coming from the connection $\gamma^{(1)}$ -- i.e., we obatin a  ``connection hair''. The full metric components can be written as
\begin{align}
	g_{tt} &= -\left(1-\frac{2M_\textsf{ren}}{r}\right) + \alpha^2\frac{\mu}{r} \ln\left(\frac{r}{r^*}\right)\notag\\
	g_{rr} &= -\frac{1}{g_{tt}}.
\end{align}
The renormalized mass $M_\text{ren}$ is now given by
\begin{equation}
    2M_\textsf{ren} := 2M +\alpha\, c_2 +\alpha^2\left( c_3-16M^2(3c_6+c_7)\right),  
\end{equation}
where the second equation holds up to order $\alpha^2$. The scale $r^*$ can be introduced by a shift in the constant $c_6\rightarrow c_6-48M^2c_7\ln(r^*)$, in order to have a dimensionless argument in the logarithm. We have also defined a new scale
\begin{align}
	\mu &:= 48 \, M^2\, c_7,
\end{align}
which characterizes the strength of the beyond GR correction -- a new ``black hole charge'' or ``connection hair''. It is important to notice that the correction terms could lead to deviations from the Schwarzschild solution even at large values of $r$. The logarithmic correction term will dominate over the renormalized Schwarzschild term for radii satisfying
\begin{equation}
	\left|\ln(r/r^*)\right| >\frac{2M_\textsf{ren}}{\alpha^2|\mu|}.
\end{equation}
This equation implies that perturbation theory breaks down at such large $r$. Only at smaller $r$ are the metric perturbations small compared to the background Schwarzschild spacetime.\\
This form of $g_{tt}$ is also particularly useful when we want to inquire the location of the Killing horizon. To that end, we just need to determine where the norm of the time-translation Killing vector field $\mathcal T := t^\alpha \partial_\alpha$ vanishes. One finds
\begin{align}
	g_{\mu\nu} t^\mu t^\nu &= g_{tt} \overset{!}{=} 0 & \Longrightarrow & & 2 M_\textsf{ren} + \alpha^2\mu \ln(r/r^*) - r \overset{!}{=} 0.
\end{align}
Numerical considerations show that this equation has generically two solutions. The first one can even be found analytically in terms of Lambert's $W$-function and we call it the internal Killing horizon. Its location is given by
\begin{equation}\label{eq:KillingHorizonEq}
	r_\textsf{Internal Killing horizon} = -\alpha^2\mu\, W\left(-\frac{r^*}{\alpha^2\mu}\e^{-\frac{2M_\textsf{ren}}{\alpha^2\mu}}\right)\approx r^* \e^{-\frac{2M_\textsf{ren}}{\alpha^2\,\mu}},
\end{equation}
where we have expanded Lambert's $W$-function to first order in its argument; $W(x) = x + \mathcal O(x^2)$. We call this the internal Killing horizon because the exponential factor is essentially zero and because a second solution, which we call the outer Killing horizon, can be found numerically, which is close to the standard Schwarzschild radius $r=2M$. There is no closed analytical expression for the outer horizon, but we can solve equation~\eqref{eq:KillingHorizonEq} to first order in $\alpha$ and we find the radius
\begin{equation}
	r_\textsf{Outer Killing horizon} = 2M + \alpha\, c_2 + \mathcal{O}(\alpha^2).
\end{equation}
Hence, the outer Killing horizon can lie below or above the standard Schwarzschild horizon. The considerations of this subsection can also be generalized to the case of an electro-vacuum and to a non-zero cosmological constant. This generalization will be the subject of the next subsection.

\subsection{Approximate electro-vacuum solution beyond GR for $f(\Q) = \Q + \alpha\, \Q^2 - 2\Lambda$}\label{ssec:ApproxSolutions2}
To generalize the approximation scheme of the previous subsection to the case of an electro-vacuum, we need to introduce the energy-momentum tensor
\begin{align}
	T^\mu{}_{\nu} = \textsf{diag}\left(\Lambda+\frac{Q^2}{r^4}, \Lambda+\frac{Q^2}{r^4} , \Lambda-\frac{Q^2}{r^4} , \Lambda-\frac{Q^2}{r^4} \right),
\end{align}
where $Q$ denotes the charge of the source. We choose again the ansatz
\begin{align}
	g_{tt} &= g_{tt}^{(0)} + \alpha\,g_{tt}^{(1)} + \alpha^2\, g_{tt}^{(2)}\notag\\
	g_{rr} &= g_{rr}^{(0)} + \alpha\,g_{rr}^{(1)} + \alpha^2\, g_{rr}^{(2)}\notag\\
	\Gamma^{r}{}_{\theta\theta} &= -r + \alpha\,\gamma^{(1)} + \alpha^2\, \gamma^{(2)},
\end{align}
but now with
\begin{align}\label{eq:SRNdS}
	g_{tt}^{(0)} &= -\left(1-\frac{2M}{r}+\frac{Q^2}{r^2}-\frac{\Lambda}{3}r^2\right) & \text{and} & & g_{rr}^{(0)} &= -\frac{1}{g_{tt}^{(0)}}.
\end{align}
Just as in the previous subsection, we can solve the field equations order by order. We do not give all details here as they are completely analogously solved as for the vacuum case. Unfortunately, the solution for $\gamma^{(1)}$ is rather implicit:
\begin{align}
	\gamma^{(1)} &= r\, c_3\left(1+\int^{r}\e^{-\frac12 \sum_{i=1}^4 \frac{(6M+7\Lambda\,\rho^3)\ln(\rho-x_i)}{3M+2\Lambda\rho^3}}\rho^3\,\ln(\rho)\,\dd\rho\right) + r\,c_4\int^{r} \e^{-\frac12 \sum_{i=1}^4 \frac{(6M+7\Lambda\,\rho^3)\ln(\rho-x_i)}{3M+2\Lambda\rho^3}}\rho^3\,\dd\rho ,
\end{align}
where $x_i$ are the four solutions to the quartic equation
\begin{equation}
	3Q^2-6M\,x_i-\Lambda\, x_i^4 = 0.
\end{equation}
However, we can discuss the cases $Q\neq 0$, $\Lambda = 0$ and $Q= 0$, $\Lambda \neq 0$ separately. We begin by setting $\Lambda=0$. Then  the metric components at second order in perturbation theory are given by
\begin{align}
	g_{tt} &= -1+\frac{2M_\text{ren}}{r}-\frac{Q_\text{ren}^2}{r^2}+\alpha^2\mu \left(\frac{2M_\text{ren}}{r}-\frac{Q_\text{ren}^2}{r^2}\right)\ln\left(\frac{r^2}{r^{*2}}\left(\frac{2M_\text{ren}}{r}-\frac{Q_\text{ren}^2}{r^2}\right)\right)\notag\\
	g_{rr} &= -\frac{1}{g_{tt}},
\end{align}
where we have defined
\begin{align}
    \mu &:=-\frac{6c_8}{M}\notag\\
    2M_\text{ren}  &:= 2M +\alpha c_2+\alpha^2(c_4-2M(8c_6-4c_7)-4c_8)\notag\\
    Q_\text{ren}^2 &:=Q^2\left(1-\alpha^2 (8c_6-4c_7)-4\alpha^2\frac{c_8}{2M}\right).
\end{align}
The $c_i$ are integration constants, where $c_2$ and $c_4$ comes from solving the $\M_{tt}$ equations for $g_{rr}$ at first and second order in $\alpha$, respectively, and $c_6,c_7,c_8$ come from solving the connection equation $\mathcal{C}_r$ at second order in $\alpha$. The integration constants in $g_{tt}$ were set to zero in order to have again $\lim_{r\to\infty}g_{tt}= -1$. The scale $r^*$ can again be introduced via a redefinition of $c_6$ and it ensures proper units.\\
Next, let us set $Q=0$ but keep $\Lambda$ different from zero. The solution for the metric is now given by
\begin{align}
    g_{tt}&= -1+\frac{2M_\text{ren}}{r}+\frac{\Lambda_\text{ren}}{3}r^2+\alpha^2\mu\left(\frac{2M_\text{ren}}{r}+\frac{\Lambda_\text{ren}}{3}r^2\right)\ln\left(\frac{r^{*2}}{r^2}\left(\frac{2M_\text{ren}}{r}+\frac{\Lambda_\text{ren}}{3}r^2\right)\right)\notag\\
    g_{rr}&=-\frac{1}{g_{tt}},
\end{align}
where we have defined
\begin{align}
    \mu &:= \frac{4c_8}{18M}\notag\\
    2M_\text{ren} &:=2M+\frac{\alpha(c_2+\alpha c_4)+24\alpha^2M(c_6-c_7)}{9}\notag\\
    \frac{\Lambda_\text{ren}}{3} &:=\frac{\Lambda}{3}\left(1+\frac{12\alpha^2(c_6-c_7)}{9}\right).
\end{align}
The integration constants and scale $r^*$ arise in a similar fashion as for the charged case above.\\
\\
We note that using the functional equation of the logarithm in the limits $Q\rightarrow0$ and $\Lambda\rightarrow 0$, respectively, we obtain the vacuum solutions from the previous subsection by absorbing the powers of $r$ in the logarithm in the prefactor $\mu$. Moreover, we see that the new correction is simply the ``background'' GR potential times the logarithm of the potential times or divided by $r^2$ for the charge and cosmological constant cases, respectively. One can draw similar conclusions for the Killing horizons as in the previous subsection, but we will not go into details here.

\subsection{Approximate solutions beyond GR from constraints on the connection}\label{ssec:VB}
In subsection~\ref{ssec:AdditionalConstraints} we have seen that the field equations can also be solved by imposing one of the constraints~\eqref{eq:ConstrOnGamma} on $\Gamma^{r}{}_{\theta\theta}$. We choose again to work with option 2 and we select the constraint
\begin{equation}
	\Gamma^{r}{}_{\theta\theta} = \pm\frac{r}{\sqrt{g_{rr}}}.
\end{equation}
We call these two cases (I$^\pm$) and we first discuss (I$^+$). As explained in subsection~\ref{ssec:AdditionalConstraints}, the constraint ensures that the connection field equation is satisfied and we are hence left with the two metric field equations $\M_{tt}$ and $\M_{rr}$. In order to find an approximate solution to these equations, we choose again the ansatz $f(\Q) = \Q + \alpha\,\Q^2$ together with
\begin{align}\label{eq:MetricAnsatz1}
	g_{tt} &= g_{tt}^{(0)} + \alpha\, g_{tt}^{(1)}\notag\\
	g_{rr} &= g_{rr}^{(0)} + \alpha\, g_{rr}^{(1)},
\end{align}
where $g_{tt}^{(0)}$ and $g_{rr}^{(0)}$ are given by the standard Schwarzschild solution \eqref{eq:SchwarzschildSolution}. First of all, we notice that the field equations $\M_{tt}$ and $\M_{rr}$ can be solved for $\partial_r g_{tt}$ and $\partial_r g_{rr}$ in full generality:
\begin{align}\label{eq:gttAndgrr}
	\partial_r g_{tt} &= \frac{g_{tt}\,\left(f(\Q) \, g_{rr} \, r^2-4\left(1+\sqrt{g_{rr}}\right)f'(\Q)\right)}{2\left(2+\sqrt{g_{rr}}\right)\, r\, f'(\Q)}\notag\\
	\partial_r g_{rr} &=	 -\frac{g_{rr}\left(f(\Q)\,g_{rr}\, r^2 - 4\left(f'(\Q)+\sqrt{g_{rr}}\right)\left(f'(\Q)+\left(2+\sqrt{g_{rr}}\right)\, r\, \partial_r \Q\,f''(\Q)\right)\right)}{2\left(2+\sqrt{g_{rr}}\right)\, r\, f'(\Q)}.
\end{align}
Moreover, it can be shown that the non-metricity scalar can be written as
\begin{equation}\label{eq:QConstr}
	\Q = \frac{2\left(1+\sqrt{g_{rr}}\right)\left(g_{tt}\,\left(1+\sqrt{g_{rr}}\right)+r\,\partial_r g_{tt}\right)}{g_{tt}\, g_{rr}\, r^2}.
\end{equation}
After inserting the first line of~\eqref{eq:gttAndgrr} into the above expression of $\Q$ and using the ansatz $f(\Q) = \Q+\alpha\, \Q^2$, we can get rid of $\partial_r \Q$ in the second line of~\eqref{eq:gttAndgrr} and any $\Q$ which appears from using the ansatz $f(\Q) = \Q+\alpha\, \Q^2$. Hence, we end up with equations which only depend on the metric components and nothing else. After inserting the ansatz~\eqref{eq:MetricAnsatz1} into these equations, we can expand them order by order. At zeroth order, we find trivially satisfied equations, as had to be expected. At first order, we find non-trivial equations which can be integrated:
\begin{align}
	g_{tt}^{(1)} &= \left(1-\frac{2M}{r}\right)c_2 +\frac{32}{3M^2}\frac{\left(r-2M\right)^{\frac32}}{r^\frac32}+\frac{1}{M^2\, r^3}\left(\ln\left(1-\frac{2M}{r}\right)r^2\,\left(r-3M\right)+M\left(2M^2 +2r^2+Mr\,\left(12+c_1\, r\right)\right)\right)\notag\\
	g_{rr}^{(1)} &= \frac{r}{(r-2M)^2}\left(c_1-\frac{1}{M}\ln\left(1-\frac{2M}{r}\right)-\frac{50 M}{r^2} + \frac{46}{r}-\frac{16\sqrt{r-2M}(M-2r)(3M-r)}{3M\, r^\frac{5}{2}}\right)
\end{align}
For the asymptotic limit we find
\begin{align}
	\lim_{r\to\infty} g_{tt} &= -1+\alpha\left(\frac{32}{3M^2}+c_2\right) &\text{and} & & \lim_{r\to\infty} g_{rr} &= 1,
\end{align}
which implies that we have to set $c_2 = -\frac{32}{3M^2}$ in order to obtain a standard asymptotically Minkowski solution. With this choice of integration constant, we find that the first order corrections for large $r$ can be written as
\begin{align}
    g_{tt}&=-1+\frac{2M_\text{ren}}{r}+\alpha\frac{32}{r^2}\notag\\
    -\frac{1}{g_{rr}}&=-1+\frac{2M_\text{ren}}{r}+\alpha\frac{96}{r^2}
\end{align}
where we have introduced the renormalized mass
\begin{equation}
    2M_\text{ren}:=2M-\alpha\left(\frac{32}{3M}+c_1\right).
\end{equation}
Hence, the first order beyond-GR corrections scale as $\frac{1}{r^2}$. As we will see shortly, for the case (I$^-$) we obtain a different scaling of the first order corrections.\\
\\
We now deal with (I$^-$) analogously to (I$^+$). The field equations are now given by
\begin{align}
    \partial_r g_{tt} &=\frac{g_{tt} \left(f(\Q)\, r^2 g_{rr}-4 f'(\Q) \left(\sqrt{g_{rr}}-1\right)\right)}{2 f'(\Q)\, r \left(\sqrt{g_{rr}}-2\right)}\notag\\
	\partial_r g_{rr} &=-\frac{g_{rr} \left(4 \left(\sqrt{g_{rr}}-1\right) \left(f''(\Q)\, r \left(\sqrt{g_{rr}}-2\right) \partial _r\Q-f'(\Q)\right)+f(\Q)\, r^2 g_{rr}\right)}{2 f'(\Q) r \left(\sqrt{g_{rr}}-2\right)},
\end{align}
and the non-metricity scalar for (I$^-$) takes the form
\begin{equation}
    \Q=\frac{2 \left(\sqrt{g_{rr}}-1\right) \left(r \partial_r g_{tt}-\sqrt{g_{rr}} g_{tt}+g_{tt}\right)}{r^2 g_{rr} g_{tt}}\ .
\end{equation}
After using  $f(\Q)=\Q+\alpha\,\Q^2$ and applying the ansatz~\eqref{eq:MetricAnsatz1}, one finds the first order solutions
\begin{align}
	g_{tt}^{(1)} &= \frac{(r-2 M)}{3r} \left(\frac{\frac{6 M}{r^2}+\frac{64}{M \sqrt{\frac{r}{r-2 M}}}+\frac{6 \ln \left(1-\frac{2 M}{r}\right)}{r-2 M}-\frac{6 \ln \left(1-\frac{2 M}{r}\right)}{M}+\frac{51-6 c_1 M}{2 M-r}+\frac{39}{r}}{2 M}+3 c_2\right)\notag\\
	g_{rr}^{(1)} &= \frac{r}{(r-2M)^2} \left(-\frac{4 \sqrt{\frac{r}{r-2 M}} \left(68 M^2 r-24 M^3-44 M r^2+8 r^3\right)}{3 M r^3}+\frac{50 M}{r^2}+\frac{\ln \left(1-\frac{2 M}{r}\right)}{M}-\frac{46}{r}+c_1\right).
\end{align}
The asymptotic limit is again easy to determine and we find
\begin{align}
	\lim_{r\to\infty} g_{tt} &=-1+\alpha\left(\frac{32}{3M^2}+c_2\right) &\text{and} & & \lim_{r\to\infty} g_{rr} &=1.
\end{align}
Upon fixing $c_2=-\frac{32}{3M^2}$, we obtain $\lim_{r\to\infty}g_{tt}=-1$, i.e., we have again an asymptotically Minkowski solution. If we use this value for the integration constant $c_2$, we find that the first order corrections at large $r$ are given by
\begin{align}
    g_{tt}&=-1+\frac{2M_\text{ren}}{r}+\alpha \frac{8M^3}{5r^5}\notag\\
    -\frac{1}{g_{rr}}&=-1+\frac{2M_\text{ren}}{r}+\alpha\frac{8M^3}{r^5},
\end{align}
where the renormalized mass now reads
\begin{equation}
    2M_\text{ren}:=2M+\alpha\left(\frac{32}{3M}-c_1\right).
\end{equation}
Hence, we find that the corrections for the (I$^{-}$) case scale like $\frac{1}{r^5}$ and the solution is therefore virtually indistinguishable from Schwarzschild at large radii.

\subsection{A comment on isotropic coordinates}
We want to mention here a different ansatz for the metric, and its connection to the solutions above, namely isotropic coordinates. In these coordinates one chooses not $g_{\theta\theta}=r^2$, as we did above in \ref{ssec:E}, but $g_{\theta\theta}=g_{RR}R^2$. We denote the isotropic radius by $R$ to distinguish it from the Schwarzschild radial coordinate $r$. The line element can then be written as
\begin{equation}
    \dd s^2 = -g_{tt}\dd t^2+g_{RR}\left(\dd R^2+R^2\dd\Omega^2\right),
\end{equation}
were $g_{tt}$ and $g_{RR}$ are functions of $R$ only. The coordinate transformation that links isotropic and Schwarzschild coordinates is given by
\begin{equation}\label{eq:IsotropicToSchwarzschild}
    \ln\left(\frac{R}{R_0}\right)=\int_{r_0}^r\dd r\, \frac{\sqrt{g_{rr}}}{r}.
\end{equation}
Of course, since the difference between the isotropic and the Schwarzschild ansatz is just a choice of coordinates, the results for the gravitational potential are the same as for the ones discussed above; so this ansatz might seem pointless. Nevertheless, we found an interesting link.\\
If we want to plug the isotropic ansatz into the equations of motion, we have to choose an ansatz for the connection. But now we have an advantage in isotropic coordinates; the spatial sections are flat. Isotropic coordinates are thus very similar to cosmological spacetimes with flat spatial sections. We can therefore try to use the coincident gauge for the connection, as it is the natural connection of flat spacetimes in flat coordinates. In spherical coordinates one then obtains the spherical connection, as used at zeroth order perturbation theory in~\ref{ssec:ApproxSolutions}, or in~\cite{Zhao:2021,Lin:2021}, but just in isotropic coordinates with $R$ instead of $r$.\\
Since the connection is fixed, we will not find any new connection hairs, but curiously enough we do find completely consistent metric equations of motions; moreover, the connection equations of motion are identically satisfied. Even better, \eqref{eq:SRNdS} is not a solution to the metric equations of motion, so it might seem that this very simple ansatz gives again new black hole corrections. But we have actually covered this already; if one performs the coordinate transformation~\eqref{eq:IsotropicToSchwarzschild} and applies this to the spherical connection, one obtains precisely the same connection as we had in case (I$^-$). Hence (I$^-$) corresponds to an ``isotropic ansatz'' to the black hole problem in $f(\Q)$ gravity.

\subsection{Exact vacuum solutions beyond GR for $f(\Q) = \Q^\kappa$}\label{ssec:ExactSolutions}
\begin{figure}[H]
    \centering
    \includegraphics[width=1\textwidth]{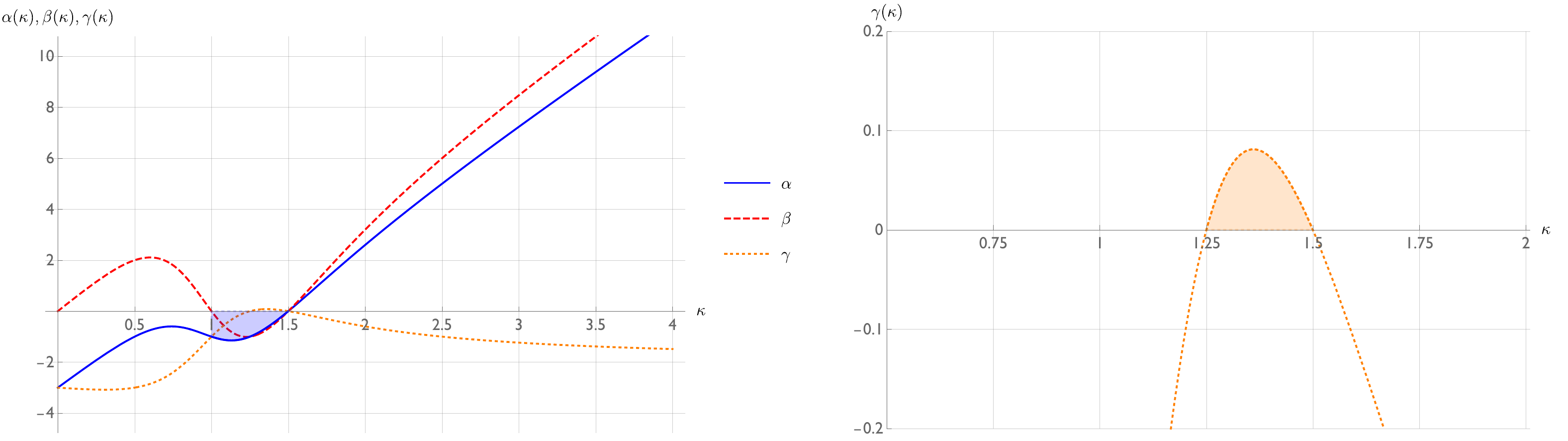}
    \caption{Left panel: The exponents $\alpha$ (solid blue curve), $\beta$ (dashed red curve), and $\gamma$ (dotted orange curve) as functions of $\kappa$. The shaded blue region indicates the range where both, $\alpha$ and $\beta$ are negative. Right panel: The function $\gamma$ is negative except in the shaded orange region which corresponds to the interval $[\frac54,\frac32]$.}
    \label{fig:ExactPowers}
\end{figure}

So far we have only discussed approximate solutions within the framework of perturbation theory and we have seen that $f(\Q)$ admits ``connection hair''. However, it is also possible to derive exact solutions of $f(\Q)$ gravity which go beyond GR and where the connection appears as ``hair'', with a new scale $r_T$ appearing in the metric components. The starting point are again the field equations~\eqref{eq:FEQ2} for option~2 since these equations are simpler and only require the specification of initial conditions, rather than the arbitrary choice of the connection components $c$ and $k$.\\
The main observation we need is that the metric field equations for option~2 imply that
\begin{align}
	g_{tt} g_{rr} = c_1\,\exp\left(2\int\dd r\,\frac{\Gamma^{r}{}_{\theta\theta}+r}{\Gamma^{r}{}_{\theta\theta}}\frac{f''(\Q)}{f'(\Q)}(\partial_r\Q) \right),
\end{align}
as we had seen at the beginning of section~\ref{sec:Solutions}. Performing integrals is in general a daunting task and it is often not possible to compute them analytically. However, if we assume that $\Gamma^{r}{}_{\theta\theta}$ is of the form
\begin{equation}\label{eq:GammaAnsatz}
	\Gamma^{r}{}_{\theta\theta} = -\lambda\, r,
\end{equation}
where $\lambda\in\mathbb R \setminus\{0\}$ is an arbitrary constant, then the integral becomes manageable. First of all, we notice that the choice $\lambda = 1$ would again give us the spherical connection of~\cite{Zhao:2021,Lin:2021}, which simply produces the GR solution for arbitrary~$f$. Hence, we have again a parametrization of $\Gamma^{r}{}_{\theta\theta}$ which allows us to ``deform'' the spherical connection and ``move away'' from the GR solution. Moreover, after inserting the ansatz~\eqref{eq:GammaAnsatz} into the integral, we easily find
\begin{equation}\label{eq:gtt_grr_Relation}
	g_{tt}\, g_{rr} = c_1\,\left(f'(\Q)\right)^{2\frac{\lambda-1}{\lambda}},
\end{equation}
where we have absorbed additional factors in the integration constant $c_1$. We can use this equation to eliminate $g_{rr}$ and we are therefore left with just $g_{tt}$ which needs to be determined. To that end, we use the metric field equation $\M_{tt}$ for the pure vacuum case and we find
\begin{equation}\label{eq:gtt_Sol}
	\partial_r g_{tt} = \frac{c_1\, \lambda}{2(2\lambda-1)}\frac{f'(\Q)^{\frac{\lambda-2}{\lambda}}}{r}\left(f(\Q)\, r^2 + 2\lambda\, f'(\Q)+ 2(2\lambda-1)\,r\,\partial_r\Q\, f''(\Q)\right) - g_{tt} \left(\frac{1}{r}+\frac{\partial_r\Q\, f''(\Q)}{\lambda\, f'(\Q)}\right)
\end{equation}
We have already made use of both metric field equations and we are thus only left with the connection field equation~$\mathcal C_r$. To analyze this equation, it is convenient to first compute the non-metricity scalar $\Q$ and then insert it into $\mathcal C_r$. For the non-metricity scalar we find
\begin{equation}
	\Q = \frac{\lambda -1}{2\lambda-1}\frac{f(\Q)\, r^2 - 2(\lambda-1)\, f'(\Q)}{r^2\, f'(\Q)}.
\end{equation}
At this point we need to make a choice for $f$ in order to continue. We choose\footnote{Unfortunately we could not proceed with the more interesting choice $f(\Q)=\Q+\alpha\,\Q^\kappa$.} $f(\Q) = \Q^\kappa$, where $\kappa\in\mathbb R\setminus\{0\}$. After inserting this ansatz for $f$ into the above expression for $\Q$, we find that $\Q$ can be written as
\begin{equation}
	\Q = -\frac{2\kappa\,(\lambda-1)^2}{1-\lambda + \kappa\,(2\lambda-1)}\frac{1}{r^2}.
\end{equation}
This is a very simple function of $r$ and when inserted into the connection field equation one finds the polynomial equation
\begin{equation}
	(\kappa-1)\left(\kappa(8-14\lambda)+5(\lambda-1)+\kappa^2(8\lambda-4)\right)\left(\frac{\kappa(\lambda-1)^2}{(1-\lambda+\kappa(2\lambda-1))r^2}\right)^\kappa=0\ .
\end{equation}
This equation is solved by
\begin{align}\label{eq:lambdaRelation}
	\lambda &= 1, & \lambda &=\frac{5-8\kappa+4\kappa^2}{5-14\kappa+8\kappa^2}.
\end{align}
The first solution simply produces the spherical connection. The second solution is more interesting, but we also need to assume that $\kappa\notin \{\frac{1}{2}, \frac{5}{4}\}$ in order for this solution to be well-defined. Moreover, it is now easy to check that all field equations, $\M_{tt}$, $\M_{rr}$, $\mathcal C_{r}$, and $\M_{\theta\theta}$, are satisfied. There is therefore no new information we can gather and we can use all results obtained thus far to integrate the equation~\eqref{eq:gtt_Sol}. We find that $g_{tt}$ is given by
\begin{align}\label{eq:gtt}
	g_{tt}= c_1\, r^{\beta} + c_2\, r^{\alpha},
\end{align} 
where $c_2$ is an integration constant and where we have defined
\begin{align}\label{eq:pqC}
	\beta &:= \frac{8(2\kappa-3)(\kappa-1)\kappa}{5+4(\kappa-2)\kappa}, & \alpha &:= \frac{(2\kappa-3)(5+4\kappa(2\kappa-3))}{5+4(\kappa-2)\kappa}.
\end{align}
The solution for $g_{rr}$ then follows from~\eqref{eq:gtt_grr_Relation} and we find
\begin{align}\label{eq:grr2}
	g_{rr} = \underbrace{\left(\frac{8\kappa^2-14\kappa+5}{4\kappa^2-8\kappa+5}\right)^2}_{=: C}\frac{c_1}{c_1 + c_2\, r^{\gamma}},
\end{align}
with $\gamma :=\alpha-\beta$. Observe that if we choose $\kappa = 1$, which corresponds to the choice $f(\Q) = \Q$ and which implies $\beta=0$, $\alpha=-1$, and $C=1$, we obtain
\begin{align}
	g_{tt} &= c_1 +\frac{c_2}{r} &\textsf{and}& & g_{rr} &= \frac{c_1}{g_{tt}}.
\end{align}
In other words: The GR solution is correctly reproduced by the equations~\eqref{eq:gtt}, \eqref{eq:pqC}, and~\eqref{eq:grr2}. This is a reassuring consistency test and we can now try to quantify how the solutions for $\kappa>1$ deviates from the Schwarzschild solution by setting $c_1 \equiv -r_s^{-\gamma}c_2$ and $c_2\equiv -r_T^{-\gamma}$, where $r_s$ is the Schwarzschild radius and $r_T$ is a new gravitational time dilation scale. The $g_{tt}$ and $g_{rr}$ components then read
\begin{align}
    g_{tt} &=-\left(\frac{r}{r_T}\right)^\beta \left(1-\left(\frac{r_s}{r}\right)^{-\gamma}\right) & \text{and} && g_{rr} &= \frac{C}{1-\left(\frac{r_s}{r}\right)^{-\gamma}}.
\end{align}
In order to compare this to the Schwarzschild solution, we need to assume $\gamma<0$. This condition is satisfied for all $\kappa\in\mathbb R\setminus[\frac{5}{4},\frac32]$; note that for $\kappa=3/2$ we have the trivial flat solution $\alpha=\beta=0$. See also Figure~\ref{fig:ExactPowers}, which shows the exponents $\alpha$, $\beta$ and $\gamma$ as functions of $\kappa$. This assumption is also reasonable since it leads to a well-behaved limit of $g_{rr}$ as $r$ goes to infinity:
\begin{equation}
	\lim_{r\to\infty} g_{rr} = C.
\end{equation}
However, the situation for $g_{tt}$ looks quite different. Its limit for $r\to\infty$ is given by
\begin{equation}
	\lim_{r\to\infty} g_{tt} = -\lim_{r\to\infty}\left(\frac{r}{r_T}\right)^\beta,
\end{equation}
This limit does not behave well because it either diverges ($\beta>0$) or it vanishes ($\beta<0$). It is only interesting for $\beta = 0$, which is achieved for $\kappa = 0$, $\kappa = 1$, and $\kappa = \frac32$. But $\kappa = 0$ is not admissible, since then $f(\Q) = 1$, $\kappa = 1$ is simply symmetric teleparallelism, and $\kappa = \frac32$ belongs to the range of excluded $\kappa$ values. Hence, the interpretation of the metric at large radii is difficult unless we are in standard symmetric teleparallelism, where the metric reduces to the Schwarzschild solution.

We conclude this subsection by noting that this solution is only close to Schwarzschild for $\kappa$ very close to unity. Even for $f=\Q^2$ one has $\alpha=2.6$ and $\beta=3.2$, which is far from Schwarzschild; Solar system tests would have revealed such large deviations. These exact solutions are thus physically not relevant. What is interesting, though, is that exact solutions can be found which go beyond GR. Finding physically interesting exact solutions is left as a challenge for future work.

\section{Application to $f(\mathbb T)$ gravity}\label{sec:f(T)}\setcounter{equation}{0}
There is yet another theory of gravity which is closely related to Symmetric Teleparallelism and which together with GR forms the trinity gravitational theories~\cite{BeltranJimenez:2019} -- Metric Teleparallelism (MT).

The underlying geometric framework of MT is closely related to the one of ST studied in this paper and we can easily transfer some of our methods to MT in order to discuss stationary and spherically symmetric solutions of $f(\T)$. Our discussion of $f(\T)$ gravity will be brief in the sense that we will only report the basic results, but we will not repeat the lengthy derivations which are analogous to the ones for ST which have been explained in great detail in this paper.\\
We begin by recalling basic definitions of MT and fixing our notation. Subsequently, we will report the symmetry reduced form of the metric, the connection, and the field equations of MT. Finally, we discuss some approximate solutions to the field equations.

We give first a brief introduction to MT, and then note the stationary and spherically symmetric Ansatz for the MT connection and the equations of motion. Results are discussed last.

\subsection{Metric Teleparallelism and $f(\T)$ gravity}\label{ssec:MTIntroduction}
Let $(\mathcal{M},g_{\mu\nu},\Gamma^\alpha_{\ \mu\nu})$ be a metric-affine geometry, where $\mathcal{M}$ is a four-dimensional manifold,  $g_{\mu\nu}$ denotes the components of the metric tensor of signature $(-,+,+,+)$, and $\Gamma^\alpha{}_{\mu\nu}$ represents an affine connection. The latter is now postulated to be flat and metric-compatible, but with non-trivial torsion. That is, we postulate 
\begin{align}\label{eq:TPPostulates}
    R^\alpha{}_{\beta\mu\nu}&\overset{!}{=} 0 &\text{ and }&& Q_{\alpha\mu\nu}&\overset{!}{=} 0.
\end{align}
As alluded to above, the only non-trivial object in this metric-affine geometry is the torsion tensor, defined by 
\begin{align}
	T^\alpha{}_{\mu\nu}:=2\Gamma^{\alpha}{}_{[\mu\nu]}.
\end{align} 
Notice that the postulates of Symmetric Teleparallelism imply that the connection is completely independent of the metric. This is no longer true in Metric Teleparallelism because the postulate of metric compatibility obviously involves the metric as well as the connection.

To define a MT theory which is equivalent to GR, one can proceed analogously to \ref{sec:SymmetricTeleparallelism}. To that end, we notice that due to the skew-symmetry of the torsion tensor, there are three independent scalars which can be constructed from $T^\alpha{}_{\mu\nu}$. One can then define the following linear combination of these scalars:
\begin{equation}
    \T := -\frac{1}{4}T_{\alpha\mu\nu}T^{\alpha\mu\nu}-\frac{1}{2}T_{\alpha\mu\nu}T^{\mu\alpha\nu}+T_\alpha T^\alpha,
\end{equation}
where we have used the trace $T_\alpha :=T^\mu{}_{\alpha\mu}$ and we refer to $\T$ as the torsion scalar. Analogously to~\ref{sec:SymmetricTeleparallelism}, one can easily derive the following relation
\begin{equation}
    -\T+2\mathcal{D}_\mu T^\mu+\mathcal{R}=0,
\end{equation}
where $\mathcal D_\mu$ denotes the covariant derivative with respect to the Levi-Civita connection and $\mathcal R$ is the Ricci scalar of the Levi-Civita connection. This shows that if we define the action of MT as $\mathcal S[g, \Gamma]:=\int\dd^4 x\,\sqrt{-g}\,\T$, we obtain an action which is equivalent to Einstein-Hilbert, up to a boundary term. Hence, in MT, just as in ST, only the Levi-Civita part of the connection contributes and everything else drops out of the field equations. The connection does not carry any physical degrees of freedom.\\
Let us now consider the non-linear extension of MT to $f(\T)$ gravity. The action defining this theory is simply given by
\begin{equation}\label{eq:TPCovAction}
	\mathcal S[g, \Gamma; \lambda, \rho] := \int_\M \dd^4 x\left(\frac12\sqrt{-g}\,f(\T) + \lambda_\alpha{}^{\beta\mu\nu} R^\alpha{}_{\beta\mu\nu} + \rho^{\alpha\mu\nu} Q_{\alpha\mu\nu}\right),
\end{equation} 
where the tensor densities $\lambda_\alpha^{\ \beta\mu\nu}$ and $\rho^{\alpha\mu\nu}$ act again as Lagrange multiplies which enforce the MT postulates~\eqref{eq:TPPostulates}. The function $f$ is again arbitrary and only subjected to the requirement that $f'(\T):=\frac{d f(\T)}{d\T}\neq 0$. As for $f(\Q)$ gravity, for generic $f$ we can no longer remove the connection from the action by a boundary term. Hence, the connection equations of motion are no longer trivial and the connection can propagate degrees of freedom. By varying~\eqref{eq:TPCovAction} with respect to the metric and the connection leads to the equations of motion of $f(\T)$ gravity~\cite{Jimenez:2018}. These are explicitly given by
\begin{align}\label{eq:MTFieldEqs}
    \M_{\mu\nu}:=(\nabla_\alpha+T_\alpha)[S_{(\mu\nu)}{}^{\alpha}f'(\T)]+f'(\T)t_{\mu\nu}-\frac{1}{2}f(\T)g_{\mu\nu} - T_{\mu\nu} &=0\notag\\
    \mathcal{C}_{\alpha\beta}:=-(\nabla_\mu+T_\mu)\left[\frac{\sqrt{-g}}{2}f'(\T) S_{[\alpha}{}^{\mu}{}_{\beta]} \right]&=0,
\end{align}
where $S_\alpha{}^{\mu\nu}$, the so-called torsion conjugate, and the symmetric tensor $t_{\mu\nu}$ are defined by\footnote{Note that $S_\alpha{}^{\mu\nu}$ is by definition antisymmetric in the last two indices, while --by using the explicit form of $S_\alpha{}^{\mu\nu}$-- one can check that $t_{\mu\nu}$ is indeed symmetric, as required by the symmetry of the metric equations of motion.}
\begin{align}
    S_\alpha{}^{\mu\nu} &:=\frac{\partial \T}{\partial T^\alpha{}_{\mu\nu}}=-\frac{1}{2}T_\alpha{}^{\ \mu\nu}-{T^{[\mu}}_{\alpha}{}^{\nu]}-2\delta_\alpha{}^{[\mu}T^{\nu]}\notag\\
    t_{\mu\nu} &:=\frac{1}{2}S_\mu{}^{\alpha\beta}T_{\nu\alpha\beta}-T^{\alpha\beta}{}_{\mu}S_{\alpha\beta\nu}.
\end{align}
Note the close similarity in the structure of these equations with the ST counterparts, except for the form of the connection equations of motion. One can then rewrite the metric equations of motion in the more useful form
\begin{equation}\label{eq:MTSimpleMetEq}
    f'(\T)G_{\mu\nu}-\frac{1}{2}g_{\mu\nu}(f(\T)-f'(\T)\T)+f''(\T)S_{(\mu\nu)}{}^\alpha\partial_\alpha \T = T_{\mu\nu},
\end{equation}
with $G_{\mu\nu}$ being again the Einstein tensor with respect to the Levi-Civita connection. For $f(\T)= \T+2\Lambda$, this reduces to the Einstein field equations with a cosmological constant.\\

In the next few subsections we will sketch the symmetry reduction of the metric, the connection, and the field equations, and finally discuss some perturbative solution. We will see that there are some similarities with $f(\Q)$.

\subsection{Symmetry reduction of the connection and metric}
Unsurprisingly, the symmetry reduced metric has the same form as in $f(\Q)$, namely, it is given by~\eqref{eq:GeneralMetric}. Moreover, we can apply the same diffeomorphisms which we described in~\ref{ssec:E} in order to bring the metric in the even simpler form~\eqref{eq:SimplerMetric}. This will not spoil the symmetry reduced form of the connection for the same reasons given in~\ref{ssec:E}: The diffeomorphism respects the symmetry of the metric-affine geometry and it can neither create curvature nor non-metricity. Thus, we choose to work with the simple form~\eqref{eq:SimplerMetric} of the metric from now on.

Finding a parametrization of the connection which is compatible with the postulates of MT can be achieved analogously to what we did in the sections~\ref{sec:SymRed}. We start with a general connection $\Gamma^\alpha{}_{\ \mu\nu}$, which has $64$ components, and first apply the conditions for stationarity and spherical symmetry. After that, we use the MT postulates~\eqref{eq:TPPostulates} to further eliminate connection components and knead the connection into the simplest possible form. We only report the main results here.

After implementing the symmetry conditions~\eqref{eq:SymmetryConditionT}, \eqref{eq:SymmetryConditionphi}, \eqref{eq:SymmetryCondition1}, and \eqref{eq:SymmetryCondition2}, we obtained the same connection as in \cite{Hohmann:2019}, but with the additional property of being stationary,  which simply means all connection components are time-independent. 

In order to implement the MT postulates, one can again split the arising equations into algebraic\footnote{Notice that the equations $Q_{\alpha\mu\nu}=0$ are purely algebraic and even linear in the connection, thus leading to unique solutions.} and differential equations for the connection. By plugging the solutions of the algebraic equations into the differential equations, one can ultimately find the following form of the connection:
\begin{align}\label{eq:MTConnection}
    \nonumber {\Gamma^t}_{rt}&=\frac{\partial_r g_{tt}}{2g_{tt}}\ ,\ {\Gamma^t}_{rr}=\pm\frac{r({\Gamma^r}_{\theta\theta}{}^2+{{\bar\Gamma}^r}_{ \theta\phi}{}^2)\partial_r g_{rr}-2g_{rr}\left({{\bar\Gamma}^r}_{\theta\phi}{}^2+{\Gamma^r}_{\theta\theta}({\Gamma^r}_{\theta\theta}-r\,\partial_r{{\Gamma^r}_{\theta\theta}})-r{{\bar\Gamma}^r}_{ \theta\phi}\partial_r{{{\bar\Gamma}^r}_{\ \theta\phi}}\right)}{2r\sqrt{g_{tt}}\sqrt{{\Gamma^r}_{\theta\theta}{}^2+{{\bar\Gamma}^r}_{\ \theta\phi}{}^2}\sqrt{r^2-g_{rr}({\Gamma^r}_{\theta\theta}{}^2+{{\bar\Gamma}^r}_{\ \theta\phi}{}^2)}}\\
    \nonumber {\Gamma^t}_{\theta\theta}&=\pm \frac{{\Gamma^r}_{\theta\theta}\sqrt{r^2-g_{rr}({\Gamma^r}_{\theta\theta}{}^2+{{\bar\Gamma}^r}_{\ \theta\phi}{}^2)}}{\sqrt{g_{tt}}\sqrt{{\Gamma^r}_{\theta\theta}{}^2+{{\bar\Gamma}^r}_{\ \theta\phi}{}^2}}\ ,\ {\Gamma^t}_{\phi\phi}=\sin^2\theta\,{\Gamma^t}_{\theta\theta}\\
    \nonumber {\Gamma^t}_{\theta\phi}&=-{\Gamma^t}_{\ \phi\theta}= \frac{{\Gamma^r}_{\theta\phi}}{{\Gamma^r}_{\theta\theta}}\sin\theta\,{\Gamma^t}_{\theta\theta}\\
    \nonumber {\Gamma^r}_{rt}&=-\frac{g_{tt}}{g_{rr}}{\Gamma^t}_{rt}\ ,\ {\Gamma^r}_{rr}=\frac{\partial_r g_{rr}}{2g_{rr}}\ ,\ {\Gamma^r}_{\phi\phi}=\sin^2\theta\,{\Gamma^r}_{\theta\theta}\ ,\ {\Gamma^r}_{\theta\phi}=-{\Gamma^r}_{\phi\theta}=\sin\theta\,{\bar\Gamma^r}_{\ \theta\phi}\\
    \nonumber {\Gamma^\theta}_{\theta t}&=-\frac{g_{tt}}{r^2}{\Gamma^t}_{\theta\theta}\ ,\ {\Gamma^\theta}_{\phi t}=-\frac{g_{tt}}{r^2}{\Gamma^t}_{\phi\theta}\ ,\ {\Gamma^\theta}_{\theta r}=-\frac{g_{rr}{\Gamma^r}_{\theta\theta}}{r^2}\ ,\ {\Gamma^\theta}_{\phi r}=\frac{g_{rr}\sin(\theta){{\bar\Gamma}^r}_{\ \theta\phi}}{r^2}\\
    \nonumber {\Gamma^\theta}_{r\theta}&= \frac{1}{r}\ ,\ {\Gamma^\theta}_{r\phi}=\sin\theta\,\frac{{\Gamma^r}_{\theta\theta}\partial_r{{\bar\Gamma}^r}_{\ \theta\phi} - \partial_r{\Gamma^r}_{\theta\theta}{{\bar\Gamma}^r}_{\theta\phi}}{{\Gamma^r}_{\theta\theta}{}^2+{{\bar\Gamma}^r}_{\theta\phi}{}^2}\ ,\ {\Gamma^\theta}_{\phi\phi}=-\cos\theta\,\sin\theta\\
    \nonumber {\Gamma^\phi}_{\theta t}&=-\frac{g_{tt}{{\bar\Gamma}^r}_{\ \theta\phi}}{r^2\sin\theta\,{\Gamma^r}_{\theta\theta}}{\Gamma^t}_{\theta\theta}\ ,\ {\Gamma^\phi}_{\phi t}=-\frac{g_{tt}}{r^2}\Gamma^t_{\ \theta\theta}\ ,\ {\Gamma^\phi}_{r\theta}=-\frac{1}{\sin^2\theta}{\Gamma^\theta}_{r\phi}\\
     {\Gamma^\phi}_{\theta r}&=-\frac{g_{rr}{{\bar\Gamma}^r}_{\theta\phi}}{r^2\sin\theta}\ ,\ {\Gamma^\phi}_{\phi r}=-\frac{g_{rr}{\Gamma^r}_{\theta\theta}}{r^2}\ ,\ {\Gamma^{\phi}}_{r\phi}=\frac{1}{r}\ ,\ {\Gamma^\phi}_{\theta\phi}={\Gamma^\phi}_{\phi\theta}=\cot\theta.
\end{align}
with all other components vanishing. This result agrees with the one reported in~\cite{Hohmann:2019nat}. As alluded to before, the connection and the metric are not completely independent, because of $Q_{\alpha\mu\nu} = 0$, and that is why the metric appears in the above expressions for the connection components. Moreover, the only free connection components are $\Gamma^r{}_{\theta\theta}$ and ${\bar\Gamma}^r{}_{\theta\phi}$,  where the latter is defined in the fourth line above, both arbitrary functions of $r$ only. In addition to being free in specifying these components, we can also freely choose the sign $\pm$ which arises from taking the square roots of the metric components. This sign has to be chosen such that it is the same for all components, either always the upper or the lower one. Thus, we obtain two distinct parametrizations for the connection which we denote by $\Gamma^\pm$.

It is not surprising that the connection now has less free components than in $f(\Q)$ gravity: While the connections both have to fulfil the symmetry and flatness conditions, in $f(\Q)$ it has to fulfil $T^\alpha_{\ \mu\nu}=0$, which are $24$ equations, while in $f(\T)$ we must have $Q_{\alpha\mu\nu}=0$, which are $40$ equations. The connection in $f(\T)$ is thus more constraint. Also, note that the connection components are not continuous at $\Gamma^r{}_{\theta\theta}={\bar\Gamma}^r{}_{\theta\phi}=0$. For instance, in the expression for $\Gamma^t{}_{\theta\theta}$, when taking such a limit one has to take care with the order of the two limits $\Gamma^r{}_{\theta\theta}\to 0$ and ${\bar\Gamma}^r{}_{\theta\phi}\to 0$.

\subsection{Symmetry reduced field equations for the metric and the connection}
After having worked out the symmetry reduced form of the connection, and having established that there are two distinct parametrizations corresponding to a choice of sign, we can now consider the symmetry reduced field equations. By plugging~\eqref{eq:SimplerMetric} and~\eqref{eq:MTConnection} into the field equations~\eqref{eq:MTFieldEqs}, one finds that they have the following structure
\begin{align}
&\text{Structure of metric field equations: }
	& &\begin{pmatrix}
		\M_{tt} & \M_{tr} & 0 & 0\\
		\M_{tr} & \M_{rr} & 0 & 0\\
		0 & 0 & \M_{\theta\theta} & 0\\
		0 & 0 & 0 & \M_{\theta\theta}\,\sin^2\theta
	\end{pmatrix} \notag\\
&\text{Structure of connection field equations: }
	&& \partial_r\T\, f''(\T) \begin{pmatrix}
		0 & \mathcal{C}_{tr} & 0 & 0 \\
		-\mathcal{C}_{tr} & 0 & 0 & 0 \\
		0 & 0 & 0 & -\sin\theta\,\mathcal{C}_{\theta\phi} \\
		0 & 0 & \sin\theta\,\mathcal{C}_{\theta\phi} & 0
	\end{pmatrix},
\end{align}
and we remark that the torsion scalar is explicitly given by
\begin{equation}\label{eq:TScalarExpl}
    \T = \frac{2}{r^2}\frac{(r+g_{rr}{\Gamma^r}_{\theta\theta})\partial_r g_{tt}+g_{tt}(1+g_{rr}+{\Gamma^r}_{\theta\theta}\partial_r g_{rr} + 2g_{rr}{\partial_r\Gamma^r}_{\theta\theta})}{g_{rr}\,g_{tt}}
\end{equation}
for both choices of sign in $\Gamma^\pm$. Observe that the metric field equations have the same structure as for $f(\Q)$ gravity. Moreover, one finds that the sign of $\Gamma^\pm$ not only has absolutely no effect on the structure of the connection field equations; it does not enter these equations at all! No matter which sign we choose, we obtain exactly the same field equations. Hence, we can drop the distinction between the $+$ and $-$ choice. \\
\\
The connection equations of motion have a very simple form, and are given by
\begin{align}
 \partial_r\T\, f''(\T) \,  \mathcal{C}_{tr}&=\M_{tr}=0\ ,\\
  \partial_r\T\, f''(\T)\,  \mathcal{C}_{\theta\phi}&= \partial_r\T\, f''(\T)\,{\bar\Gamma}^r_{\ \theta\phi}=0.
\end{align}
In particular, we note that no derivatives of the connection --apart from the $\partial_r\mathbb{T}$ term-- appear in the connection equations of motion. The connection is thus \textit{not dynamical}, and the connection equations of motion are mere constraints.\\
\\
Let us now look more closely at the off-diagonal metric field equation, which reads
\begin{equation}
    \M_{tr}=-\frac{2\sqrt{g_{tt}}}{r^2}\Gamma^r{}_{\theta\theta}\sqrt{\frac{r^2-g_{rr}(\Gamma^r{}_{\theta\theta}{}^2+{\bar\Gamma}^r{}_{\theta\phi}{}^2)}{\Gamma^r{}_{\theta\theta}{}^2+{\bar\Gamma}^r{}_{\theta\phi}{}^2}}\,\partial_r\T \,f''(\T)=0\ .
\end{equation}
This equation is again structurally similar to the equation we obtained in $f(\Q)$ gravity, To solve it, we either have $\T_\text{sol} = $ const., or $f''(\T_\text{sol}) = 0$, or we end up with a constraint equation for the connection. 

Just as in $f(\Q)$ gravity, the first two options will immediately lead to trivially satisfied connection field equations. Using a similar argument as in~\ref{sec:SymRedFieldEq}, one can show that $f''(\T_\text{sol})=0$ immediately implies $f(\T) = a\,\T+b$ and hence one can only obtain the Schwarzschild-deSitter-Nordstr\"{o}m solution of GR. Also, if $\T_\text{sol}$ is a constant, we find from~\eqref{eq:MTSimpleMetEq} the GR field equations
\begin{equation}
	G_{\mu\nu} + \Lambda_\textsf{eff}\,g_{\mu\nu} = \bar{T}_{\mu\nu},
\end{equation}
where $G_{\mu\nu}$ is the standard Einstein tensor with respect to the Levi-Civita connection and where we have defined
\begin{align}
	\Lambda_\textsf{eff} &:= \frac{1}{2}\frac{f(\T_\textsf{sol})-f'(\T_\textsf{sol})\T_\textsf{sol}}{f'(\T_\textsf{sol})}\notag\\
	\bar{T}_{\mu\nu} &:= \frac{1}{f'(\T_\textsf{sol})} T_{\mu\nu}.
\end{align}
This is exactly the same results obtained in~\ref{sec:SymRedFieldEq} for $f(\Q)$ gravity! Notice that in the case of $f(\T)$ gravity, we can immediately establish that solutions with $\T_\text{sol} = $ const. are not an empty set. To that end, assume $\T_\text{sol} = $const. and solve~\eqref{eq:TScalarExpl} for ${\Gamma^r}_{\theta\theta}$:
\begin{equation}
    {\Gamma^r}_{\theta\theta}=-\frac{1}{\sqrt{g_{tt}\,g_{rr}}}\left(c+\int^r \dd \rho\ \frac{g_{tt}(2+\rho(2+\mathbb{T}_\sol\, \rho^2))+2\rho\, \partial_\rho g_{tt}}{4\sqrt{g_{tt}\,g_{rr}}}\right).
\end{equation}
Hence, the connection can, in principle, be fixed such that $\T$ becomes a constant.\\

Finally, the only choice left for solving the off-diagonal metric field equation which does not force GR solutions on us is the constraint equation
\begin{equation}\label{eq:ConstrF(T)}
	\Gamma^r{}_{\theta\theta}\sqrt{\frac{r^2-g_{rr}(\Gamma^r{}_{\theta\theta}{}^2+{\bar\Gamma}^r{}_{\theta\phi}{}^2)}{\Gamma^r{}_{\theta\theta}{}^2+{\bar\Gamma}^r{}_{\theta\phi}{}^2}} = 0.
\end{equation}
Once $\M_{tr}$ is fulfilled $\mathcal{C}_{tr}$ is also fulfilled. Since the connection field equation $\mathcal C_{\theta\phi} = 0$ uniquely implies $\bar{\Gamma}^{r}{}_{\theta\phi} = 0$, we find that the constraint equation~\eqref{eq:ConstrF(T)} has the two solutions
\begin{equation}
	\Gamma^{r}{}_{\theta\theta} = \pm\frac{r}{\sqrt{g_{rr}}}.
\end{equation}
With this, the off-diagonal metric field equation and all connection field equations are satisfied and we find that the connection~\eqref{eq:MTConnection} reduces to
 \begin{align}
        \nonumber \Gamma^t_{\ rt}&=\frac{\partial_r g_{tt}}{2g_{tt}}\ ,\ \Gamma^r_{\ rr}=\frac{\partial_r g_{rr}}{2g_{rr}}\ ,\ \Gamma^r_{\ \theta\theta}=\pm\frac{r}{\sqrt{g_{rr}}}\ ,\ \Gamma^r_{\ \phi\phi}=\Gamma^r_{\ \theta\theta}\sin(\theta)^2\ ,\\
        \nonumber \Gamma^\theta_{\ r\theta}&=\frac{1}{r}\ ,\ \Gamma^\theta_{\ \theta r}=\mp \frac{\sqrt{g_{rr}}}{r}\ ,\ \Gamma^\theta_{\ \phi\phi}=-\cos(\theta)\sin(\theta)\ ,\\
        \Gamma^\phi_{\ r\phi}&=\frac{1}{r}\ ,\ \Gamma^\phi_{\ \phi r}=\mp\frac{\sqrt{g_{rr}}}{r}\ ,\ \Gamma^\phi_{\ \theta\phi}=\Gamma^\phi_{\ \phi\theta}=\cot(\theta)\ ,
    \end{align}
with all other components vanishing. We call these two solutions (III$^\pm$). The sign ambiguity here comes only from the choice of $\Gamma^r_{\ \theta\theta}$; the connection is the same for both $\Gamma^\pm$ for this connection choice.

Note especially that the two choices (III$^\pm$) are in complete analogy to the choices of connection for $f(\Q)$, namely (I$^{\pm}$), in \ref{ssec:VB}. The important difference --apart from the obvious differences in the forms of the whole connections-- is of course that in $f(\T)$ the connection is not dynamical. Its equations of motion completely fix the connection from the start, and leave no room for a dynamical evolution. We can still have beyond GR solutions for (III$^{\pm}$), but they will not involve any connection hairs in this sense.\\
\\
We now have only three equations of motion left, namely $\M_{tt},\ \M_{rr}$, and $\M_{\theta\theta}$. One can check analogously to \ref{ssec:RemainingEqs} that $\M_{\theta\theta}$ follows from $\M_{tt}$ and $\M_{rr}$ so we have only two equations left for the two metric components $g_{tt}$ and $g_{rr}$.\\
The analogy between $f(\T)$ and $f(\Q)$ gravity for stationary and spherically symmetric spacetimes comes to its conclusion when one notes that for each case (III$^{\pm}$) the \textit{exact} equations of motion are identical for any $f$ with those from $f(\Q)$ gravity for the cases (I$^{\pm}$); in particular we have $\T=\Q$. These $f(\T)$ black hole solutions are thus merely a subset of $f(\Q)$ solutions, where the connection is fixed and given by (I$^{\pm}$). This is the main result of this section. As we have discussed the equations and their (approximate) solutions already in \ref{ssec:VB}, we are done with these cases. \\

Note especially that the (approximate) beyond GR solutions found in \cite{DeBenedictis:2016aze} are the same as we found for the case (III$^-$). The solution of \cite{Ruggiero:2015oka} corresponds to the case (III$^+$). In these references the $f(\T)$ theory was discussed using tetrads instead of the full connection, but one can check that the connections that were constructed there are precisely the same as the ones we derived for (III$^\pm$).

\section{Conclusion}\label{sec:Conclusion}\setcounter{equation}{0}
In this paper we have systematically derived and studied symmetry reduced field equations for $f(\Q)$ gravity and we have sketched how a similar approach can be applied to $f(\T)$ gravity. We began our analysis by performing a systematic symmetry reduction of the metric affine geometry described by $(\mathcal M, g_{\mu\nu}, \Gamma^{\alpha}{}_{\mu\nu})$. The main results, which have been extensively discussed in section~\ref{sec:SymRed}, are the following: (a) There are two classes of parametrizations for the connection which guarantee that the connection is stationary, spherically symmetric, torsionless, and flat, as required by the postulates of Symmetric Teleparallelism (cf. subsections~\ref{ssec:Set1} and~\ref{ssec:set2}); (b) the connection in coincident gauge fails to be spherically symmetric, which in part explains why~\cite{Zhao:2021, Lin:2021} were not able to find beyond-GR solutions in $f(\Q)$ which describe black holes; (c) the first parametrization class of the connection (aka solution set 1) can be obtained from the second class (aka solution set 2) by a double scaling limit (see subsection~\ref{ssec:D2}); (d) the metric can be brought into a simple diagonal form, which is parametrized by only two arbitrary functions of $r$, without spoiling the structure of the solution sets for the connection, as explained in~\ref{ssec:E}. Hence, we have constructed the simplest, and yet most general metric-affine geometry which is stationary, spherically symmetric, torsionless, and flat.

In section~\ref{sec:SymRedFieldEq} we have discussed the implications of the simple form of the metric and the two parametrization classes of the connection for the field equations of $f(\Q)$ gravity. In subsection~\ref{ssec:SymRedFieldEq2} we have formulated precise conditions under which $f(\Q)$ either reduces to Symmetric Teleparallelism, gives rise to GR solutions for generic choices of $f$, or produces beyond-GR solutions. In particular, this subsection fully explains why~\cite{Zhao:2021, Lin:2021} were not able to find any beyond-GR solutions.

Additionally, we have shown in subsection~\ref{ssec:SymRedFieldEq1} that solution set 1, while attractive because of its simplicity, is not viable when looking for beyond-GR solutions. In the subsections~\ref{ssec:VB} and~\ref{ssec:AdditionalConstraints} we have extensively discussed the self-consistency of the field equations, the number of degrees of freedom they propagate, the initial data which needs to be specified, and additional constraints on the connection which can appear. We have seen that the connection becomes dynamical, in stark contrast with Symmetric Teleparallelism, where the connection is unphysical, or that it can be completely fixed by additional constraints. The latter option leads nevertheless to beyond-GR solutions for the metric.

In section~\ref{sec:Solutions} we finally constructed explicit beyond-GR solutions. In the subsections~\ref{ssec:ApproxSolutions}, \ref{ssec:ApproxSolutions2}, and~\ref{ssec:VB} we used a perturbative approach to construct solutions for the pure vacuum as well as the electro-vacuum case and non-zero cosmological constant case for the ansatz $f(\Q) = \Q + \alpha\, \Q^2$, where $\alpha$ is assumed to be small. We have done so for a dynamical connection as well as for one which is fixed by the additional constraint described in~\ref{ssec:AdditionalConstraints}. 

In subsection~\ref{ssec:ExactSolutions} we even succeeded in finding exact solutions which go beyond GR. These solutions ultimately turned out to have undesirable properties, which make them physically unattractive. But it is nevertheless interesting that exact solutions can be found, given the complexity of the $f(\Q)$ field equations.

Finally, in section~\ref{sec:f(T)} we sketched how the same methods which have been described in detail for $f(\Q)$ can be applied to $f(\T)$ in order to perform a systematic symmetry reduction of the metric, the connection, and the field equations. There are many (perhaps surprising) structural similarities between $f(\Q)$ and $f(\T)$ which facilitate the analysis of $f(\T)$ gravity. Moreover, we have also reported some beyond-GR solutions for $f(\T)$.

In conclusion, we have succeeded in showing the consistency of the symmetry reduced field equations of both, $f(\Q)$ and $f(\T)$ gravity, we have formulated precise criteria under which beyond-GR solutions can exist, and we have shown that the GR solutions can exist for arbitrary choices of $f$. Moreover, we have discussed a few perturbative beyond-GR solutions to the $f(\Q)$ and $f(\T)$  field equations. Whether these solutions are stable or whether they lead to instabilities is beyond the scope of the current analysis and will be left for future work. We have also not discussed the question of formation processes in the context of $f(\Q)$ or $f(\T)$ gravity. It would be interesting to understand whether a realistic formation process could give rise to one of the solutions discussed here or whether it leads to beyond-GR solutions at all. This question is also left for future work.

Finally, we note that a similar symmetry reduction analysis to the one carried out here can be performed for cosmological models. A detailed discussion will be given elsewhere.

\section*{Acknowledgements}
LH is supported by funding from the European Research Council (ERC) under the European Unions Horizon 2020 research and innovation programme grant agreement No 801781 and by the Swiss National Science Foundation grant 179740.

\newpage

\appendix
\section{Connection transformation when diagonalizing the metric}
\label{app:A}
For the sake of eliminating doubt about whether the diffeomorphism described in~\ref{sec:SymRed} preserves the structure of the solution sets of the connection, we explicitly determine the transformation of the connection components. The untransformed components are denoted by $\Gamma^\alpha{}_{\mu\nu}$, the components after having applied $\phi_2$ are $\bar\Gamma^{\alpha}{}_{\mu\nu}$, and after applying $\phi_1$ we call them $\bb\Gamma^\alpha{}_{\mu\nu}$. The same for the metric components $g_{\mu\nu}$. The order in which we apply $\phi_1$ and $\phi_2$ is irrelevant. We then find that the components transform as
\begin{align}
        \overline c&=c\notag\\
        \overline k&=k\notag\\
        {\overline \Gamma^t}_{rr}&={\Gamma^t}_{rr}-\frac{1}{g_{tt}{}^3\,{\Gamma^r}_{\theta\theta}}\left[c(k-2c)g_{tr}{}^3\, {\Gamma^r}_{\theta\theta}{}^2+g_{tr}{}^2\, g_{tt}\, {\Gamma^r}_{\theta\theta}\left(3c-k+3c(2c-k){\Gamma^t}_{\theta\theta}\right)-\right.\notag\\
        &\left.-g_{tt}{}^2\, \partial_r g_{tr}\,{\Gamma^r}_{\theta\theta}+g_{tt}g_{tr}\left(g_{tt}{\Gamma^r}_{rr}{\Gamma^r}_{\theta\theta}-2(2c-k){\Gamma^t}_{\theta\theta}(1+c{\Gamma^t}_{\theta\theta})+\partial_r g_{tt}{\Gamma^r}_{\theta\theta}\right)\right]\notag\\
        {\overline\Gamma^t}_{\theta\theta}&={\Gamma^t}_{\theta\theta}-\frac{g_{tr}{\Gamma^r}_{\theta\theta}}{g_{tt}}\notag\\
        {\overline\Gamma^r}_{rr}&={\Gamma^r}_{rr}+\frac{c(k-2c)g_{tr}{}^2{\Gamma^r}_{\theta\theta}}{g_{tt}{}^2}+\frac{2cg_{tr}\left(1+(2c-k){\Gamma^t}_{\theta\theta})\right)}{g_{tt}}\notag\\
        {\overline\Gamma^r}_{\theta\theta}&={\Gamma^r}_{\theta\theta},
\end{align}
and
\begin{align}
        \bb c&=\overline c\notag\\
        \bb k&=\overline k\notag\\
        {\bb\Gamma^t}_{rr}&=\frac{{\overline\Gamma^t}_{rr}}{\partial_r\overline g_{\theta\theta}{}^2}\notag\\
        {\bb\Gamma^t}_{\theta\theta}&={\overline\Gamma^t}_{\theta\theta}\notag\\
        {\bb\Gamma^r}_{rr}&=\frac{{\overline \Gamma^r}_{rr}\partial_r\overline g_{\theta\theta}-\partial^2_r \overline g_{\theta\theta}}{\partial_r\overline g_{\theta\theta}{}^2}\notag\\
        {\bb\Gamma^r}_{\theta\theta}&={\overline\Gamma^r}_{\theta\theta}\partial_r \overline g_{\theta\theta}.
\end{align}
All other components are zero. The derivative relations coming from the flatness condition remain, as they come from the coordinate invariant condition $R^\alpha{}_{\beta\mu\nu}=0$. The constants are thus unchanged, and the diffeomorphisms can be absorbed in the arbitrary components $\Gamma^t{}_{\theta\theta}$ and $\Gamma^r{}_{\theta\theta}$. The structure is preserved, as expected.

\section{Approximate solutions for $f(\Q)=\Q+\alpha\,\Q^\kappa$}
\label{app:B}
The perturbative solutions for $f(\Q)=\Q+\alpha\,\Q^2$ derived in subsection~\ref{ssec:ApproxSolutions} can be generalize to the ansatz $f(\Q)=\Q+\alpha\,\Q^\kappa$ for an integer $\kappa$ which satisfies $\kappa\geq 2$. The solutions are derived in complete analogy to those for $f(\Q)=\Q+\alpha\,\Q^2$, so we only report the final results for the metric components here.

\subsection{Connection hair solutions}
This is the generalization of \ref{ssec:ApproxSolutions}, which we explicitly did for $\kappa=3,4$ for vacuum. It turns out one has to go to perturbation order $\kappa$ in $\alpha$ in the metric, at which the first order correction of the connection $\Gamma^r_{\ \theta\theta}=-r+\alpha\gamma^{(1)}$ enters in the form
\begin{align}
    g_{tt}&=-\left(1-\frac{2M_\text{ren}}{r}\right)+\alpha^\kappa\frac{\mu}{r}\ln\left(\frac{r}{r^*}\right)\ ,\\
    g_{rr}&=-\frac{1}{g_{tt}}\ .
\end{align}
$\mu$ and $r^*$ are new scales coming from the connection integration constants in $\gamma^{(1)}$ and $M_\text{ren}$ is the renormalized mass. We suspect that this formula holds for all integer $\kappa\geq 2$.

\subsection{Constraint solutions}
This is the generalization of (I$^\pm$) of \ref{ssec:VB}, which we explicitly did for $\kappa=3,4,5,10$ for vacuum. It turns out that we only need to go to first order metric perturbations. We then have found the formulas
\begin{align}
    g_{tt}&=-1+\frac{2M_\text{ren}}{r}+\alpha\frac{2^{3\kappa-1}}{(2\kappa-3)r^{2\kappa-2}}+\mathcal{O}(r^{-2\kappa+1})\ ,\\
    -\frac{1}{g_{rr}}&=-1+\frac{2M_\text{ren}}{r}+\alpha\frac{(2\kappa-1)(\kappa-1)2^{3\kappa-1}}{(2\kappa-3)r^{2\kappa-2}}+\mathcal{O}(r^{-2\kappa+1})
\end{align}
for (I$^+$), and
\begin{align}
    g_{tt}&=-1+\frac{2M_\text{ren}}{r}+\alpha\frac{(-1)^\kappa 2\kappa (2M_\text{ren})^{2\kappa-}}{2^\kappa (4\kappa-3)r^{4\kappa-3}}+\mathcal{O}(r^{-4\kappa+2})\ ,\\
    -\frac{1}{g_{rr}}&=-1+\frac{2M_\text{ren}}{r}+\alpha\frac{(-1)^\kappa 2\kappa (2M_\text{ren})^{2\kappa-1}}{2^\kappa r^{4\kappa-3}}+\mathcal{O}(r^{-4\kappa+2})
\end{align}
for (I$^-$). We again suspect these formulas to hold for all integer $\kappa\geq 2$.

\bibliographystyle{utcaps}
\bibliography{Bibliography}

\end{document}